\documentclass{emulateapj}
\usepackage[varg]{txfonts}
\usepackage{booktabs}
\newcommand{\pcc}{\ensuremath{\mbox{cm$^{-3}$}}}

\let\url\relax
\usepackage[hypertex]{hyperref}
\usepackage[all]{hypcap}
\makeatletter
\renewcommand{\fps@figure}{tbp}
\makeatother
\setkeys{Gin}{width=\linewidth,height=0.9\textheight,keepaspectratio=true}
\shorttitle{Hydrodynamics of cometary HII regions}
\shortauthors{Arthur \& Hoare}
\begin{document}
\title{ Hydrodynamics of Cometary Compact HII Regions}
\author{ S. Jane Arthur\altaffilmark{1}}
\affil{Centro de Radioastronom\'{\i}a y Astrof\'{\i}sica, UNAM,
  Campus Morelia, Apartado Postal 3-72, 58090 Morelia, M\'exico}
\email{j.arthur@astrosmo.unam.mx}
\and
\author{M. G. Hoare}
\affil{School of Physics and Astronomy, University of
Leeds, LS2 9JT, UK}
\email{mgh@ast.leeds.ac.uk}
\altaffiltext{1}{Work partially carried out while on sabbatical leave at the School of Physics and
Astronomy, University of Leeds, UK}

\begin{abstract}
We present numerical radiation-hydrodynamic simulations of cometary
\ion{H}{2} regions for a number of champagne flow and bowshock models. For the
champagne flow models we study smooth density distributions with both
steep and shallow gradients. We also consider cases where the ionizing
star has a strong stellar wind, and cases in which the star
additionally has a proper motion within the ambient density
gradient. We find that our champagne flow plus stellar wind models
have limb-brightened morphologies and kinematics which can see the
line-of-sight velocities change sign twice between the head and tail
of the cometary \ion{H}{2} region, with respect to the rest frame
velocity. Our bowshock models show that pressure gradients across and
within the shell are very important for the dynamics, and that simple
analytic models assuming thin shells in ram pressure balance are
wholly inadequate for describing the shape and kinematics of these
objects at early times in their evolution. The dynamics of the gas
behind the shock in the neutral material ahead of the ionization front
in both champagne flow and bowshock type cometary \ion{H}{2} regions
is also discussed. We present simulated emission-measure maps and
long-slit spectra of our results. Our numerical models are not
tailored to any particular object but comparison with observations
from the literature shows that, in particular, the models combining
density gradients and stellar winds are able to account for both the
morphology and general radial velocity behavior of several observed cometary
\ion{H}{2} regions, such as the well-studied object G29.96$-$0.02.
\end{abstract}

\keywords{\ion{H}{2} regions --- ISM: kinematics and dynamics ---
shock waves --- stars:formation --- stars: winds}

\section{Introduction}
\label{sec:Intro}
Massive stars are born in dense molecular clouds and modify their
environment by the action of their ionizing photons and stellar
winds. In a uniform medium a spherical \ion{H}{2} region, or
Str\"omgren sphere, will form \citep{1939ApJ....89..526S}, which then
expands due to the overpressure of the photoionized gas with respect to
its surroundings \citep{1954BAN....12..187K}. The earliest stages in
the lives of high mass stars are generally invisible at optical
wavelengths, since the star and its photoionized region remain
embedded in the molecular cloud. These clouds are, however, optically
thin to radio frequencies and infrared radiation and observations at
these wavelengths have shown that uniform, spherical \ion{H}{2}
regions are the exception.

The survey at radio frequencies by \citet{1989ApJS...69..831W}
developed \textit{IRAS} color criteria for embedded compact and
ultracompact \ion{H}{2} regions, which were used in later surveys
\citep{1994ApJS...91..659K}. These surveys show that
around 16\% of the embedded \ion{H}{2} regions have a cometary (i.e.,
head-tail) morphology, while an \textit{IRAS}-selected methanol maser
survey shows that 30\% of ultracompact \ion{H}{2} regions with methanol masers
are cometaries \citep{1998MNRAS.301..640W}. Around 14\% of much more
evolved, optically visible
\ion{H}{2} regions, such as the Orion Nebula, also demonstrates
cometary shapes at radio frequencies \citep{1993ApJS...86..475F}. In
fact, the Orion Nebula gives a clue as to the possible origin of the
morphology of these objects since it is the archetype of the so-called
``blister''
\ion{H}{2} regions identified by \citet{1978A&A....70..769I}. Its
close proximity and availability for study at all wavelengths show us
that the brightest part of the nebula corresponds to the ionization
front on the surface of a molecular cloud, from which material is
being photoevaporated. In the direction away from the cloud the nebula
is open and ionized gas flows out into the diffuse intercloud medium.

The hydrodynamics of ``blister'' \ion{H}{2} regions was modeled by 
Tenorio-Tagle and collaborators in a series of papers \citep{1979A&A....71...59T,
1979ApJ...233...85B, 1981A&A....98...85B, 1983A&A...127..313Y}, who
renamed them ``champagne'' flows. Later,
the effects of including a stellar wind in this scenario were studied
by \citet{1997A&A...326.1195C}. In both of these studies an \ion{H}{2}
region evolving in a uniform, dense cloud breaks out into the diffuse
intercloud medium once the ionization front reaches the edge of the
cloud. Ionized gas accelerates out of the \ion{H}{2} region reaching velocities
up to $\sim 30$~km~s$^{-1}$.

For \ion{H}{2} regions still embedded in their natal clouds, the
classification ``compact'' is applied to those whose size scales are
in the range 0.1--0.3~pc and ``ultracompact'' to those whose sizes are
$< 0.1$~pc. It is thought that these regions, because of their small
size,  represent early stages in the lifetime of the star ($<
10^5$~yrs) before the \ion{H}{2} region has expanded out of the
cloud. If a compact \ion{H}{2} region is still wholly contained within
a molecular cloud, a ``blister'' scenario cannot strictly be
applied since the ionized region has not reached the edge of the
cloud. Instead, it could be envisaged that the champagne flow  occurs
due to a density gradient within the cloud. (The density gradient
could be a relic of the cloud-collapse process that produced the
massive star.) \citet{1996ApJ...469..171G} studied the morphologies
that arise when ionization fronts form in stratified media and
spherical density distributions, and the
instabilities to which they are subject when cooling and stellar winds
are included. The hydrodynamics of photoevaporated champagne flows in density
gradients was investigated by \citet{2005ApJ...627..813H}, who found
that long-lived quasi-stationary flows are set up in which the ionized
flow is steady in the frame of the ionization front.

Observational studies of molecular clouds show that they are centrally
condensed, with moderate-density envelopes ($10^3$--$10^5$~cm$^{-3}$)
surrounding high-density cores ($10^7$~cm$^{-3}$), suggesting strong
density gradients within the clouds
\citep{2000ApJ...536..393H}. Molecular line studies of dark clouds
show that the density gradients can be approximated by power-laws
$\rho(r) \propto r^{-\alpha}$, with exponents ranging from 1 to about
3 \citep{1985ApJ...297..436A}. Near-infrared dust extinction studies
of dark clouds show similar results
\citep[e.g.,][]{2005ApJ...629..276T}. Studies of the extended sub-millimeter dust
emission around ultracompact \ion{H}{2} regions also display power-law density
structures, with the exponent varying between 1.25 and 2.25
\citep{2003A&A...409..589H}. In these studies, $\alpha$ equal to 2
is the most commonly found value for the power-law exponent, which
corresponds to an isothermal, self-gravitating sphere. Analysis of the radio continuum
spectra of three highly compact \ion{H}{2} regions by
\citet{2000ApJ...542L.143F} also imply negative power-law density distributions
with exponent $\alpha \ge 2$, possibly even as steep as $\alpha = 4$
in one case. These hyper and ultracompact \ion{H}{2} regions are
considered to represent the
earliest stages of \ion{H}{2} region evolution, and so provide
information about the natal cloud density distribution. Once the
\ion{H}{2} region has begun to expand, the density distribution of the
ionized gas will become shallower as structure within the \ion{H}{2}
region is smoothed out on the order of a sound crossing time. Thus,
the density gradients reported by \citet{2000ApJ...542L.143F} should
be treated as lower limits to the initial density gradients in the
star-forming cores. For the largest values of $\alpha$ reported, such
a steep radial, power-law density distribution is unlikely and it is
possible that the density stratification is better represented by a
Gaussian or other exponential function. Power-law density gradients
with exponents $1.6 < \alpha < 2.4$ are also reported for giant
extragalactic \ion{H}{2} regions \citep{2000ApJ...544..277F}, showing
that density gradients should be taken into consideration at all
stages of \ion{H}{2} region evolution.

An alternative explanation for the cometary shape of some compact
\ion{H}{2} regions is that the ionizing star is moving with respect to
the dense cloud gas. \citet{1991ApJ...369..395M}  and
\citet{1992ApJ...394..534V} developed a bowshock interpretation for
cometary compact \ion{H}{2} regions, in which the ionizing star with
its strong stellar wind is moving supersonically with respect to the
cloud gas and the ionization front becomes trapped in the swept-up
shell behind the bow shock that forms ahead of the star. Numerical
models of supersonically moving high mass stars have been carried out by
\citet{1997RMxAA..33...73R} and
\citet{1998A&A...338..273C} in the context of runaway O stars in the
diffuse interstellar medium. Even though the parameters are not
appropriate for young massive stars moving in dense molecular clouds,
these numerical models show that instabilities complicate the
dynamics. Recent numerical models of \ion{H}{2} regions around massive stars orbiting in the
gravitational potential of structured molecular clouds (but without
including the stellar winds) have been done
by \citet{2005arXiv.0508467.F}, who show that a wide variety of
unpredictable morphologies can result. 

Bowshock models are attractive because they easily reproduce the
limb-brightened morphology seen in radio continuum images of cometary
compact \ion{H}{2} regions, whereas champagne flows have no limb
brightening. However, it is in the kinematics of the ionized gas that
these two scenarios can really be distinguished. In the bowshock model
the highest ionized gas velocities are at the head and are equal to
the star's velocity. In the champagne flow model the highest ionized
gas velocities are found in the tail and can reach a few times the
sound speed.  Unfortunately, observational kinematical studies are
difficult to perform given the complete extinction that exists in the
optical due to the dense environments in which these objects are
found. Radio recombination line studies
\citep{{1991ApJ...372..199W},{1994ApJ...437..697A},{1994ApJ...429..268G},{2003ApJ...596..344C}}, infrared
recombination line observations \citep{{1996ApJ...464..272L},
{1999MNRAS.305..701L},{2003RMxAC..15..172H},{2003A&A...405..175M}},
and mid-infrared fine-structure line observations
\citep{{2003ApJ...596.1053J},{2005ApJ...631..381Z}} 
have, however, been able to extract some kinematical information from
these regions, which should enable us to distinguish between the
various theoretical models that have been put forward to explain this
phenomenon. It is important to obtain a good
determination of the background molecular cloud velocity, which can
have a major effect on the interpretation of the observations.

The kinematical observations available suggest that neither a bowshock
model nor a champagne flow adequately explain the velocity variation
between head and tail of many observed cometary \ion{H}{2} regions, or
between the center and the wings. \citet{1994ApJ...432..648G}
highlighted the need for more realistic models, incorporating stellar
winds and more complicated ambient density distributions in the
surrounding molecular cloud. \citet{1999MNRAS.305..701L} proposed an
empirical model for the well-studied object G29.96$-$0.02 in which a
champagne flow is modified by the presence of a strong stellar wind
within the \ion{H}{2} region which traps the \ion{H}{2} region in the
denser gas into a swept-up shell and diverts the photoevaporated flow
around the hot stellar wind bubble. They also required a component of
expansion into the molecular cloud of $\sim
10$~km~s$^{-1}$. \citet{2005ApJ...631..381Z} discuss the need for
pressure gradients within a paraboloidal shell to explain the velocity
range for this object.

In addition to the \ion{H}{2} region dynamics, hot, massive stars
possess strong stellar winds, which will also affect the velocity
structure of these regions. Work by \citet{1996ApJ...469..171G} and
\citet{2003ApJ...594..888F} has shown that the interaction of the
photoionized region with a swept-up stellar wind shell produces all
sorts of interesting structures, such as spokes, shells, clouds and
fingers. These arise due to cooling, thin shell, Rayleigh-Taylor and
other instabilities, which have been analyzed by
\citet{1983ApJ...274..152V}, \citet{1988ApJ...331..350R} and \citet{1994ApJ...428..186V}.
The shapes of stellar wind bubbles forming in \ion{H}{2} regions with
density gradients was studied analytically by \citet{1977A&A....59..161D}.
That cometary, compact \ion{H}{2} regions also contain stellar winds is evident
from observations of our nearest compact \ion{H}{2} region, the Orion
Nebula. The size scale of the Orion nebula is a few tenths of a parsec
in diameter. \textit{Hubble Space Telescope} (\textit{HST}) observations
of the nebula reveal the presence of bowshocks ahead of the
photoevaporating circumstellar material around the ``proplyds''
(protoplanetary disks) in the direction directly towards the main
ionizing star of the nebula, $\theta^1$~C~Ori. These bowshocks are
direct evidence of an interaction between the photoevaporated flows
within the nebula and the strong stellar wind from the ionizing star
and enable us to constrain $\dot{M}V_w$ of the stellar wind
\citep{{2001ApJ...561..830G},{2002RMxAA..38...51G}}. 

In this paper we investigate the formation and evolution of cometary
compact \ion{H}{2} regions both with and without stellar winds in a
variety of ambient density distributions and compare these results to
kinematic data from the literature. The organization of the paper is
as follows: in \S~\ref{sec:nummodel} we describe our numerical model
and test cases, while in \S~\ref{sec:basic} we describe a set of basic
reference models comprising a simple champagne flow, a bowshock in a
uniform medium, and a champagne flow with a stellar wind. In
\S~\ref{sec:modified} we describe modifications to these basic models,
including varying the steepness of the density gradient and the
strength of the stellar wind, and adding stellar motion to the
champagne flow plus stellar wind models. We discuss our results in the
context of other models and briefly compare our results to
observations taken from the literature in \S~\ref{sec:discussion}.
Finally, in \S~\ref{sec:conclusion} we conclude our findings. 

\section{Numerical Model}
\label{sec:nummodel}
\subsection{Method}
We need to calculate both the hydrodynamics and the radiative transfer
within the cometary \ion{H}{2} regions. This is a non-trivial problem
involving a variety of timescales.

For the radiative transfer we adopt a similar procedure to that used
in \citet{1999RMxAA..35..123R}, which uses the method of short characteristics
\citep{{1978ApJ...220.1001M},{1988JQSRT..39...67K}}  to calculate the column
density from every source point to each point on the numerical grid,
and thence the photoionization rate at each point. We have adapted
this method, described in \citet{1999RMxAA..35..123R} for a three-dimensional
Cartesian system, to a two-dimensional cylindrically symmetric
system. In the models presented in this paper we consider only a
single source point (i.e., the star) for the ionizing radiation.  The
ionizing flux is calculated from the number of photons with energies
above the hydrogen ionization threshold of 13.6~eV produced by a
simple black body of a given effective temperature and stellar
radius. Only one frequency of radiation is included in the radiative
transfer and the diffuse field is treated by the on-the-spot
approximation \citep{2005ApJ...627..813H}.

The hydrodynamics is treated with a two-dimensional, cylindrically
symmetric hybrid Godunov-van Leer second order method
\citep{{1959MatSb..Godunov}, {1993MNRAS.261..681A},{1982LNP...170..507V}}. A
hybrid method is used in order to avoid the ``carbuncle'' instability
seen behind shocks parallel to grid lines with purely Riemann solver
codes \citep[see, e.g.,][]{1994IJNMF..18..555Q}. Van Leer and Riemann
solver steps are alternated. The Van Leer scheme is not strictly
upwind according to the definition of \citet{1998JCPHY.145..511S} and puts a
small shear term into the momentum equation. This shear dissipation is
sufficient to overcome the carbuncle instability. On the other hand, a
Riemann solver does a more accurate job in general of representing the
flow, particularly in transonic regions, and so the combined scheme
has the benefits of both methods without their respective
disadvantages.

The hydrodynamics and radiative transfer are coupled through the
energy equation, where the source term depends on the photoionization
heating and the radiative cooling rates. The radiative cooling
includes collisional excitation of Lyman alpha, collisional excitaton
of neutral and ionized metal lines (assuming standard ISM abundances),
hydrogen recombination, collisional ionization and bremsstrahlung,
with each cooling process being characterized by an analytic
approximation which depends on temperature and ionization fraction.
The energy equation is solved explicitly using a second order
Runge-Kutta method. In photoionization equilibrium the gas temperature
in our models is $10^4$~K. The time step for the calculations is the
minimum of the hydrodynamic (Courant condition) and cooling times,
evaluated at each time step. The present models do not include thermal
conduction nor absorption of ionizing photons by dust
grains. The presence of dust within an \ion{H}{2} region can
have a dramatic effect on the size of the Str\"omgren sphere. However,
the distribution of dust within the \ion{H}{2} region depends strongly
on the gas density and stellar spectrum and there can even be quite a
large volume devoid of grains due to sublimation
\citep{2004ApJ...608..282A}. Such complications are beyond the scope
of this paper.

\subsection{Test cases}
\subsubsection{Expansion of an \ion{H}{2} region in a uniform density
medium}
\begin{figure}
\centering
\includegraphics{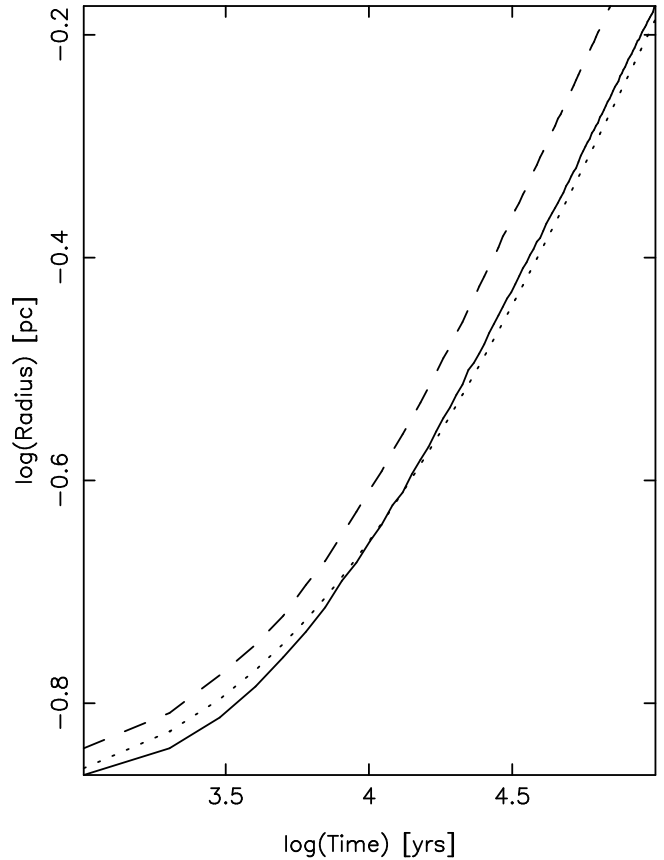}
\caption{Expansion as a function of time of the ionization front
(solid line) and preceding shock front (dashed line) for an
\protect{\ion{H}{2}} region in a uniform density medium. The dotted
line represents the analytical expansion law (Spitzer 1978) for the
same parameters.
}
\label{fig:uniexp}
\end{figure}
We modeled the expansion of the \ion{H}{2} region produced by a
30,000~K black body in a uniform medium of density 6000~\pcc\ and
initial temperature $300$~K. The
ionizing photon rate of the source is $Q_{\rm H0} = 2.2\times
10^{48}$~s$^{-1}$ and the theoretical initial Str\"omgren sphere has a radius of
$R_{\rm s} = 0.124$~pc for an ionized gas temperature of $T =
10,000$~K. Our numerical model also has a Str\"omgren
radius of $R_{\rm ns} = 0.124$~pc, for an ionized gas temperature of 
$T \approx 10,500$~K. Once formed, 
this overpressured sphere of gas begins to expand into the
surrounding medium. An analytical approximation to the expansion
\citep{1978ppim.book.....S} is 
\begin{equation}
R_{\rm i} = R_{\rm s}\left(1 + \frac{7}{4} \frac{c_{\rm II} t}{R_{\rm s}} \right)^{4/7}
\label{eq:expand}
\end{equation}
where $c_{\rm II}$ is the sound speed in the ionized gas, $R_{\rm i}$
is the ionization front radius and $t$ is time,
 under a set of simplifying assumptions, including (i)
that the swept-up neutral shell between the shock wave and ionization
front is thin such that the shock and ionization front move with the
same velocity (ii) the pressure across the neutral shell is
uniform. At late times the expansion law is characterized by $R_{\rm
i} \propto t^{4/7}$.  

In Figure~\ref{fig:uniexp} we show the numerical results of a 2D
cylindrically symmetric
simulation with grid resolution $512\times1024$ cells for the shock
front and ionization front radii as a function of time, together with
the analytic result for the same initial parameters. At early times
the analytic and numerical solutions differ because the rarefaction
wave that propagates back through the \ion{H}{2} region when the
expansion first starts in the numerical model reduces the pressure in
the ionized gas and hence the numerical solution lags behind the
analytical one to start with. After 10,000~yrs the numerical and
analytic results agree but then begin to deviate slightly, with the
numerical solution being ahead of the analytical one at these
times. This is because the assumptions made for the analytic expansion
break down. The neutral shell ceases to be thin because the shock
propagates faster than the
ionization front. In particular,
the density just ahead of the ionization front decreases as a fraction
of the average density in the neutral shell as time goes on, meaning
that the ionization front is propagating into a successively less
dense medium with time, and hence the expansion is faster than that
predicted by the analytic formula, which does not allow for any
evolution of conditions in the neutral shell.

\subsubsection{Expansion of an \ion{H}{2} region in the presence of a
stellar wind} 
\label{subsubsec:expansion}
The expansion of a stellar wind into a surrounding environment has
been studied numerically by many authors \citep[e.g.,][]{{1975A&A....43..323F},
{1995ApJ...455..160G},{1996A&A...305..229G}}. In one-dimensional
studies the textbook regions of unshocked wind, shocked wind, swept-up
shocked ISM are easily distinguished \citep{{1997pism.book.....D},{1977ApJ...218..377W}}. However,
once studies are extended to two dimensions, thin shell instabilities
in the cooling swept-up ISM lead to a more complex appearance, which
could influence the \ion{H}{2} region outside
\citep{{1995ApJ...455..160G},{1996A&A...305..229G},{1996ApJ...469..171G},
{1997A&A...326.1195C},{2003ApJ...594..888F}}.  In particular, the
knotty appearance of the unstable thin shell leads to shadowing by
clumps of the \ion{H}{2} region. The resulting pressure variations
between shadowed and unshadowed regions lead to complicated, low
velocity ($\leq 20$~km~s$^{-1}$) kinematics in the ionized gas.

We used a \citet{1985Natur.317...44C} stellar wind model to input the
initial stellar wind conditions as a volume-distributed source of mass and energy
\citep[this method for dealing with the stellar
wind was also used by][]{{1985A&A...143...59R},{1997A&A...326.1195C}}. This avoids the
problem of reverse shocks that occur when the stellar wind is input
simply as a region of high density, high velocity flow near the
origin. The mass and energy source terms were calculated so as to give
the correct mass-loss rate and terminal velocity for the stellar
wind. The volume occupied by stellar wind source terms was chosen to
be sufficiently small so that the stellar wind always reached its
terminal velocity and shocked outside of the source region, but
sufficiently large so as to be a reasonable geometrical representation
of a sphere on a cylindrical grid. Of course, this thermal wind is not
the same as a pressureless stellar wind at the center of the inner
unshocked wind region, but it does have the same ram pressure. The
Mach number of the thermal wind just before the inner stellar wind
shock is $\sim 10$, and so is highly supersonic, meaning that any
thermal pressure in this region is not dynamically important.

We performed a number of studies of the evolution of \ion{H}{2}
regions around stellar wind bubbles with the aim of establishing the
optimum resolution for our calculations. The high densities ($\approx
6000$~\pcc) in the ambient gas imply extremely short cooling times,
requiring a short time step for adequate resolution. In addition, a
fast stellar wind also requires a short time step due to the high
temperatures (and hence sound speeds) in the shocked wind region. 

On our two-dimensional grids we could not hope to achieve the
resolution necessary to fully resolve the cooling shocked swept-up ISM
shell. However, by reducing the cooling rate and heating due to
photoionization by a given factor we can maintain the temperature in
the ionized gas while artificially enhancing the resolution in the
cooling shell (by increasing the cooling length). This means that
radiative cooling in the shell occurs more slowly than it would
otherwise have done so, but given the densities involved, cooling is
still rapid in these simulations. We then examined a number of
simulations at different grid resolutions in order to determine a
consistent qualitative behaviour in the cooling region. Note that in
these simulations there is no need to trigger instabilities in the
swept-up shell with density fluctuations \citep[e.g.,][]{1996A&A...305..229G}
because the cylindrical symmetry means that the stellar wind
source volume introduces its own numerical noise simply by not being
perfectly spherical. Furthermore, we artificially broaden the ionization front (by
reducing the ionization cross-section), in order to avoid artificial
numerical instabilities generated at the grid scale \citep{1999MNRAS.310..789W}.
 
\begin{figure*}
\centering
\setkeys{Gin}{width=0.3\linewidth}
\includegraphics{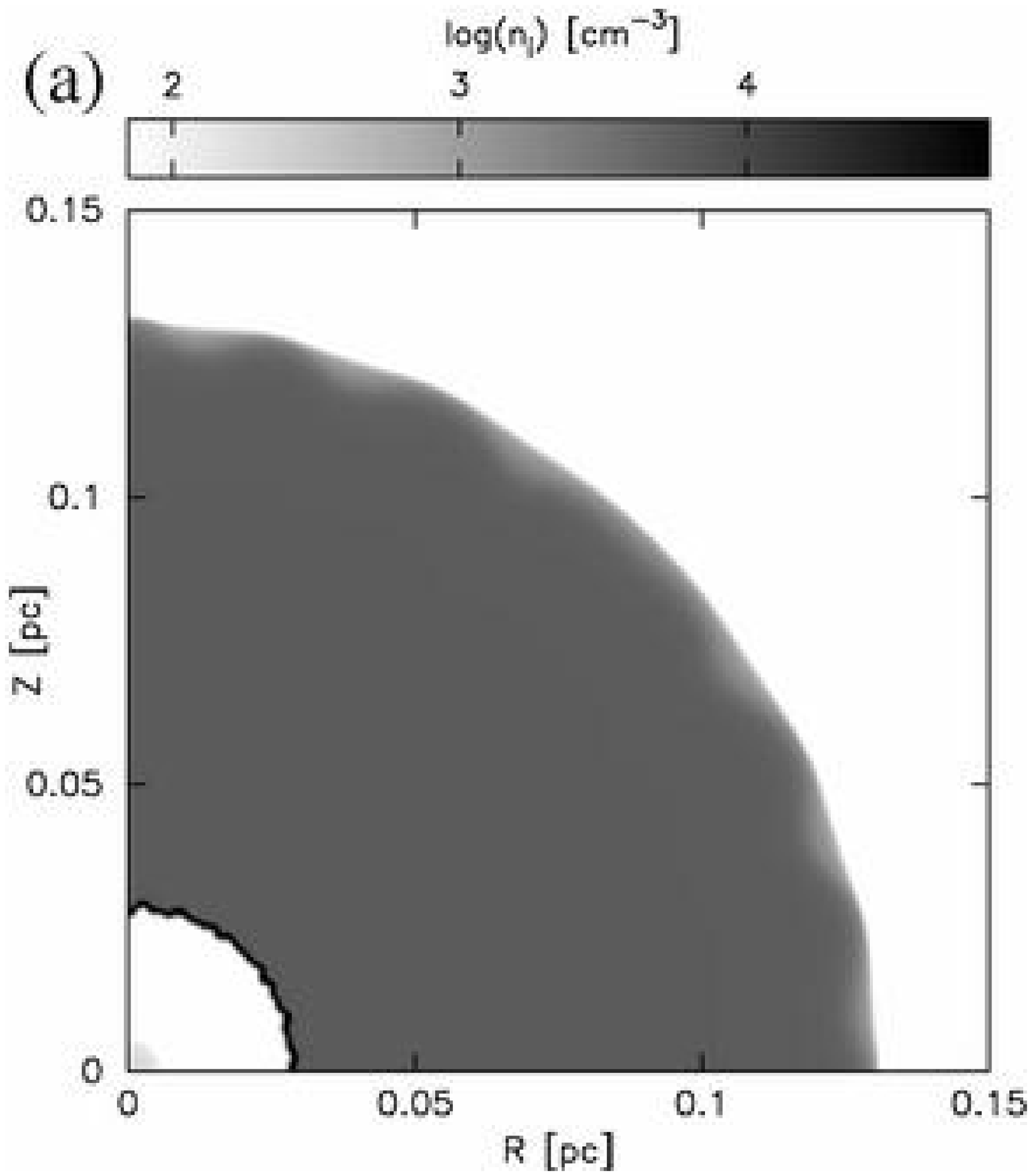}\hfill
\includegraphics{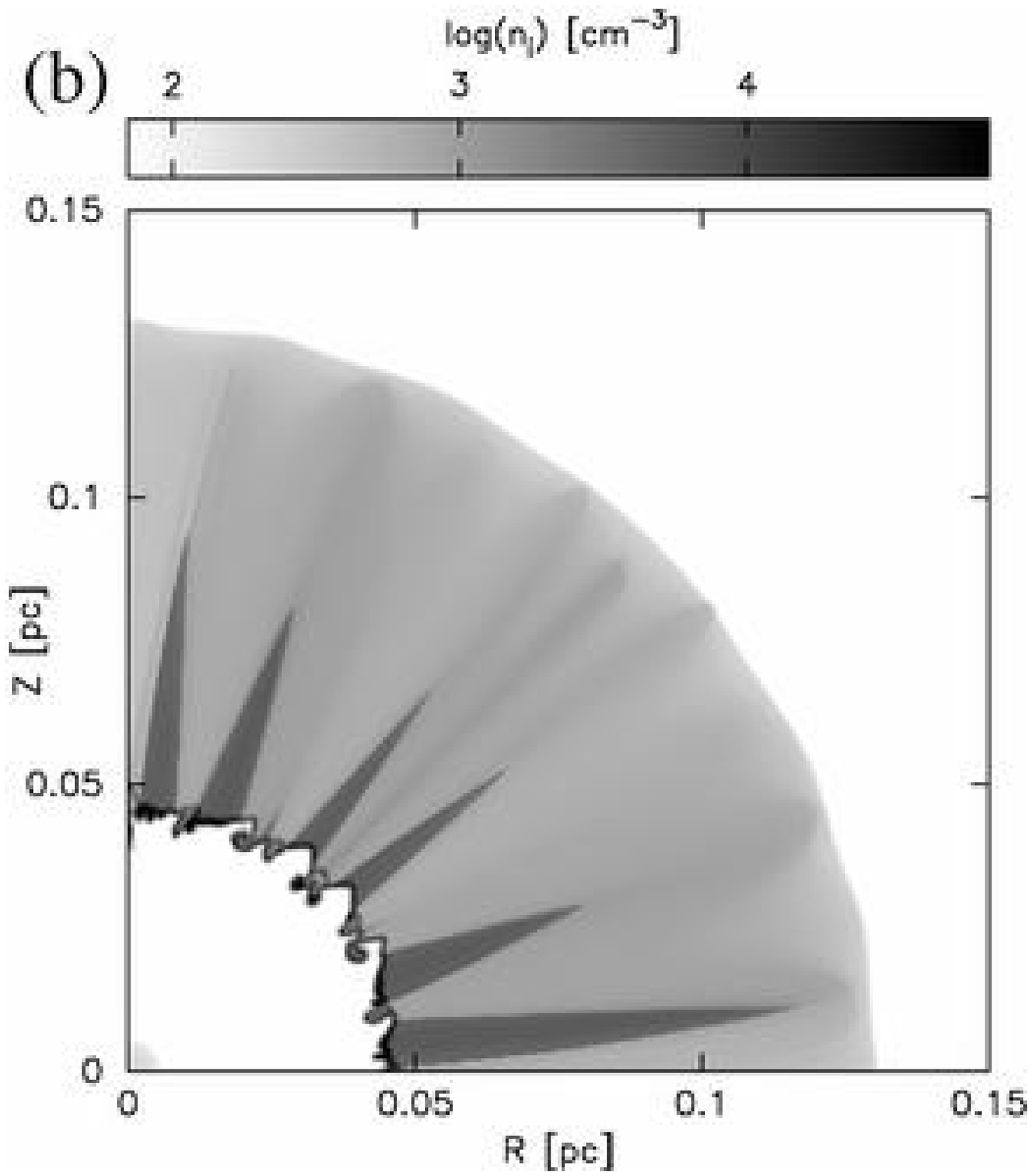}\hfill
\includegraphics{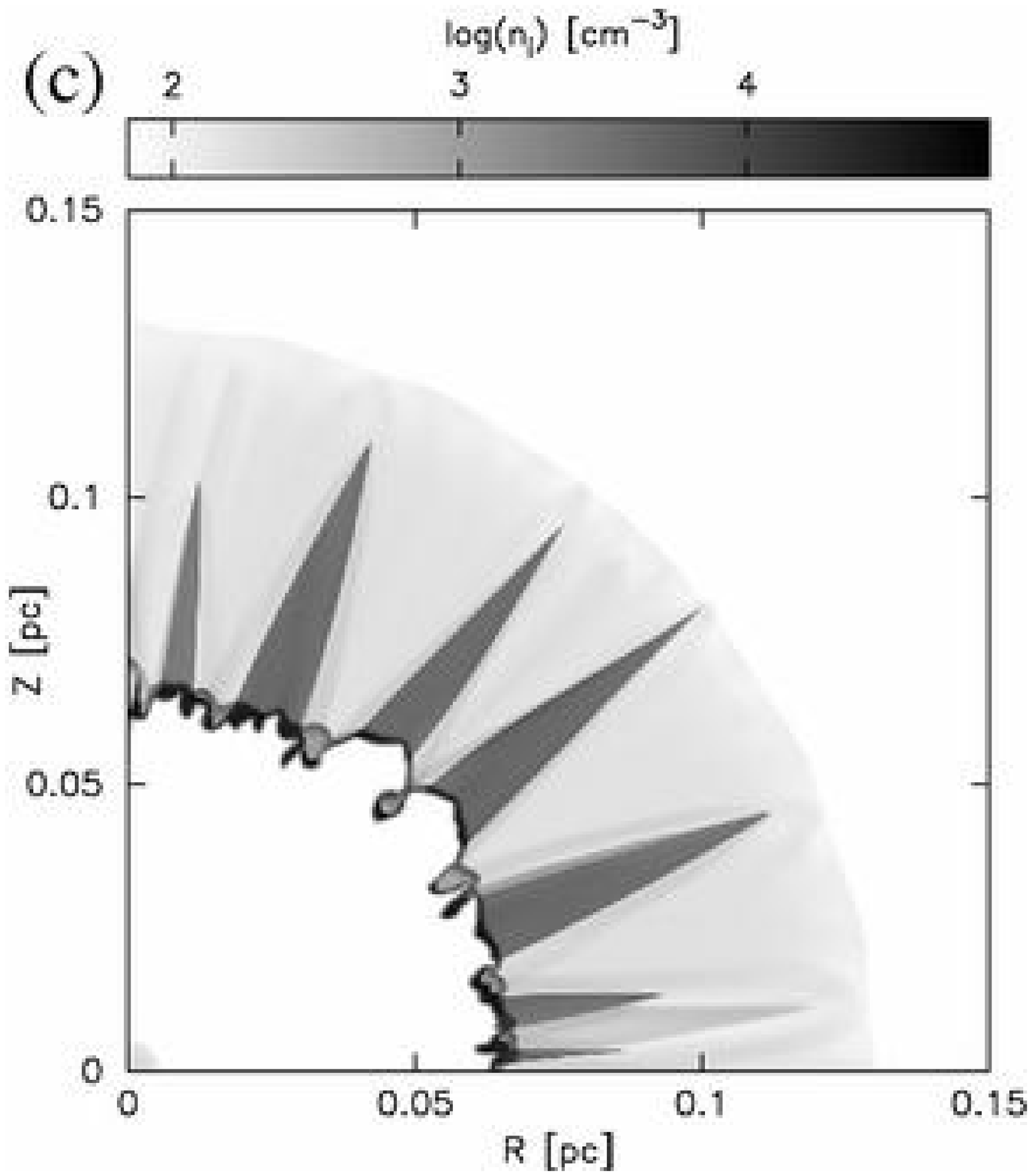}\\
\includegraphics{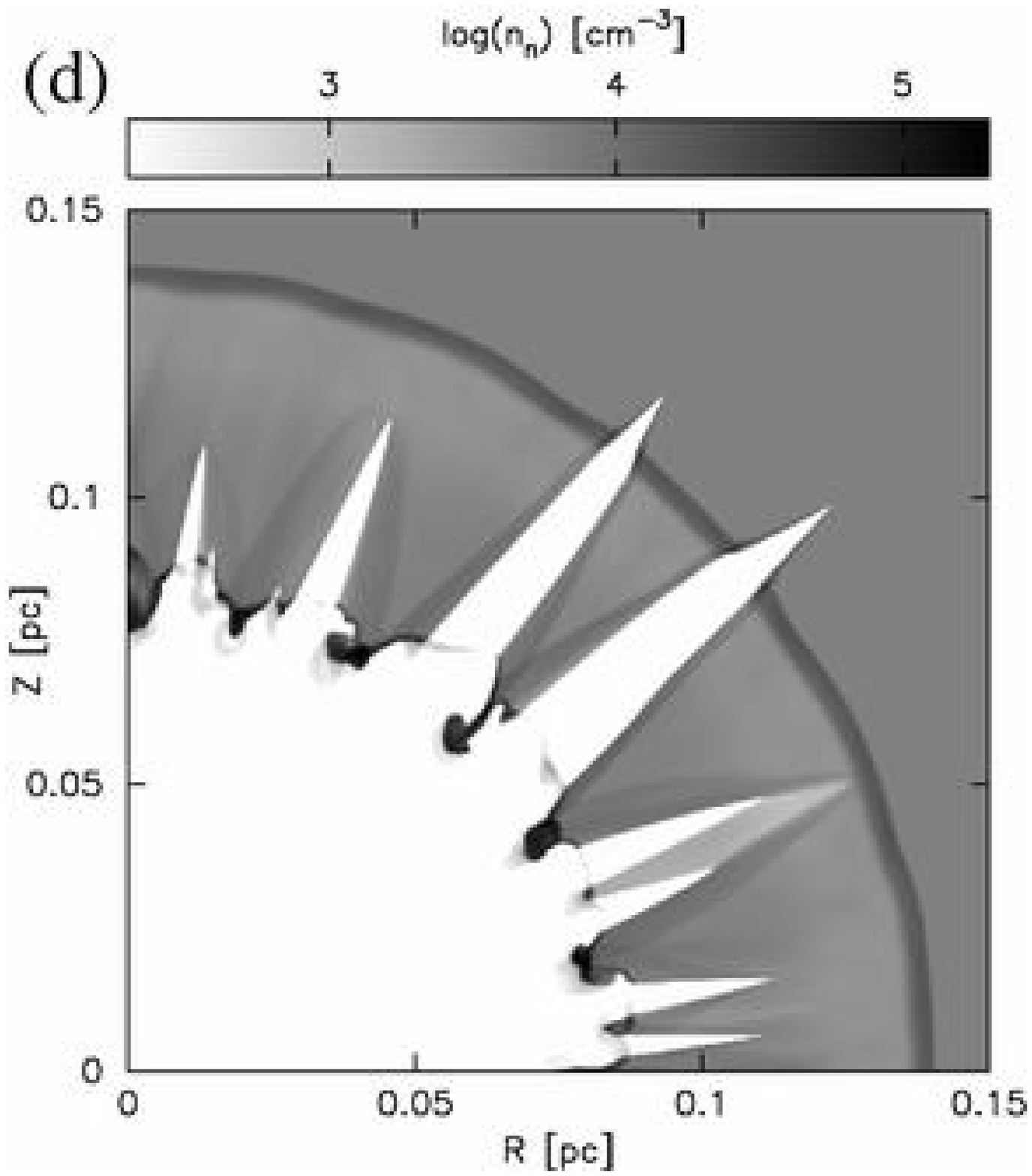}\hfill
\includegraphics{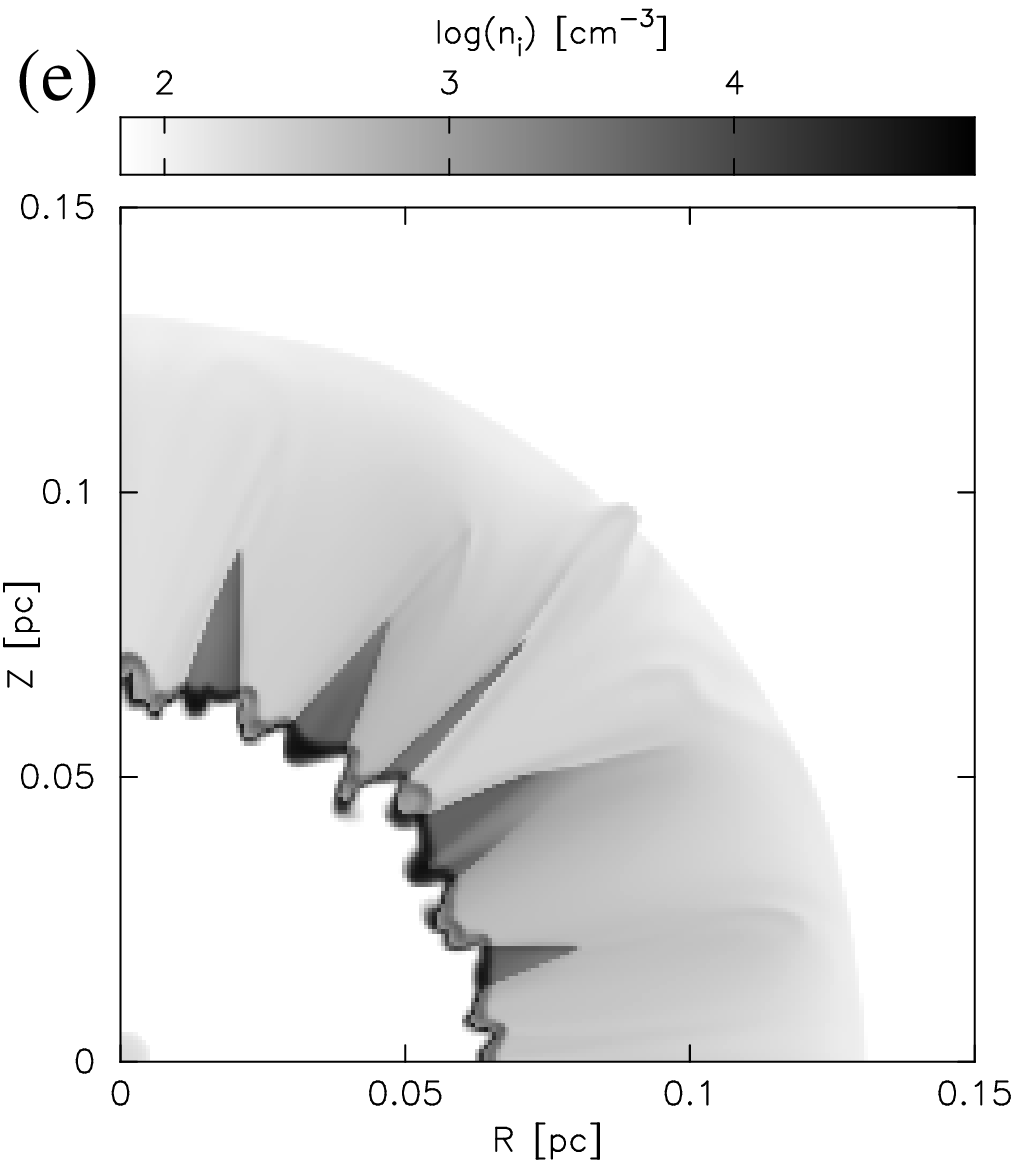}\hfill
\includegraphics{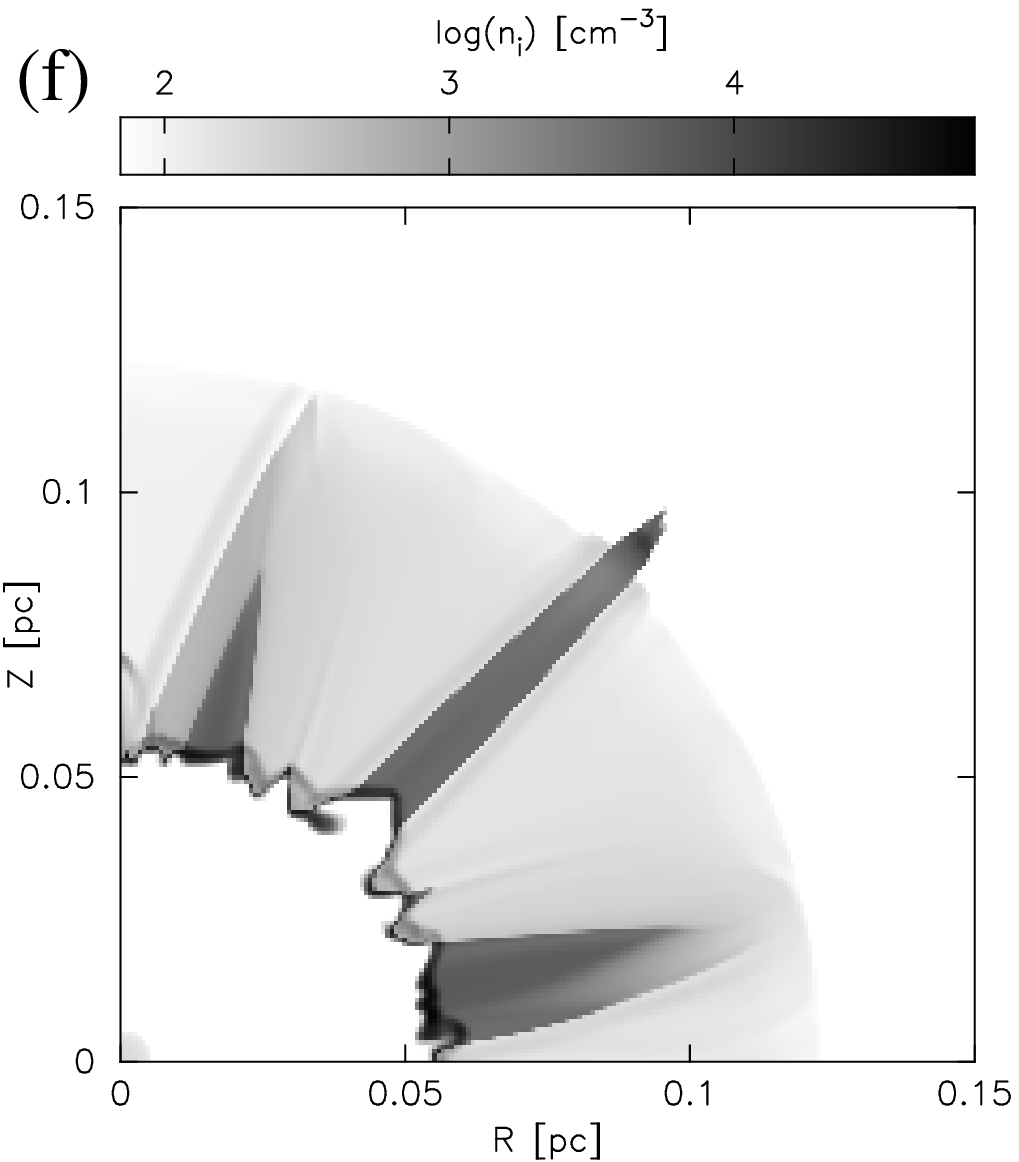}
\caption{Evolution of $\log_{10}(\mbox{density})$ in an \protect{\ion{H}{2}} region with a stellar
wind. Grayscale for the ionized gas covers the range $70 < n_i <
70000$~cm$^{-3}$, while for the neutral gas the grayscale range is $200 < n_n <
200000$~cm$^{-3}$. (a) Ionized gas density after 200 years (full range
$0 < n_i < 11700$~cm$^{-3}$), showing the formation of ripples in the
swept-up thin shell due to cooling instabilities and corrugations in
the ionization front due to shadowing. (b) Ionized gas density after
500 years ($0 < n_i < 92400$~cm$^{-3}$) showing the growth
of the shadowing instability. (c) Ionized gas density after 1000 years
($0 < n_i < 61100$~cm$^{-3}$). (d) \textit{Neutral} gas
density after 1500 years ($0 < n_n < 210000$~cm$^{-3}$)
showing the formation of dense neutral clumps at the base of the
ionized spokes. (e) Same as (c) but at half the grid resolution ($0 <
n_i < 45200$~cm$^{-3}$). (f) Same as (e) but with a 
larger cooling coefficient ($0 < n_i < 63000$~cm$^{-3}$).}
\label{fig:h2wind}
\end{figure*}
In Figure~\ref{fig:h2wind} we show the evolution of the ionized
density in the region surrounding a (blackbody) star of effective
temperature $T_{\rm eff} = 30,000$~K, stellar wind mass-loss rate
$\dot{M} = 10^{-6}$~M$_\odot$~yr$^{-1}$ and terminal velocity $V_{\rm
w} = 2000$~km~s$^{-1}$, where the initial ambient medium has number
density $n_0 = 6000$~\pcc\ and temperature $T_0 = 300$~K.  The
ionizing photons and the stellar wind are started simultaneously. The
top four panels show the ionized density at times 200, 500, 1000 and
1500 years since the start of the calculation with grid resolution
$400 \times 400$ cells. The grid shown represents a spatial size of
$0.15 \times 0.15$ parsecs. The \ion{H}{2} region forms essentially
immediately, on a recombination timescale, with a radius of
0.13~parsecs. The stellar wind bubble takes longer to form and will
grow until balance is achieved between the ram pressure of the stellar
wind and the thermal pressure of the surrounding ISM.

By the time of the first panel (200 years), instabilities have begun to form in
the swept up shell of ionized gas, at 0.03~parsecs radius. The rapid
cooling in this shell means that it is thin and subject to the thin
shell and Vishniac
instabilities
\citep{{1983ApJ...274..152V},{1987ApJ...313..820R},{1988ApJ...331..350R},{1989ApJ...337..917V},
{1994ApJ...428..186V}}, which have been shown to
occur in 2D numerical stellar wind bubble models
\citep{{1995ApJ...455..160G},{1996ApJ...469..171G},{2003ApJ...594..888F}}.
The ionized densities in the shell reach $10^5$~cm$^{-3}$, i.e.,
compression factors much higher than the factor 4 expected for an
adiabatic shock, confirming that this is a region of strong cooling.
The increased opacity due to the swept-up shell leads to
recombination of the ionized region beyond. The inhomogeneities in
the shell represent opacity variations, which lead to variations in
the recombination rate that can be detected as ripples in the outer rim. 

In the second panel (500 years), the instability has saturated and the
knotty thin shell at 0.045~parsecs is being pushed outwards by the
hot, shocked stellar wind bubble. The degree of ionization in the
\ion{H}{2} region beyond the swept-up shell is very small as this
region has now essentially completely recombined.  The knotty nature
of the thin shell causes variations in the optical depth and this
leads to the shadowing instability discussed by
\citet{1999MNRAS.310..789W}. Spokes of ionized gas are characteristic
of this instability and represent the propagation of R-type ionization
fronts in collimated beams through gaps (regions of underdensity) in
the knotty shell. The bases of the beams are wider than the tips
because the ionized gas at the base is pressurized and the gas can
expand laterally.  The freely flowing and shocked stellar wind regions
interior to the swept-up shell are of such low density that they do
not contribute to the map of ionized density despite being fully
ionized regions.
\begin{table*}\centering
\caption{Model parameters.}
\setlength\tabcolsep{0.5cm}
\begin{tabular}{clclccr}
\toprule
Model &   & $n_0$ & $H$ & $\dot{M}$ & $V_{\rm w}$ & $V_*$ \\
     &  $n$    & cm$^{-3}$ & pc & $M_\odot$~yr$^{-1}$ & km~s$^{-1}$ &
km~s$^{-1}$ \\
\midrule
A & $n_0 \exp(z/H)$ & 8000 & 0.05 & \nodata & \nodata & \nodata\\ 
B & $n_0$ & 6000 & \nodata & $10^{-6}$ & $2000$ & $20$ \\
C & $n_0 \exp(z/H)$ & 8000 & 0.05 & $10^{-6}$ & $2000$ & \nodata \\
D & $n_0 \exp(z/H)$ & 8000 & 0.05 & $10^{-7}$ & $2000$ & \nodata \\
E & $n_0 \exp(z/H)$ & 8000 & 0.2 & $10^{-6}$ & $2000$ & \nodata \\
F & $n_0 \left[1 + (R/0.01)^2\right]^{-1}$ & $9.8\times 10^6$ &
\nodata & $10^{-6}$ & $2000$ & \nodata \\
G & $n_0 \exp(z/H)$ & 8000 & 0.05 & $10^{-6}$ & $2000$ & 5\\
H & $n_0 \exp(z/H)$ & 8000 & 0.05 & $10^{-6}$ & $2000$ & 10\\
I & $n_0 \exp(z/H)$ & 8000 & 0.2 & $10^{-6}$ & $2000$ & 10\\
\bottomrule
\end{tabular}
\label{tab:models}
\medskip
\end{table*}

In the third (1000 years) and fourth (1500 years) panels the knotty
shell has expanded outwards pushed by the hot shocked stellar wind,
reaching a mean radius of 0.08~parsecs in panel four. The dense clumps
of neutral gas that form at the base of the ionized spokes are the
result of the shadowing instability \citep{1999MNRAS.310..789W}. Their
density continues to increase as time goes on, unlike the ionized gas
density in the swept-up thin shell, which decreases since the pressure in
this region decreases as the expansion proceeds (the temperature of
this gas remains constant at $10^4$~K). The main ionization front is
trapped in the swept-up stellar wind shell at these times. As the
expansion of the stellar wind bubble proceeds, the movement of the
knots, plus the contributions to the opacity of the material
photoevaporating from the neutral clumps, lead to changes in the
pattern of spokes.

The shadowing instability discussed by \citet{1999MNRAS.310..789W}
results from the effect on an R-type ionization front of
inhomogeneities in the upstream density. First, corrugations in the
ionization front are formed, which become amplified by the
instability, leading to the formation of characteristic spokes of
ionized gas with clumps of dense, neutral material at the base. In the
simulations presented in Figure~\ref{fig:h2wind}, the shadowing effect
is produced by clumps formed by the thin-shell instability in the
swept-up stellar wind shell, within the \ion{H}{2} region itself. Although
corrugation of the outer ionization front is seen at early times
(Fig.~\ref{fig:h2wind}\textit{a}), this is not important for the
growth of the instability, since the whole region outside of the thin
shell is recombining by this point. The shadowing instability takes
hold within the thin shell itself, where the main ionization front has
become trapped. Because the whole structure is expanding due to the
high pressure of the shocked stellar wind, the pattern of ionized spokes
in our simulations changes with time. 

The fifth panel of Figure~\ref{fig:h2wind} shows the effects
of reducing the grid resolution. Reducing the
resolution by a factor 2 in each dimension gives qualitatively similar
results to the standard case. 

In the sixth panel we have used the
full value of the cooling rate and heating rate with the lower grid
resolution. Increasing the cooling and heating rates means that cooling
occurs sooner and more quickly, leading to a thinner, denser swept-up
shell. The recombining region has receded to 0.12~parsecs by
1000~years of evolution while the swept-up wind shell is located at
0.05~parsecs from the central star, as compared to 0.06~parsecs in the
standard case. However, for this grid resolution the shell is certainly not
resolved and would require much higher resolution. In these
calculations it is desirable to resolve the shell as well as possible
because of the effect this has on the ionized region beyond. The device of
decreasing the cooling and heating rates permits us to better resolve
the swept-up shell without having to use an unpractical number of grid
cells. 

\section{Basic Models}
\label{sec:basic}
Our models are motivated by a desire to explain features of the
spectra of cometary ultracompact \ion{H}{2} regions such as G29.96$-$0.02,
whose morphology and kinematics cannot satisfactorily be explained by
either a simple champagne flow or a simple bowshock model. In
particular, we are interested in explaining the turnover from
blueshifted to redshifted to blueshifted velocity seen going from the head to the
tail along the symmetry axis
in observations of this object in hydrogen recombination lines \citep[see,
e.g.,][]{{1996ApJ...464..272L},{1999MNRAS.305..701L},{2003A&A...405..175M}}
and fine-structure lines of the Ne$^+$ ion \citep{2005ApJ...631..381Z}.

Instead of tailoring a model to fit a particular observation, in this
paper we wish to identify features in the simulated spectra that
correspond to different configurations of density distribution,
stellar wind parameters and stellar motion. To this end, we consider
only one set of parameters for the ionizing star, namely a 30,000~K
black body with an ionizing photon rate of $2.2 \times
10^{48}$~s$^{-1}$. This corresponds approximately to spectral type O9V
\citep{{1973AJ.....78..929P},{2005A&A...436.1049M}}. The ambient
medium is taken to be at rest and neutral, with a temperature of
300~K. For the majority of models, the grid resolution is
$750\times1000$~cells, representing a total grid size of
$0.6\times0.8$~parsecs.

To begin with we discuss three ``canonical models'', Models A, B, and
C of Table~\ref{tab:models}, which correspond to a simple
champagne
flow, a stellar bowshock in a uniform medium, and a champagne flow
with a stellar wind, respectively.  The model parameters are
summarized in Table~\ref{tab:models}. Model A is followed for
20,000~yrs, while Models~B and C are evolved for 40,000~yrs.

\subsection{Model Parameters}\nobreak
\subsubsection{Champagne flow: Model A}
We set up a photoevaporated flow in an exponential density distribution,
where the density at the position of the source is $n_0 =
8000$~cm$^{-3}$ and the scale height is $H = 0.05$~pc.

\subsubsection{Bow Shock: Model B}
The source is assumed to move with a velocity of 20~km~s$^{-1}$ in a
uniform medium of density $n_0 = 6000$~cm$^{-3}$. This velocity is
motivated by previous empirical models
\citep[e.g.,][]{{1994ApJ...437..697A},{1996ApJ...464..272L},{2003A&A...405..175M},{2005ApJ...631..381Z}}
for the ultracompact \ion{H}{2} region 
G29.96$-$0.02. This star has a
strong stellar wind with mass loss rate $\dot{M} = 10^{-6}
M_\odot$~yr$^{-1}$ and terminal velocity $V_{\rm w} = 2000$~km~s$^{-1}$, typical of an
early type star. After the 20,000~yrs of evolution followed by the
simulation, the star has traveled a distance of 0.4~pc. The resolution
of this model is $512\times1024$~cells, representing a spatial size of
$0.4\times0.8$~parsecs.

\subsubsection{Champagne flow with stellar wind: Model C}
We investigate the effect of including a stellar wind in the champagne
flow described in Model~A. The stellar wind has a mass loss rate $\dot{M} = 10^{-6}
M_\odot$~yr$^{-1}$ and terminal velocity $V_{\rm w} =
2000$~km~s$^{-1}$, i.e., the same wind parameters as used in the
standard bowshock model.
\subsection{Basic model results}
\subsubsection{Velocity field}
\begin{figure}
\centering
\includegraphics{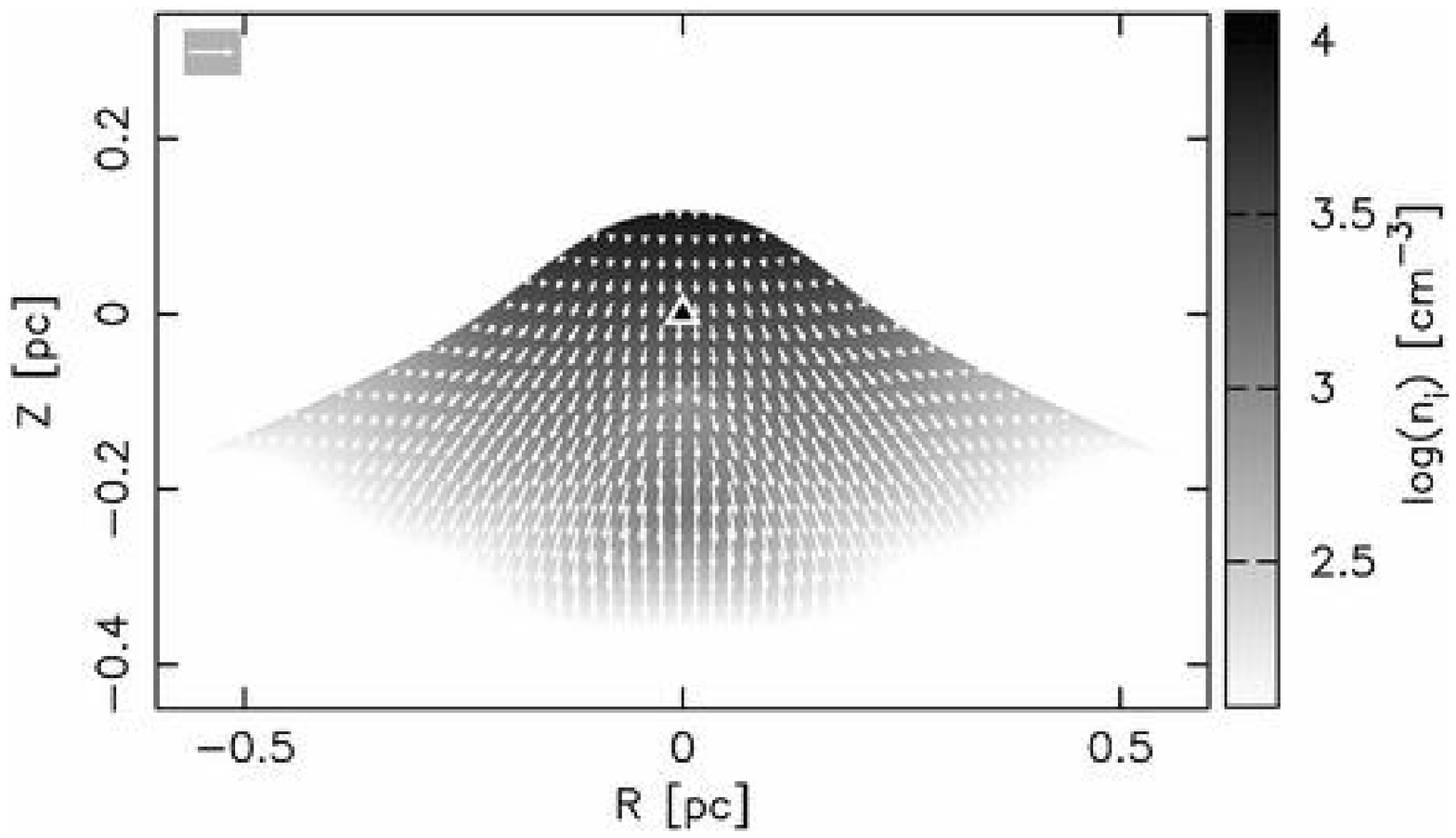}\\
\includegraphics{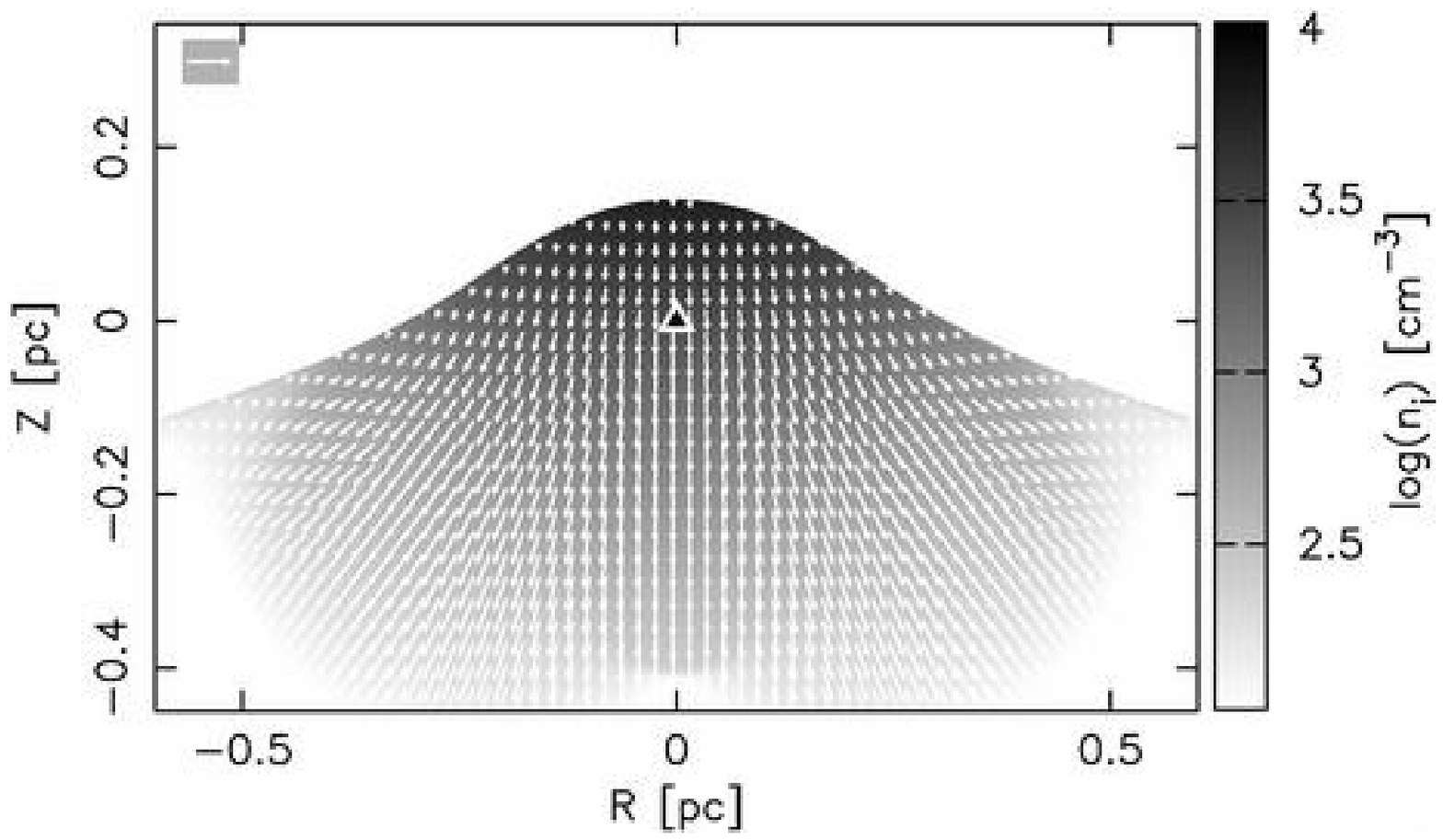}
\caption{Logarithm of ionized density for a cut through
the $r-z$ plane of the numerical simulation of Model~A (simple champagne flow) after elapsed times of
10,000~yrs (top panel) and 20,000~yrs (bottom panel).  The arrows show the  
velocity field, with the scale arrow representing
30~km~s$^{-1}$ in these models. The position of the source is marked by a black
triangle.}
\label{fig:ev1a}
\end{figure}
\begin{figure}
\centering
\includegraphics{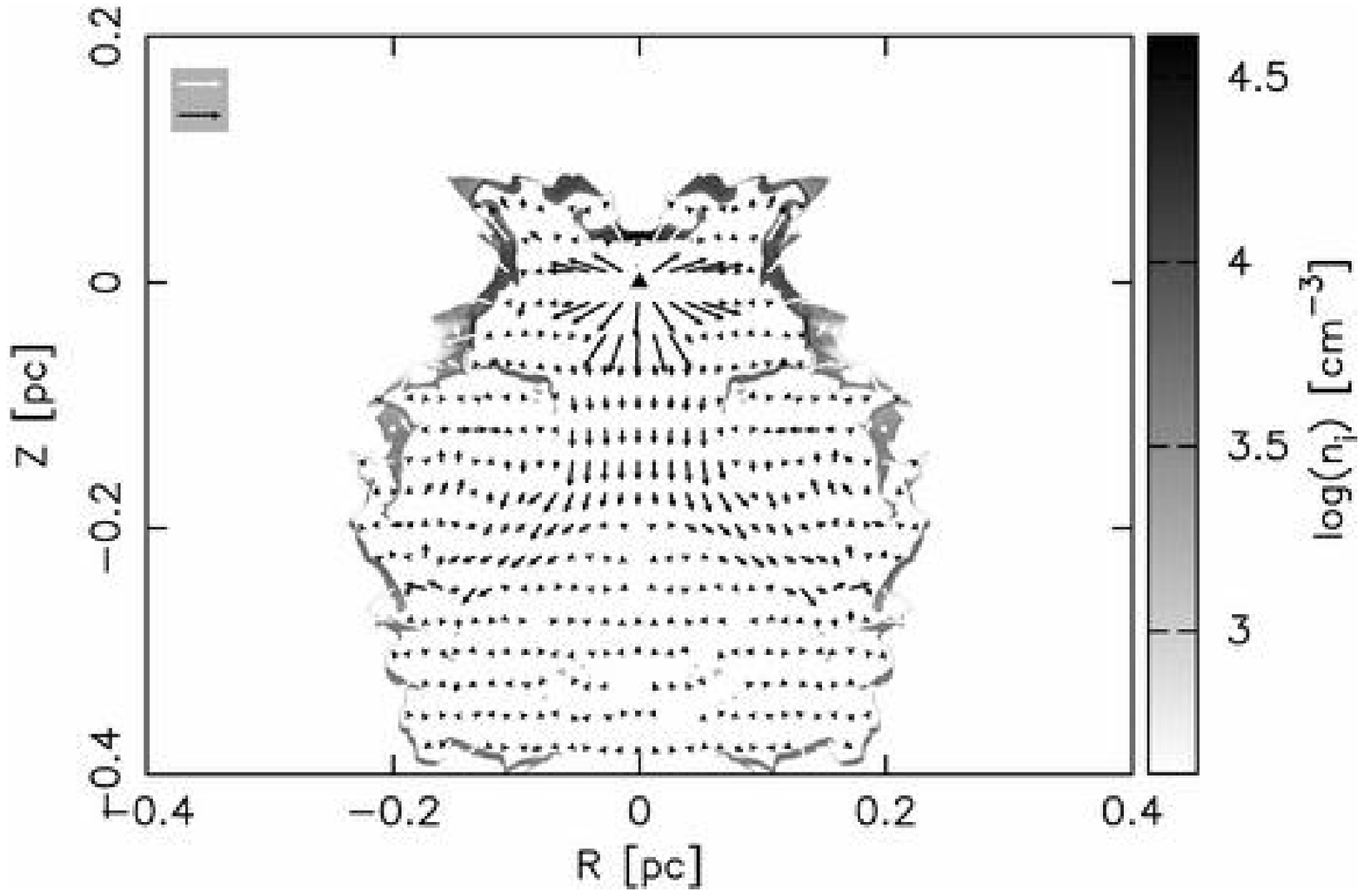}\\
\includegraphics{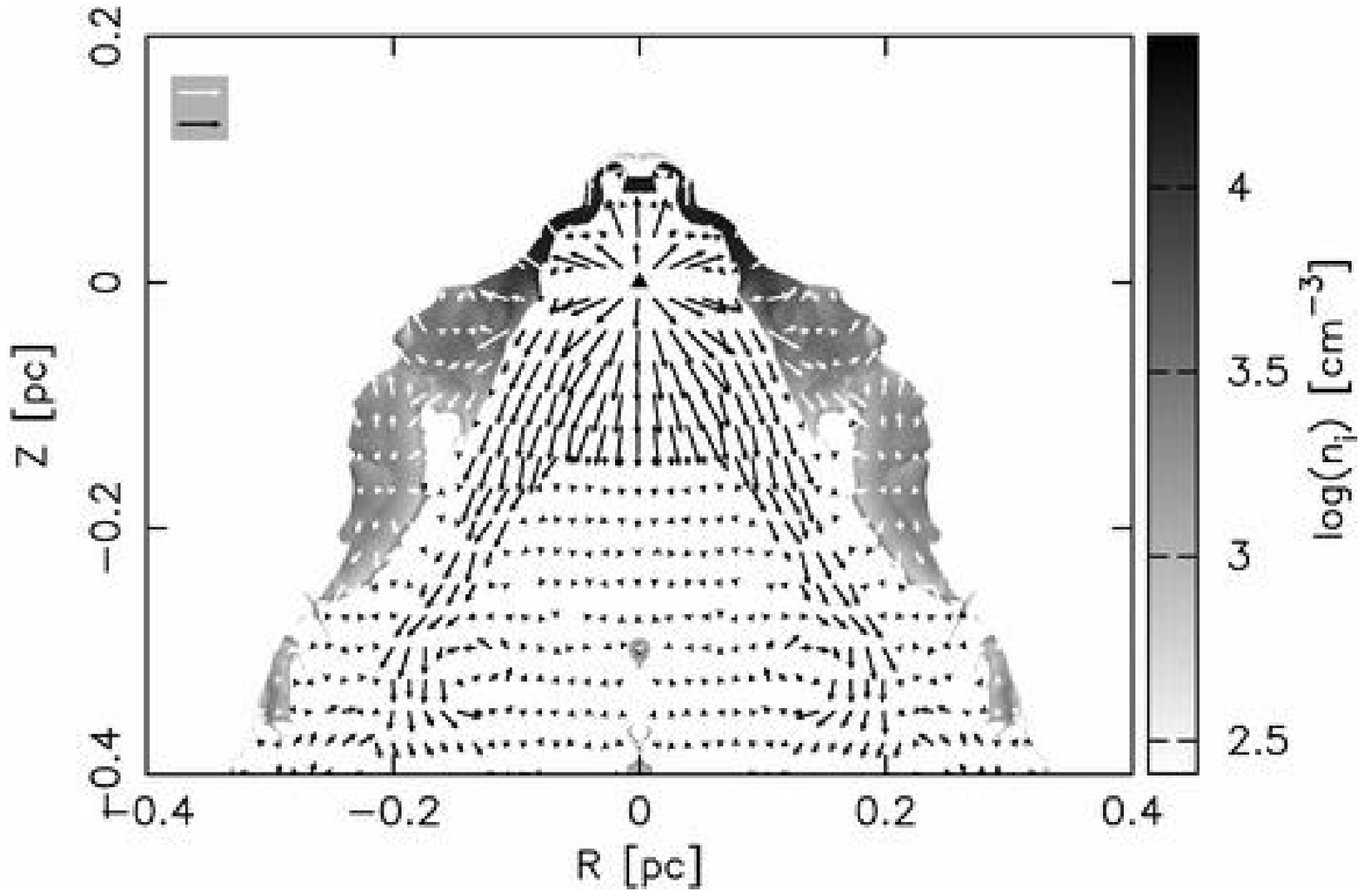}\\
\includegraphics{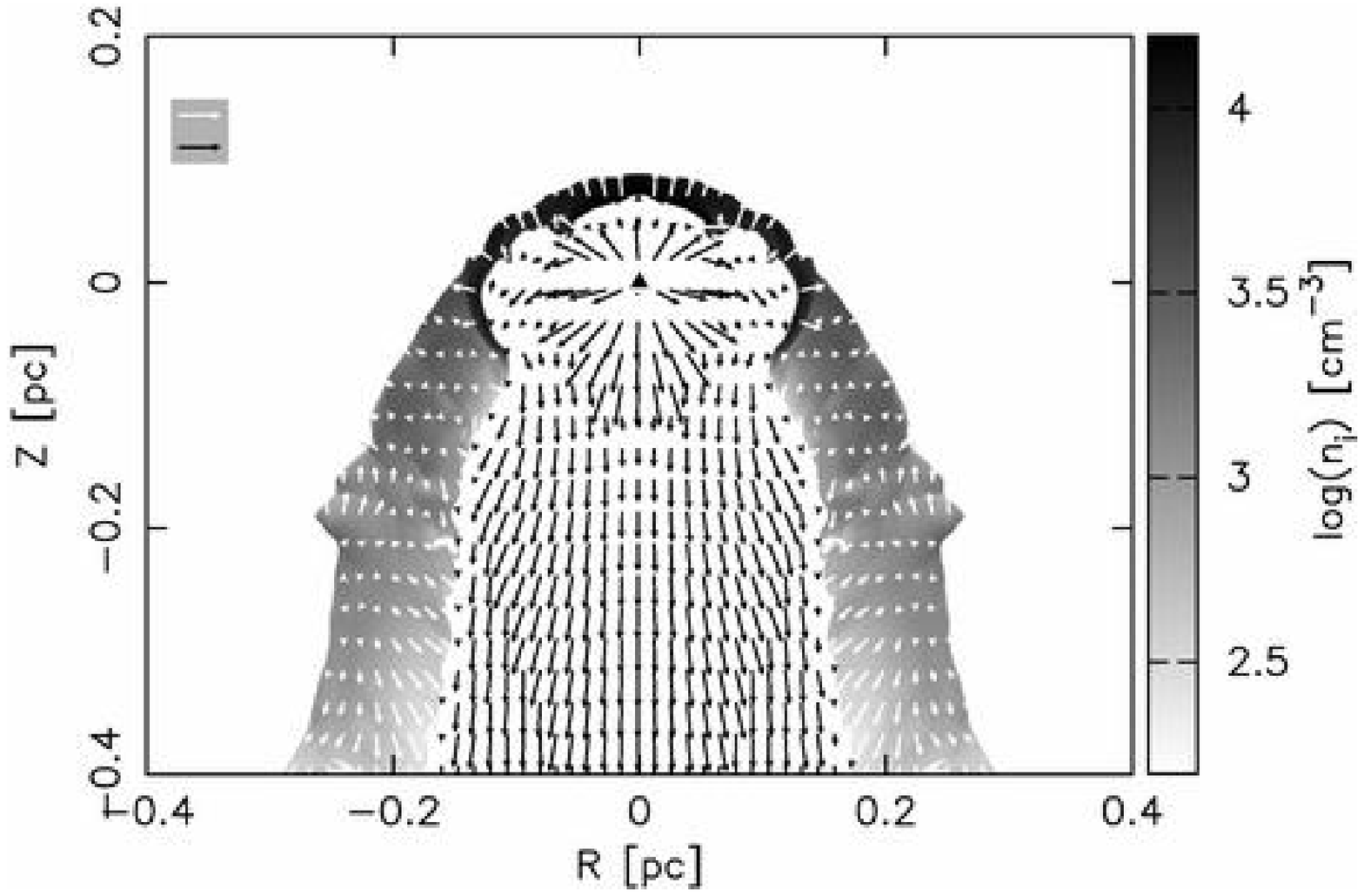}
\caption{Same as Figure~\ref{fig:ev1a} but for Model~B (simple 20~km~s$^{-1}$ bowshock model) and showing
evolution after 10,000~yrs (top panel), 20,000~yrs (middle panel) and
40,000~yrs (bottom panel). The black arrows represent velocities
of $30 < v \le 2000$~km~s$^{-1}$ in this figure, while the white arrows 
represent velocities of $v \le 30$~km~s$^{-1}$.}
\label{fig:ev1b}
\end{figure}
\begin{figure}
\centering
\includegraphics{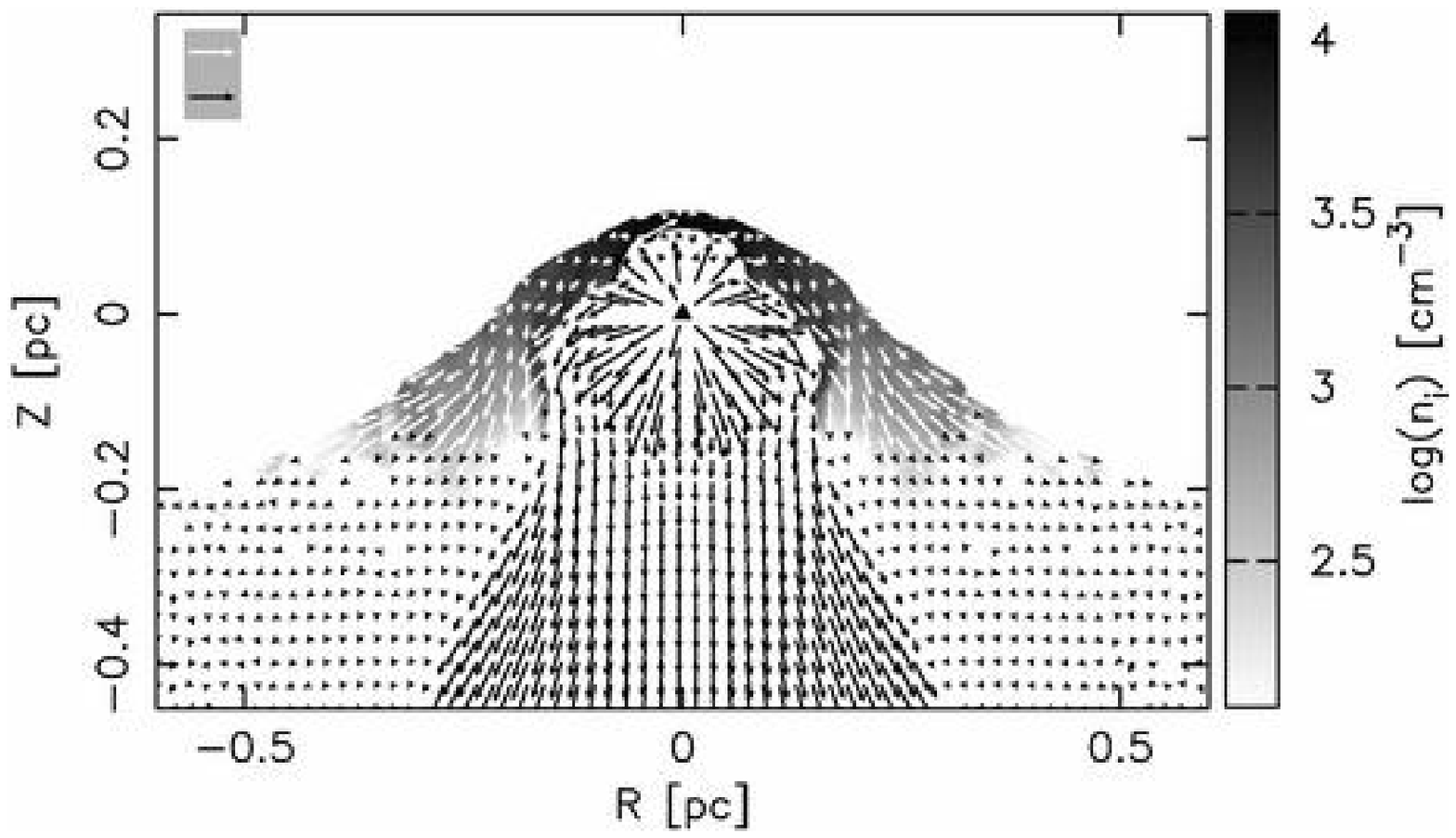}\\
\includegraphics{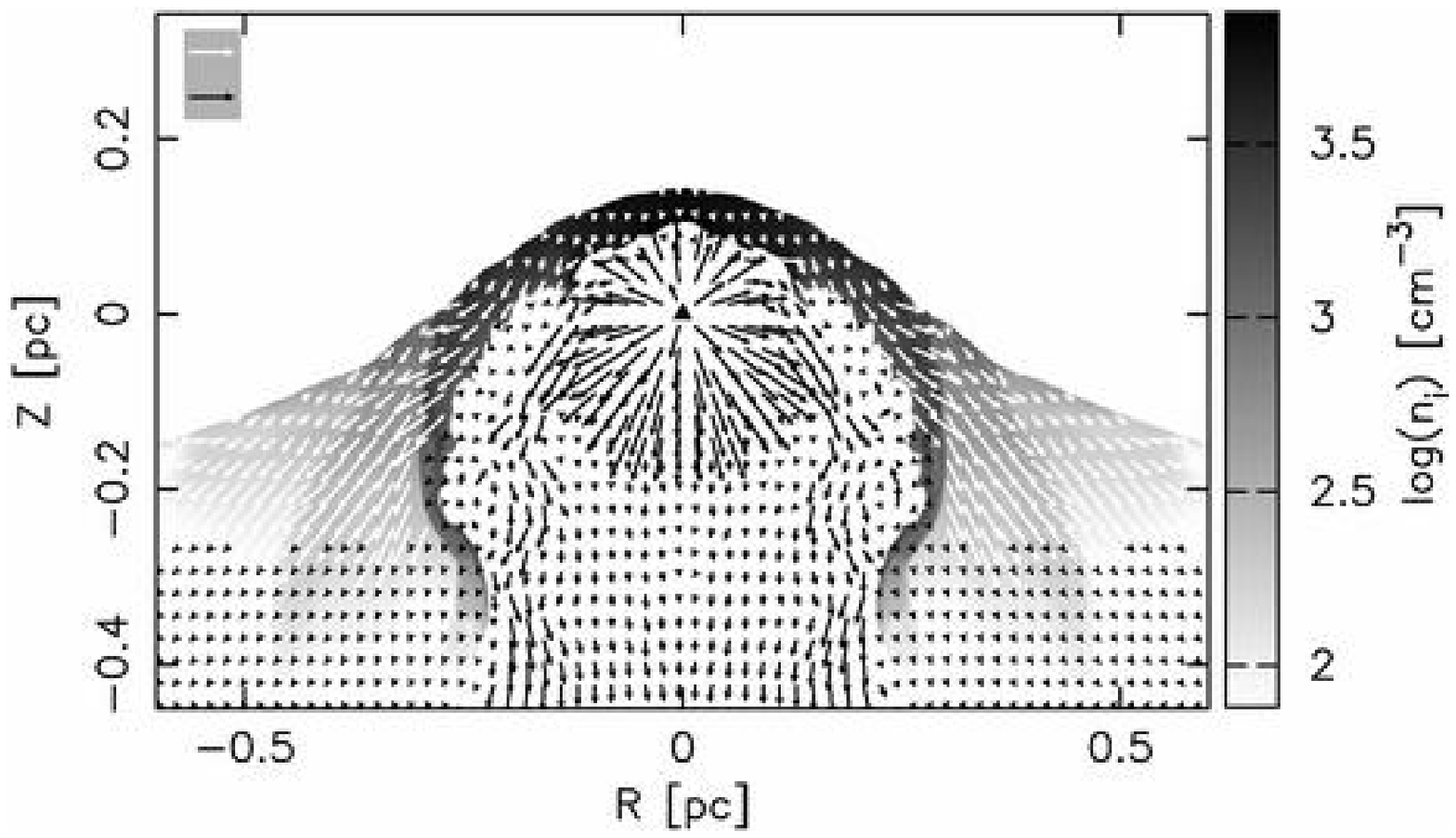}\\
\includegraphics{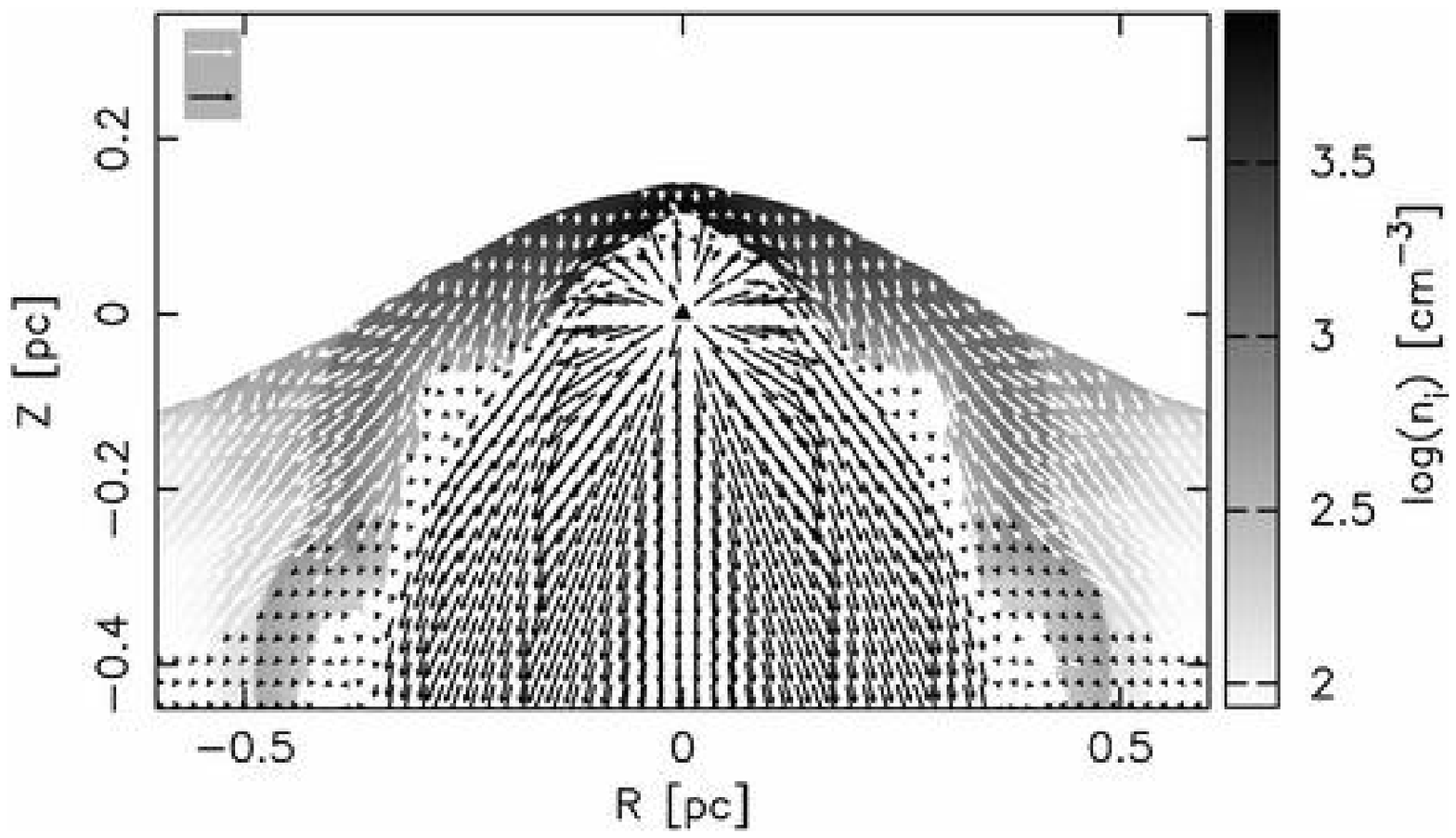}
\caption{Same as Figure~\ref{fig:ev1b} but for Model~C (champagne flow plus strong stellar wind).}
\label{fig:ev1c}
\end{figure}
Figures~\ref{fig:ev1a}, \ref{fig:ev1b} and \ref{fig:ev1c} show the
logarithm of the ionized density (greyscale) and 
velocity field in the $r-z$ plane for each of the canonical models A
(champagne flow), B (bowshock), and C (champagne flow plus stellar
wind) described in Table~\ref{tab:models} after 10,000~yrs, 20,000~yrs
and (for Models~B and C) 40,000~yrs of
evolution. The velocity field is represented by a split scale: the
black arrows show the highest velocity gas $v > 30$~km~s$^{-1}$
(generally representing stellar wind gas), while the white arrows are
scaled to the lower velocity gas $v \le 30$~km~s$^{-1}$ (present in the
photoevaporated flows and swept up shells). The grayscale shows the 
densest ionized gas in black with a dynamic range of 2 orders of
magnitude. Neutral gas, or very low density ionized gas (i.e., stellar
wind) appears white. 

From these figures we see that in the case of the pure
champagne flow, Model~A, the ionization front is smooth, whereas that of
Model~C, which has a stellar wind, has a rippled appearance due to the
formation of instabilities in the swept-up wind shell, such as those
presented in Figure~\ref{fig:h2wind} and discussed
above. In Model~B, the ionization front becomes trapped in the
swept-up wind shell, which forms a bowshock ahead of the moving
star. In Model~C the ionization front is trapped ahead of the star in
the denser gas but escapes at the sides where the density is lower and
forms a champagne flow here. 
\begin{figure}
\centering
\includegraphics{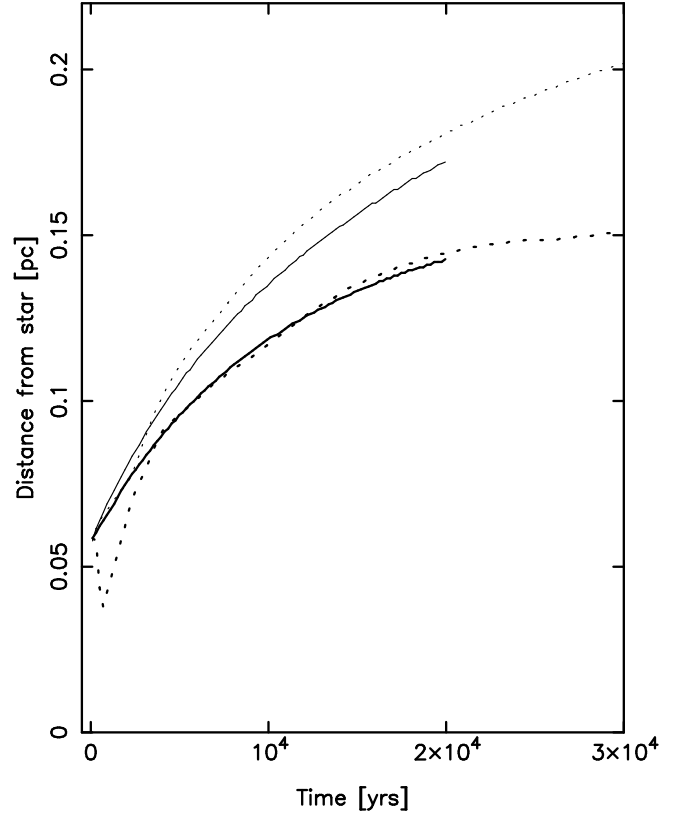}
\caption{Distance from star of ionization front propagating up the
density gradient against time for a champagne flow (thick solid line) and a
champagne flow plus stellar wind (thick dotted line). Also plotted are
the positions of the respective shock fronts in the neutral material
ahead of the ionization front: thin solid line---champagne flow,
thin dotted line---champagne flow plus stellar wind.}
\label{fig:distance}
\end{figure}
As regards temporal evolution, we see from Figure~\ref{fig:ev1a} that
the champagne flow model (Model A) does not appreciably
change in appearance between the two times depicted. This
quasi-stationary phase was described by \citet{2005ApJ...627..813H} and
the ionization front advances very slowly ($\sim 2$~km~s$^{-1}$) up
the density gradient. At the earlier time, the champagne flow has not
completely cleared the low-density material from the grid: the
rarefaction wave moving back from the initial Str\"omgren surface
through the ionized gas when the \ion{H}{2} region begins to expand
hydrodynamically reaches the symmetry axis earlier where the
ionization front shape is narrower. This shows up as a ``bump'' in the
velocity distribution of the upper panel of Figure~\ref{fig:ev1a}. By
the time 20,000~yrs have passed, the rarefaction wave has reached the
symmetry axis at all $z$ heights and the champagne flow fills the
whole grid. We note that for this model we do not see the shadowing
instability discussed in \S~\ref{subsubsec:expansion} because the
initial conditions are totally smooth and there is no stellar wind
shell to form clumps.  

In Figure~\ref{fig:distance} we can see how the presence of a strong
stellar wind affects a champagne flow. The initial ionization front
distance ahead of the star for Models~A and C is the same, reflecting
the fact that the ionized region is set up within one recombination
time of the star switching on. The ionization front distance in
Model~C then retreats, which corresponds to the formation of the dense
swept-up stellar wind shell that traps the ionization front. At these
very early times ($t < 4000$~yrs), before the density gradient has become important,
the shadowing instability produces spokes of ionized gas protruding
beyond the swept-up shell. However, once the bubble diameter is
greater than about one density scale height, the strong advection
around the shell due to the density gradient means that any clumps
that could cause shadowing are moving too fast for the instability to
take effect. The initial shadowing does open up ``holes'' in the thin
shell, through which the hot, shocked stellar wind can flow,
generating a turbulent interface. As the hot, pressurized, energy-driven stellar wind
bubble expands supersonically, the ionization front advances up the
density gradient and after 5000~yrs has caught up with the simple
champagne flow ionization front. Eventually, in the direction of
decreasing density the stellar wind bubble blows out, and there is a
transition to a momentum-driven flow. Meanwhile, a champagne flow
starts to establish itself within the swept-up shell. Initially, the
champagne flow is blocked from expanding down the density gradient
because the stellar wind is strong enough that the contact
discontinuity between the shocked wind and the swept-up material is
pushed past the position where the champagne flow would go through a
sonic point \citep[see, e.g., case~A of ][]{2005ApJ...627..813H} and
so the champagne flow stays subsonic on the axis. The champagne flow
first establishes itself off-axis, where the advancing ionization
front opens up and distances itself from the contact discontinuity,
allowing the flow to become supersonic. The opening up of the
ionization front appears to be driven by Kelvin-Helmholtz
instabilities, which form at the head and travel down the structure,
broadening as they reach regions of lower density and pressure. This
leads to variations in the opacity and allows the ionization front to
expand. Even though the external shock in Model~C is the stellar wind
shock, rather than the ionization front shock of Model~A, the shock
velocity is so small (see Figure~\ref{fig:distance}) that the density
enhancement is negligible and so the ionization front in Models~A and
C follows the same trajectory.  The majority of the opacity in this
part of the
\ion{H}{2} region comes from just behind the ionization front, since
this is where the densest gas is to be found, hence even though the
stellar wind has evacuated a cavity behind the ionization front, the
conditions in the photoionized gas behind the ionization front are
almost the same as in the pure champagne flow case and so the
expansion is the same.

\begin{figure*}
\centering
\includegraphics[width=0.75\textwidth]{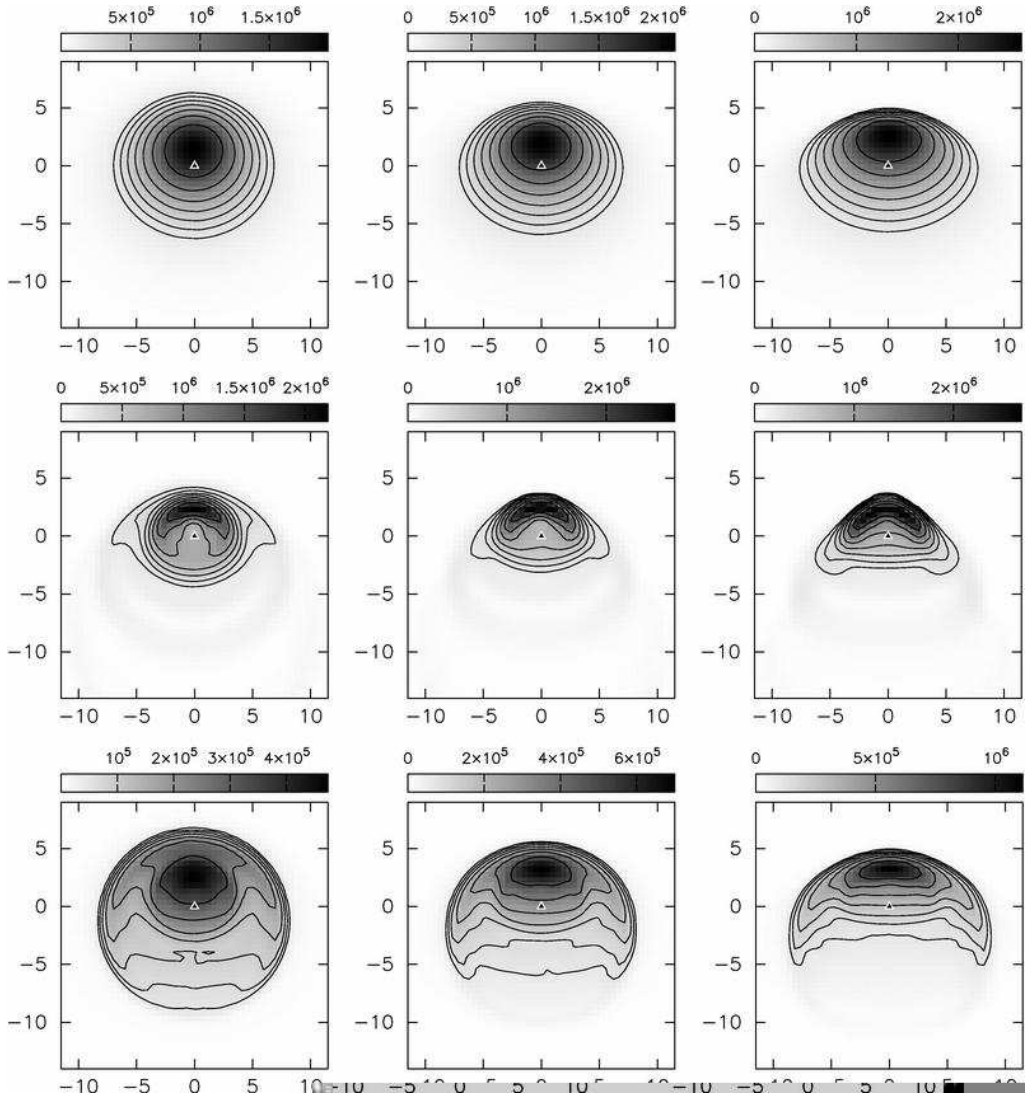}
\caption{Linear greyscale and logarithmic contours with $\sqrt{2}$
interval of the emission measure (units pc~cm$^{-6}$) for Models~A (top), B (center) and C (bottom)
at different projection angles after 20,000~years of evolution. For
each row: left panel---projection angle
$30\arcdeg$, center panel---projection angle $45\arcdeg$, right
panel---projection angle $60\arcdeg$. The scale for each panel is given in arcsec. The position
of the star is marked by a black triangle in each figure.}
\label{fig:emissmeasure}
\end{figure*}
\begin{figure*}
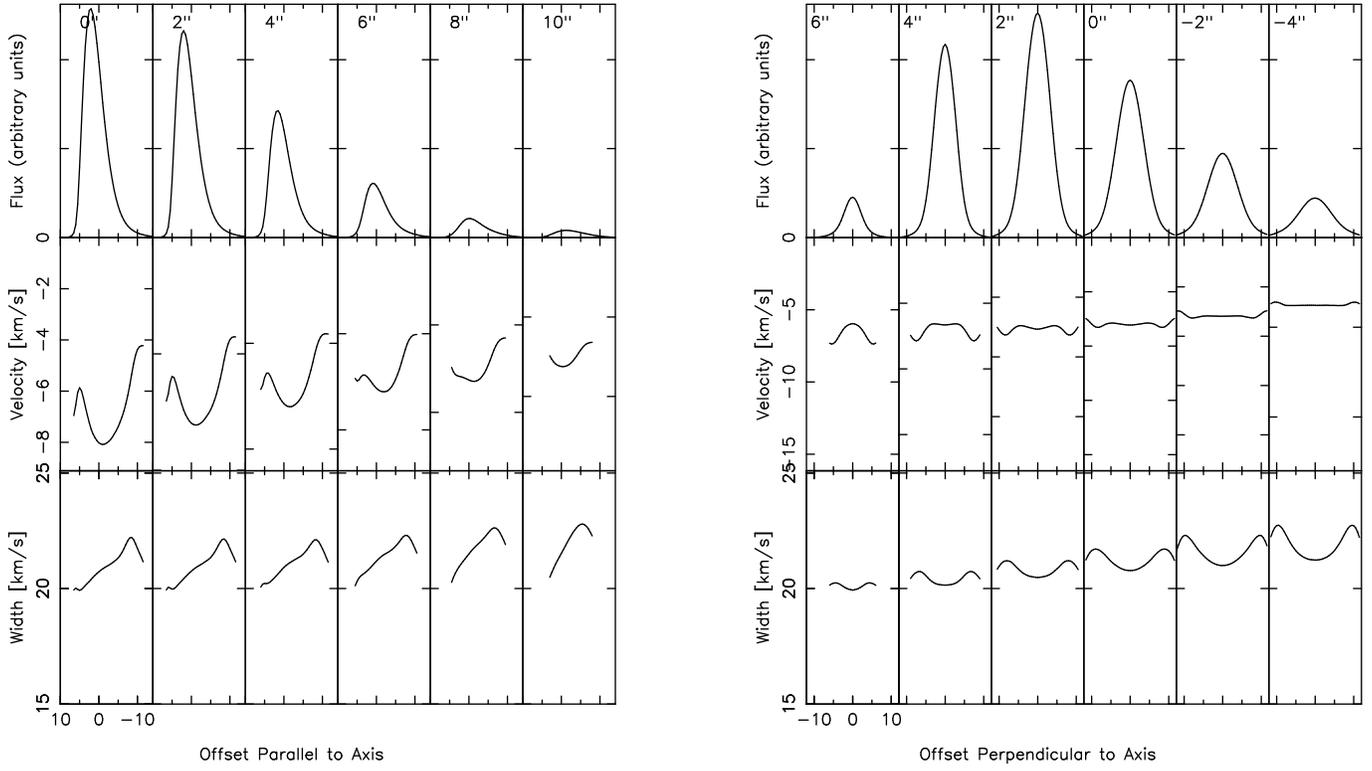

\centering
\includegraphics[width=0.45\textwidth]{f8a}\hfill
\includegraphics[width=0.45\textwidth]{f8b}
\caption{Simulated emission line moments for Model~A (simple champagne flow), inclination angle
$45\arcdeg$---line intensity, mean velocity and RMS velocity width
(multiplied by scale factor $2\sqrt{2\log 2}$) as a function of
position along the slit. Left: slits parallel to symmetry axis; Right:
slits perpendicular to symmetry axis. The offsets of the slits from
the star are indicated in the upper left corner.}
\label{fig:specA}
\end{figure*}
\begin{figure*}
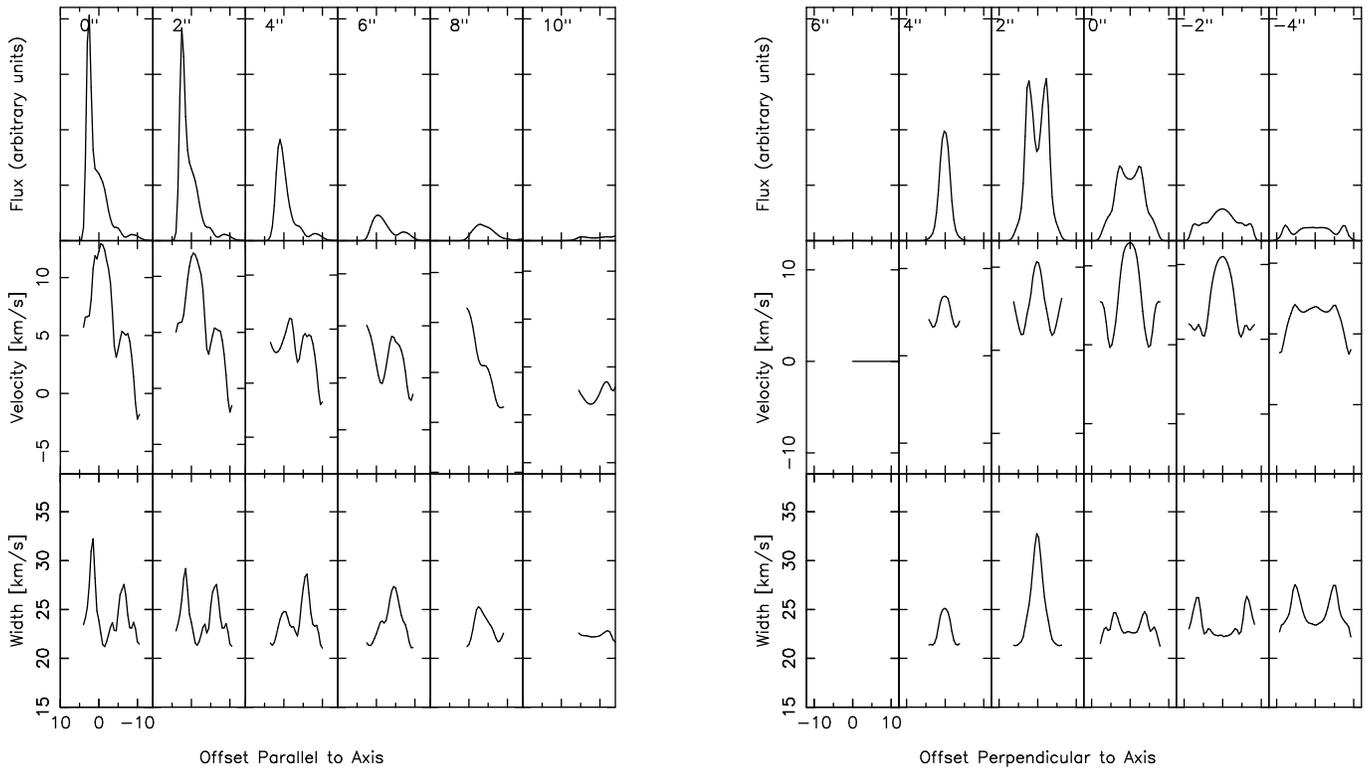

\centering
\includegraphics[width=0.45\textwidth]{f9a}\hfill
\includegraphics[width=0.45\textwidth]{f9b}
\caption{Same as Fig.~\protect\ref{fig:specA} but for Model~B
(20~km~s$^{-1}$ bowshock).}
\label{fig:specB}
\end{figure*}

In Model~B, at very early times, the shadowing instability is
important (not shown here) and is triggered by a cooling instability
in the swept-up thin shell (as discussed in
\S~\ref{subsubsec:expansion}), which produces shadowing clumps. At the
base of the spokes of ionized gas, the gas pressure widens the ``gap''
in the shell and allows the shocked stellar wind to flow through. This
can be seen in the top panel Figure~\ref{fig:ev1b} as ``ears'' on the
bowshock structure. Although strong advection around the bowshock
after $t > 10,000$~yrs leads to the disappearance of the
characteristic spokesof ionized gas, the initial instability is the
seed for a very turbulent interface between the hot, shocked stellar
wind and swept-up material.  This turbulent interface leads to the
shedding of large eddies from the head of the bowshock and affects the
thickness of the trapped ionized region. The eddies contain large
amounts of dense, ionized material. At later times, Kelvin-Helmholtz
instabilities at the interface region seem to be the dominant
instability \citep[cf.][]{1997RMxAA..33...73R,
  1998A&A...338..273C}. The time sequence of images shown in
Figure~\ref{fig:ev1b} shows that the shape of the bowshock varies
considerably. At 10,000~yrs the shell is very thin, but with a large
bowshock opening angle, at 20,000~yrs the shell is thin at the head
but thicker at the sides, with a narrow bowshock opening angle, while
at 40,000~yrs, the shell is quite thick and the opening angle of the
bowshock is also quite large. The dense ``blobs'' on the axis in the
20,000~yrs figure are a result of the imposed cylindrical symmetry
focussing the turbulent flow onto the symmetry
axis. \citet{1997RMxAA..33...73R} state that, for late times of their
models of high-velocity runaway O stars in low-density media, the
bowshock flow reaches a ``statistically stationary'' regime. Our
models have yet to achieve this regime but it could be expected that
they will, since by 40,000~yrs all the transient features due to the
initial instabilities have been advected a long way downstream. The
eddies that are shed periodically from the apex of the bowshock cause
pressure variations that alternately pinch then open up the region of
shocked stellar wind.

The velocity fields in these models are quite complex. The
photoevaporated flows in the champagne models (A and C) start at low
velocities at the ionization front ($v \sim 5$~km~s$^{-1}$, see case~A of
\citet{2005ApJ...627..813H}) and are initially accelerated rapidly to
the sound speed ($v\sim 10$~km~s$^{-1}$), and then more gradually due to
the density gradient, reaching velocities of around $v \simeq
30$~km~s$^{-1}$ before leaving the numerical grid (this gas would be
below the detection limit due to its low density). In Model~C, with a
stellar wind, we also see thin shell instabilities, which form first
in the densest part at the head of the cometary-shaped thin shell
(because of faster cooling here) and later round the sides. The
resulting knots are then deformed both by advection within the
swept-up shell and also by Kelvin-Helmholtz instabilities due to the
shear between the shell flow and the photoevaporated flow. Large knots
of dense ionized material formed at the head of the cometary flow can
thus be advected off the grid during the simulation timescale. 

In the bowshock case (Model B), velocities at the head are roughly
equal to the stellar velocity. Towards the tail of the cometary
\ion{H}{2} region, velocities in the dense, ionized gas are low and
randomly oriented due to eddies formed at the shear interface.
Although the highest velocities are to be found in the stellar wind
itself, this gas has such low density that it is not detectable.

\subsubsection{Simulated emission measure maps}
Figure~\ref{fig:emissmeasure} shows the simulated emission measure
smoothed by a Gaussian beam of $1.0\arcsec$. The numerical grid was
projected onto a pixel grid whose longest dimension of 80 pixels
represents a spatial size of $2R\cos i + Z\sin i$, where $R$ and $Z$
are the maximum extents of the radial, $r$, and longitudinal, $z$,
directions of the numerical model (see e.g., Figure~\ref{fig:ev1a}),
and $i$ is the inclination angle to the line of sight. In
Figure~\ref{fig:emissmeasure} we have arbitrarily assigned an angular
size of $40\arcsec$ to be equivalent to 80 pixels. This would place
all Models A and C at distances of approximately 7~kpc, depending
slightly on inclination angle, while Model~B would be at approximately
6~kpc. In Figure~\ref{fig:emissmeasure} we plot only a $24\arcsec
\times 24\arcsec$ subset, with the projected stellar position assigned
to $(0,0)$ in each case. For reasons of space, we show results only
for inclination angles $i = 30\arcdeg$, $45\arcdeg$ and
$60\arcdeg$. Emission measure maps for other inclination
angles are available from the authors on request.

The emission measure maps show a range of morphologies. Model~A
(champagne flow) gives a clump or fan-shaped emission, as expected,
while the addition of a stellar wind (Model~C) flattens this
emission and gives a more arc-like appearance. The bowshock
model (Model~B) results in an arc of limb-brightened emission,
as expected.

\subsubsection{Simulated spectra}
\begin{figure*}
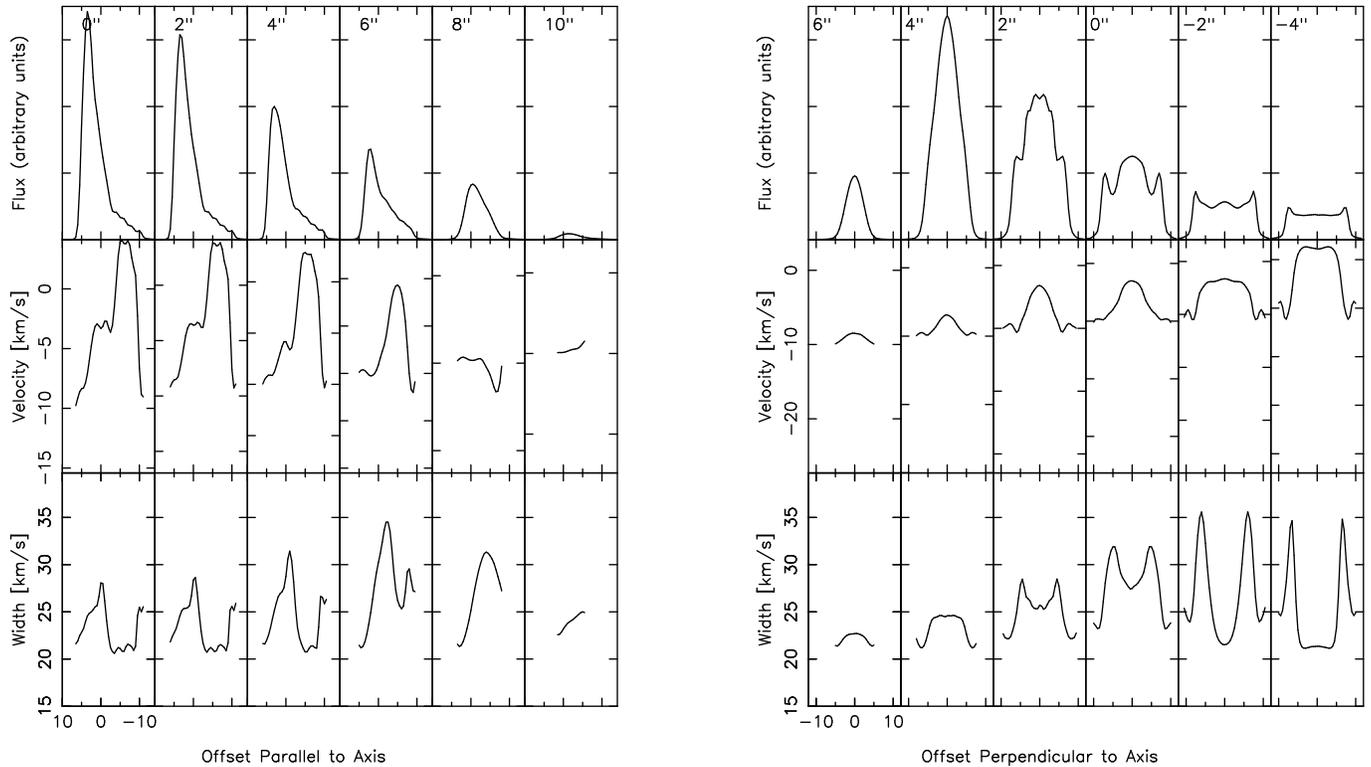

\centering
\includegraphics[width=0.45\textwidth]{f10a}\hfill
\includegraphics[width=0.45\textwidth]{f10b}
\caption{Same as Fig.~\protect\ref{fig:specA} but for Model~C
(champagne flow plus stellar wind).}
\label{fig:specC}
\end{figure*}

In Figures~\ref{fig:specA}--\ref{fig:specC} we plot the line
intensity, mean velocity and velocity dispersion for slits parallel
and perpendicular to the symmetry axis for the models shown in
Figure~\ref{fig:emissmeasure} for inclination angle
$45\arcdeg$. Results for other angles are available from the
authors on request. These quantities are evaluated using the emission
line moments, where the velocity dispersion is the RMS width scaled by
a factor $2\sqrt{2\log 2}$, which would give the FWHM in the case of a
Gaussian profile. In each case the slit width is one arcsec and we
have included thermal broadening using the temperatures which result
from the numerical simulation. In the gas in
photionization equilibrium this temperature is $10^4$~K but the
partially ionized gas (either recently photoionized or in the process
of recombining) will be at a lower temperature, and gas in the
ionized, shocked stellar wind will be at a much higher
temperature. The flux intensity for each slit is relative to the
maximum intensity of the whole object. The slit offsets indicated are
relative to the symmetry axis (for the parallel slits) and stellar
position (perpendicular slits).

The peak flux intensity occurs on the symmetry axis ahead of the
star. For the slits parallel to the symmetry axis, the flux intensity
decreases steeply along the slit for all models. For the slits
perpendicular to the symmetry axis, there are differences between
Models~A, B, and C. For Model~A (champagne flow), the peak flux in
each slit always occurs on the symmetry axis and falls off to the
sides. For Models~B (bowshock) and C (champagne flow plus wind), the
flux intensity has a flat-topped shape as the slits move down the tail
of the cometary \ion{H}{2} region, with peaks at the sides due to the
emitting mass being concentrated in a shell around the low-density
stellar wind bubble.

The velocity structures of Models~A, B, and C are very different. In
Figure~\ref{fig:specA} we see that the flow in Model~A leaves the head
of the cometary \ion{H}{2} region with blueshifted velocities. It then accelerates and reaches maximum
(blueshifted) speed at the projected position of the star. Thereafter,
the velocities become more redshifted because of the contribution from
the higher density, lower velocity, redshifted gas on the other side
of the object in projection.

In Figure~\ref{fig:specB} (Model~B) the velocities at the head are redshifted
and correspond to the bowshock region being pushed forward ahead of
the moving star. The noisiness of this spectrum is due to the eddies
that form in the shear flow interface between the shocked stellar wind
and the swept-up ambient medium. The drop in velocity just ahead of
the projected position of the star in the slits parallel to the
symmetry axis is because the line-of-sight cuts through a large region
of low-density, low-random-velocity gas as well as through the more
ordered, high blue-shifted velocity, dense material in the swept-up
shell in the bow region. The velocities fall off steeply towards the
tail region because the influence of the stellar motion on the gas is
very small by this point.

Figure~\ref{fig:specC} (Model~C) shows a combination of blue-shifted and
red-shifted material. As for the simple champagne flow (Model~A), the
photoevaporated flow leaving the ionization front is
blue-shifted. However, the stellar wind confines the champagne flow
into a shell, causing it to flow around the stellar wind bubble. The
line of sight through the shell at offset $\sim -7 \arcsec$ cuts
through a thick, dense region of redshifted velocities, which weight
the spectrum and lead it to turn over.

The velocity widths of Model~A increase from head to tail of the
champagne flow but are uniform across it for a given slit, and do
not differ much from the value of the thermal width. In Model~B,
the velocity widths are noisy for slits parallel to the symmetry axis,
due to the eddies formed in the shear flow. The widths of
Model~C increase towards the wings of the cometary
\ion{H}{2} region, due to the ionized gas being channeled into a shell
and accelerated around the stellar wind zone.

Although we do not show them here, the spectra for Models~B and C at
40,000~yrs are very similar to those at 20,000~yrs, particularly as
regards the velocity range and pattern. There are differences in the
line widths, with the widths after 40,000~yrs being generally
smoother, reflecting the fact that eddies formed earlier in the
simulation at the head of the object are advected down the structure
where they expand and so their influence diminishes. 

\section{Modified models}
\label{sec:modified}
Taking Models~A, B, and C as our reference models, we vary parameters
such as the stellar wind strength and the density scale height to
examine their effect on the morphology and kinematics of the resultant
cometary \ion{H}{2} regions. We also study the effect of stellar
motion up a density gradient, considering stellar speeds of 5 and
10~km~s$^{-1}$. The parameter
space to explore is huge, and we restrict ourselves to a few
representative cases.
\subsection{Model parameters}
\subsubsection{Champagne flow plus stellar wind: Models D, E, and F}
These models are similar to Model~C, except that in Model D the stellar wind mass
loss rate is an order of magnitude smaller, $\dot{M} = 10^{-7}
M_\odot$~yr$^{-1}$, in Model~E the scale height of the density
distribution is much larger, $H = 0.2$~parsec, and hence the density
gradient is shallower. In Model~F the
density distribution is spherical, rather than plane parallel, and is
given by
\begin{equation}
	n = n_{\rm 0}\left[1 + \left(\frac{R}{r_{\rm c}}\right)^2 \right]^{-\alpha}
\end{equation}
where $n_{\rm 0} = 9.8\times 10^6$~cm$^{-3}$ is the number density at
the center of the core of this distribution, where the core is assumed
to have a radius $r_{\rm c} = 0.01$~pc, and $R^2 = r^2 + z^2$ is the
spherical radius. The exponent $\alpha = 1$ in
this model, giving $n \propto R^{-2}$ at large radii. The star is
offset from the center of the distribution such that $n = 8000$~\pcc\ at its
position.
 
\subsubsection{Champagne flow plus stellar wind plus stellar motion: Models
G, H, and I}
These models combine a density gradient, stellar wind and stellar
motion.  Models~G and H
have an exponential density distribution with scale height $H =
0.05$~parsec. In Model~G the star's velocity is 5~km~s$^{-1}$ and the
star starts from 0.05~pc below the position where the number density
is $n = 8000$~\pcc, while
for Model~H the velocity is 10~km~s$^{-1}$ and the star starts
from 0.1~pc below this reference position. Finally, Model~I also has an
exponential density distribution but with a larger scale height, $H =
0.2$~pc and the star's velocity is $V_* = 10$~km~s$^{-1}$, where the
star again starts its motion a distance 0.1~pc below the position
where the number density is $n = 8000$~\pcc.

\subsection{Modified model results}
\begin{figure}
\centering
\includegraphics{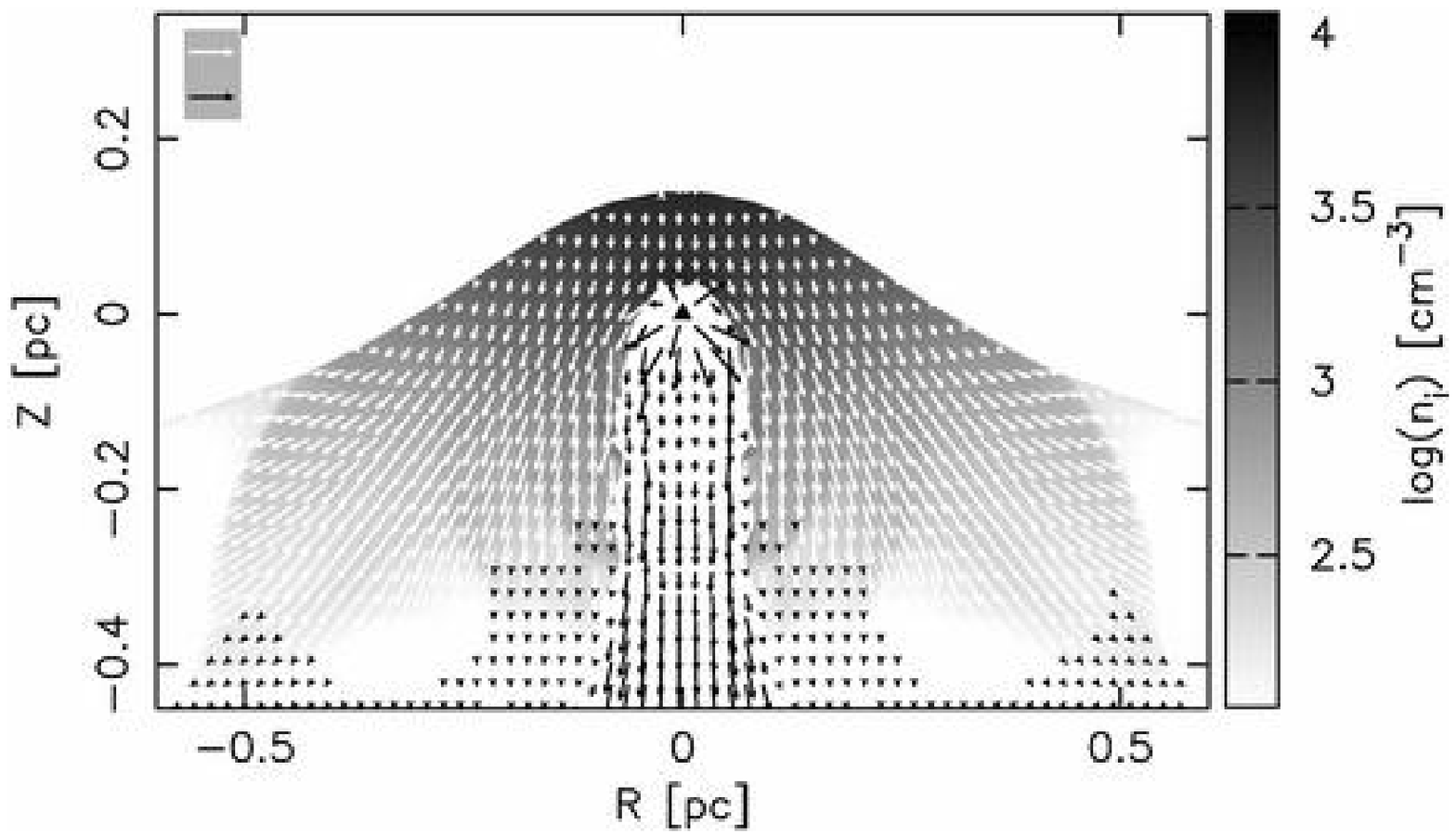}\\
\includegraphics{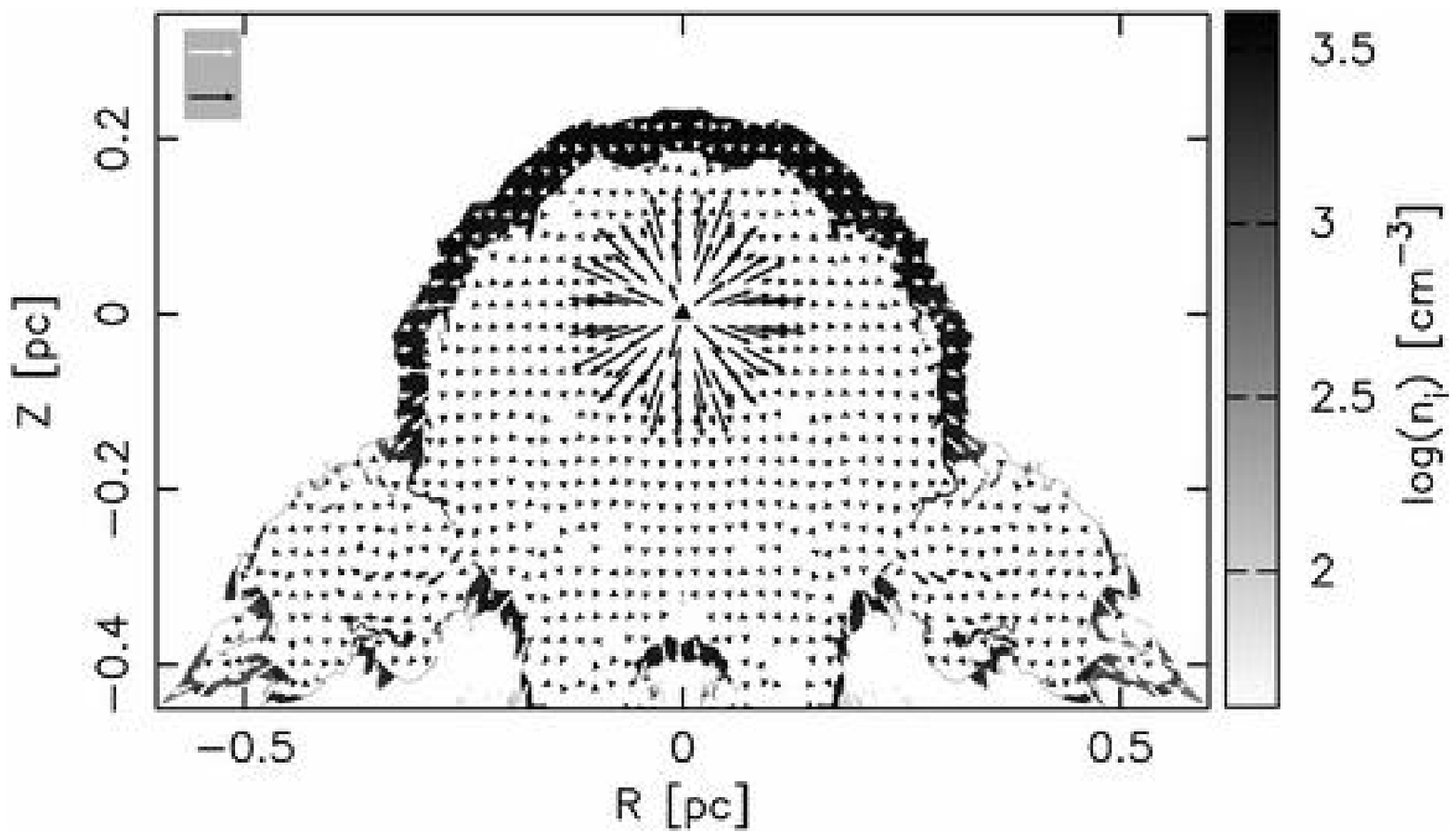}\\
\includegraphics{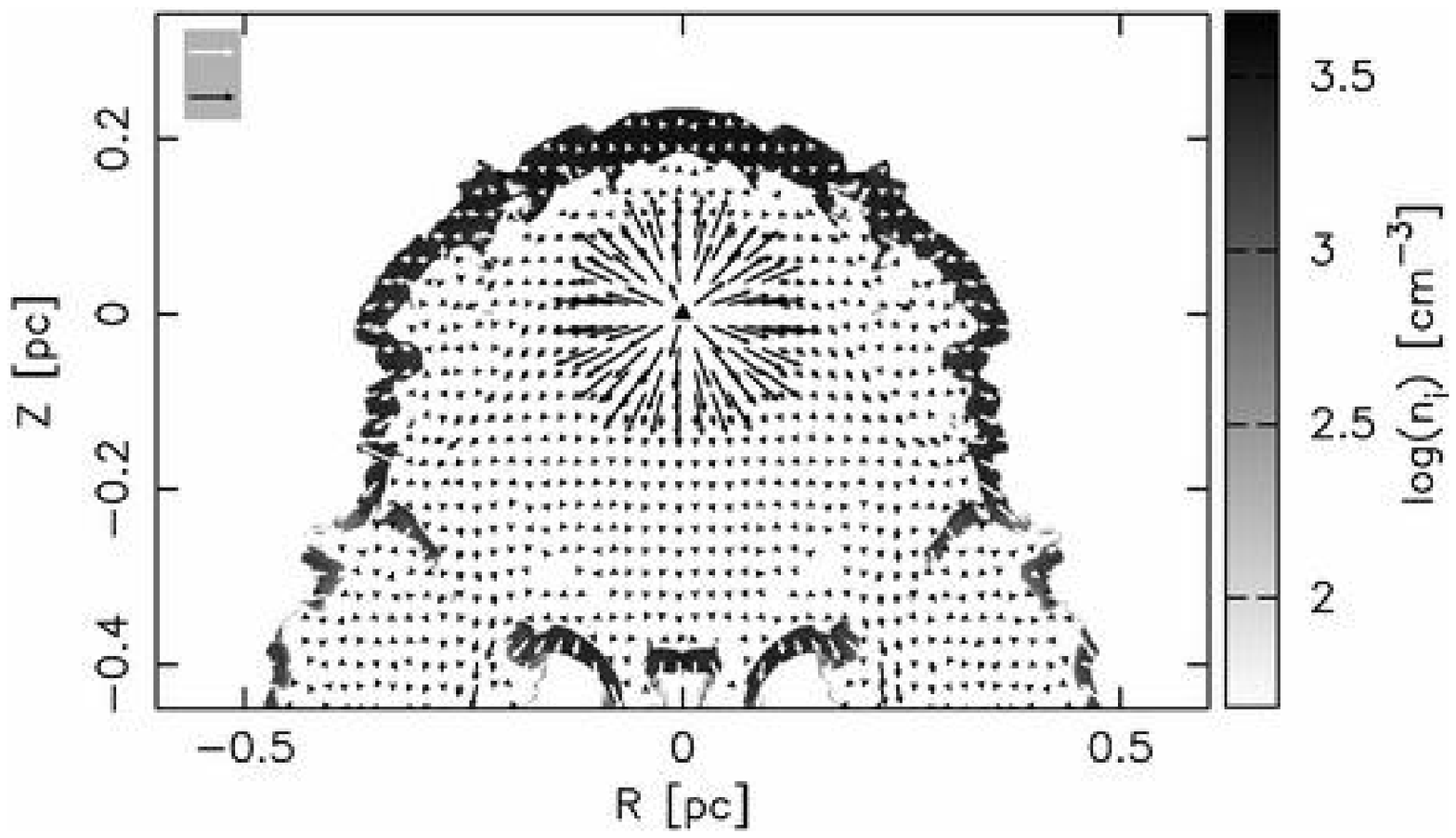}
\caption{Same as Fig.~\protect\ref{fig:ev1b} but for models (from top to bottom)  
(a) Model D: champagne flow plus stellar wind, $H = 
0.05$~pc, $\dot{M} = 10^{-7} M_\odot$~yr$^{-1}$  (b) Model E:
champagne flow plus stellar wind, $H = 0.2$~pc, $\dot{M} = 10^{-6}
M_\odot$~yr$^{-1}$ (c) Model F: champagne flow, spherical density
distribution. All models shown after 20,000~years of evolution.}
\label{fig:ev1x}
\end{figure}
\begin{figure}
\centering
\includegraphics{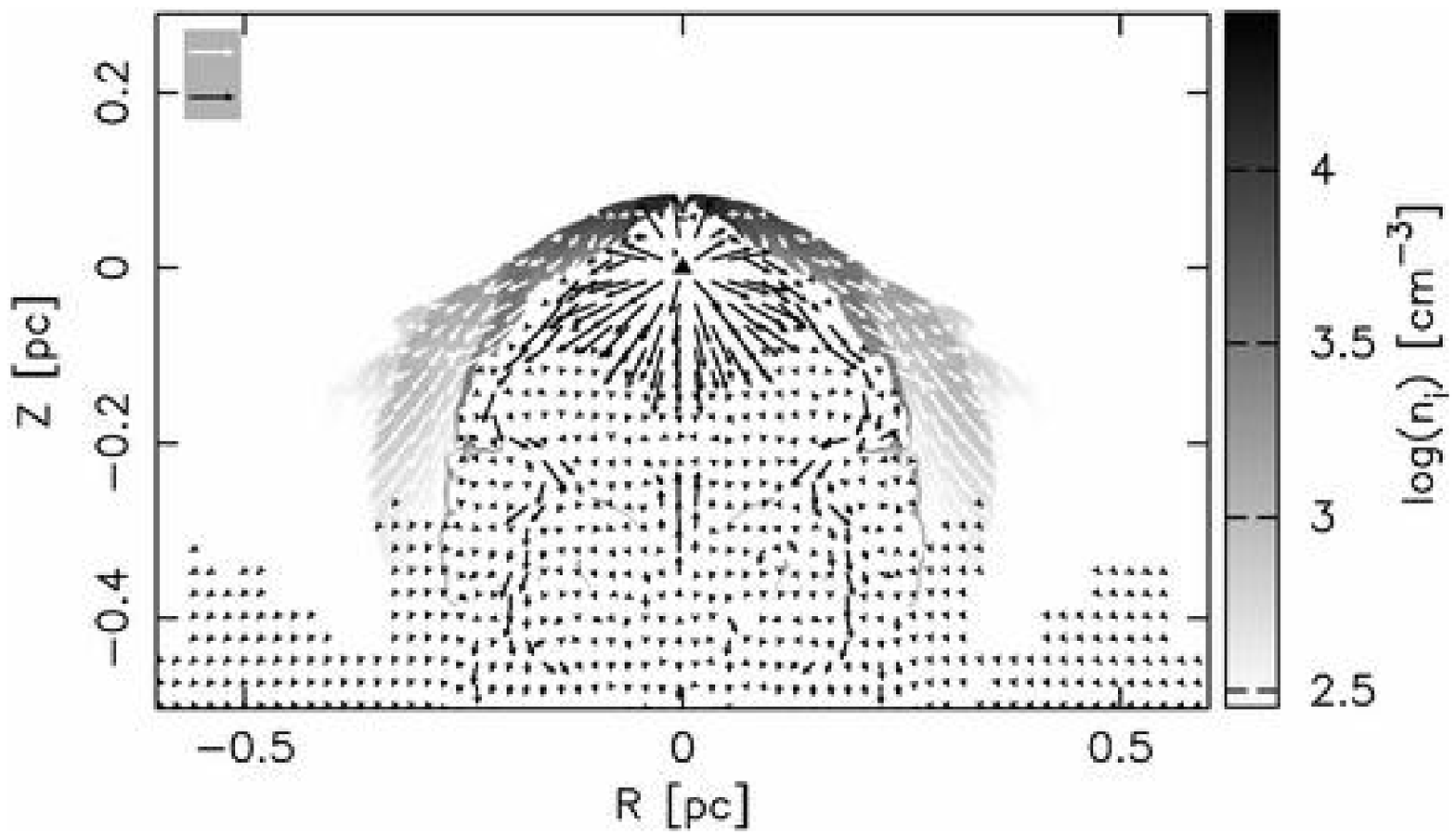}\\
\includegraphics{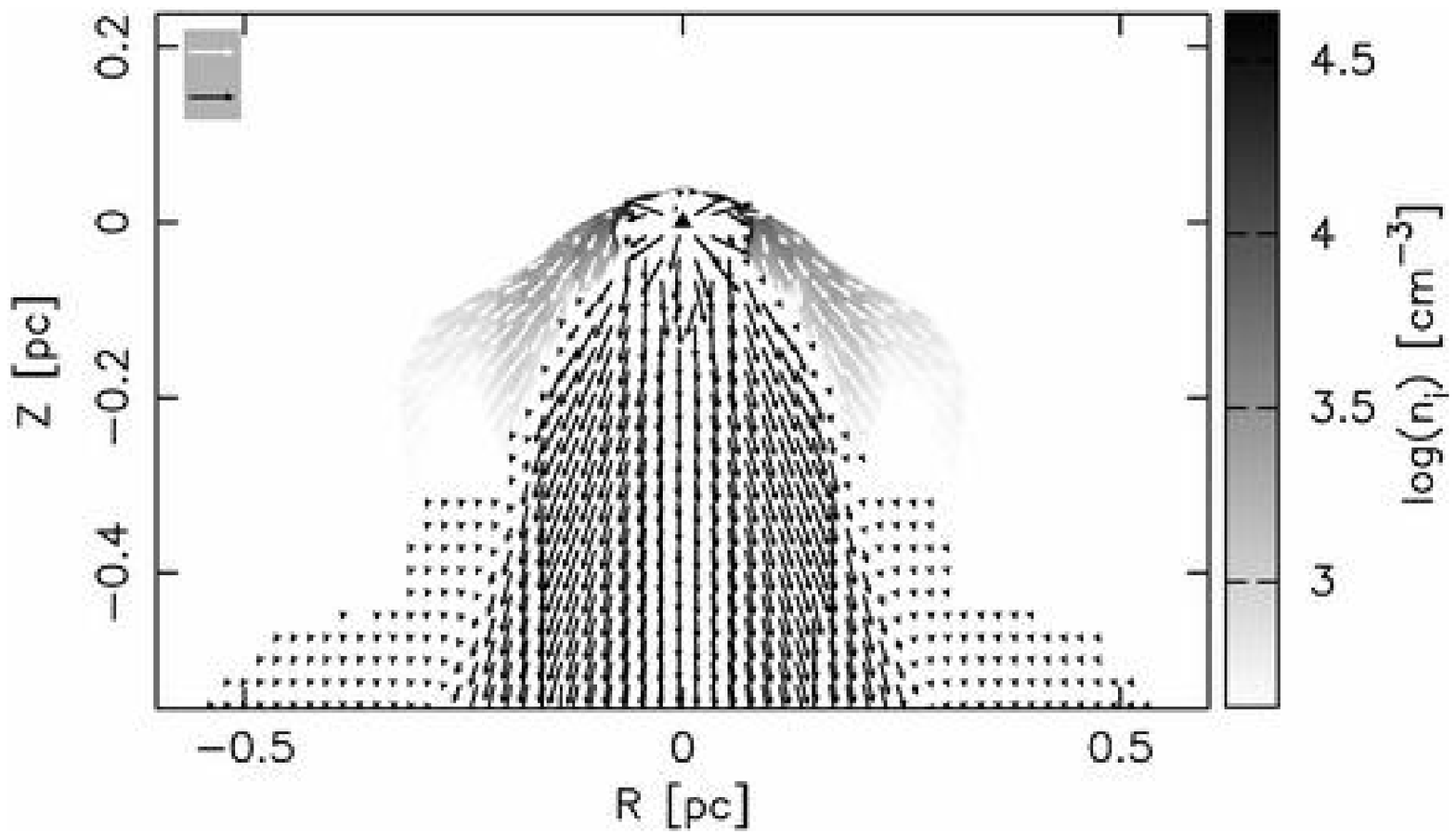}
\caption{Same as Figure~\protect\ref{fig:ev1b} but for models with
density gradient plus stellar wind plus stellar motion. (a) Top---Model G: $V_* =
5$~km~s$^{-1}$, $H = 0.05$~parsec. (b) Bottom---Model H: $V_* = 10$~km~s$^{-1}$,
$H = 0.05$~parsec. Both models shown after 20,000~yrs
of evolution.}
\label{fig:ev2x}
\end{figure}
\begin{figure}
\centering
\includegraphics{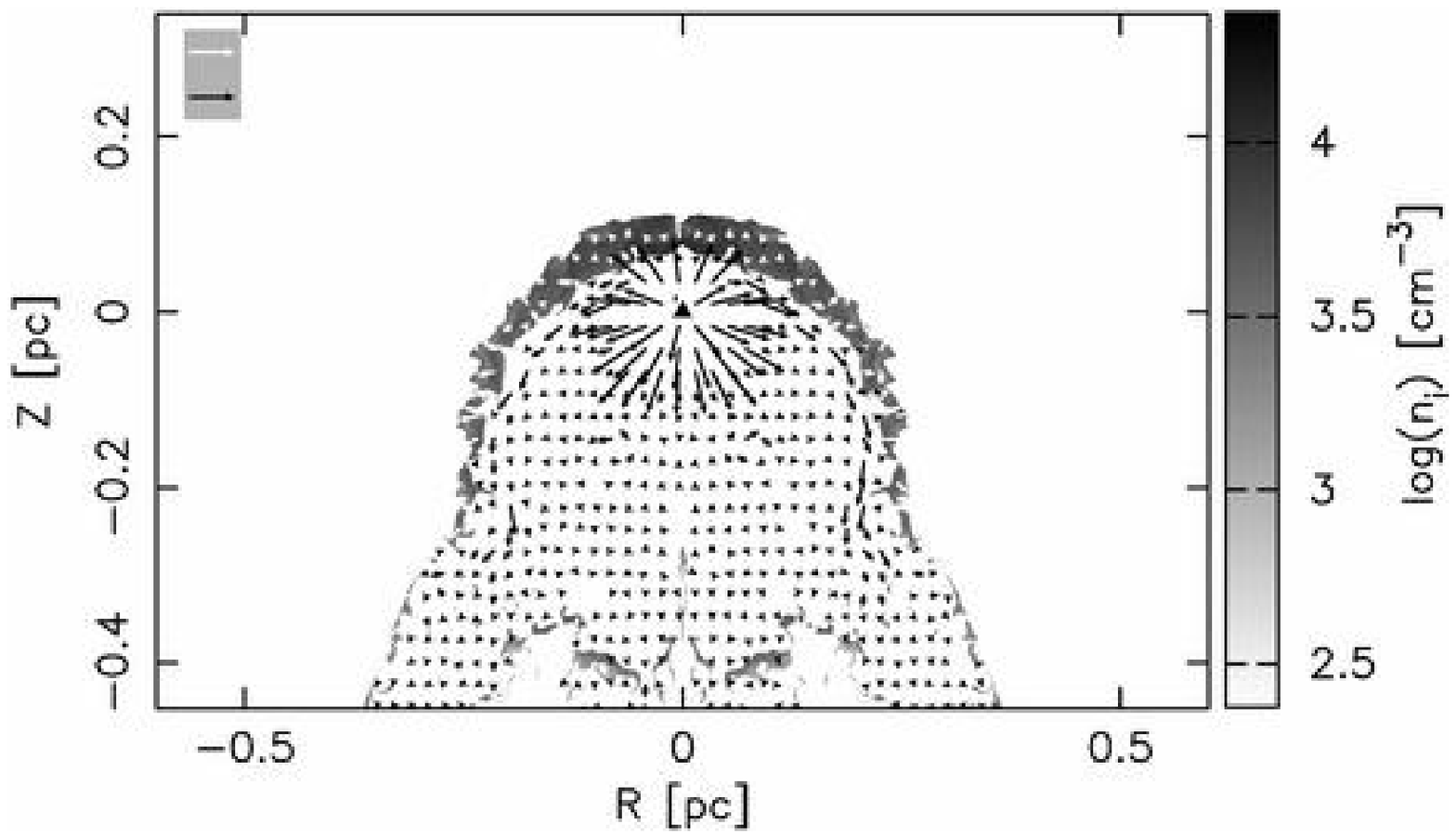}\\
\includegraphics{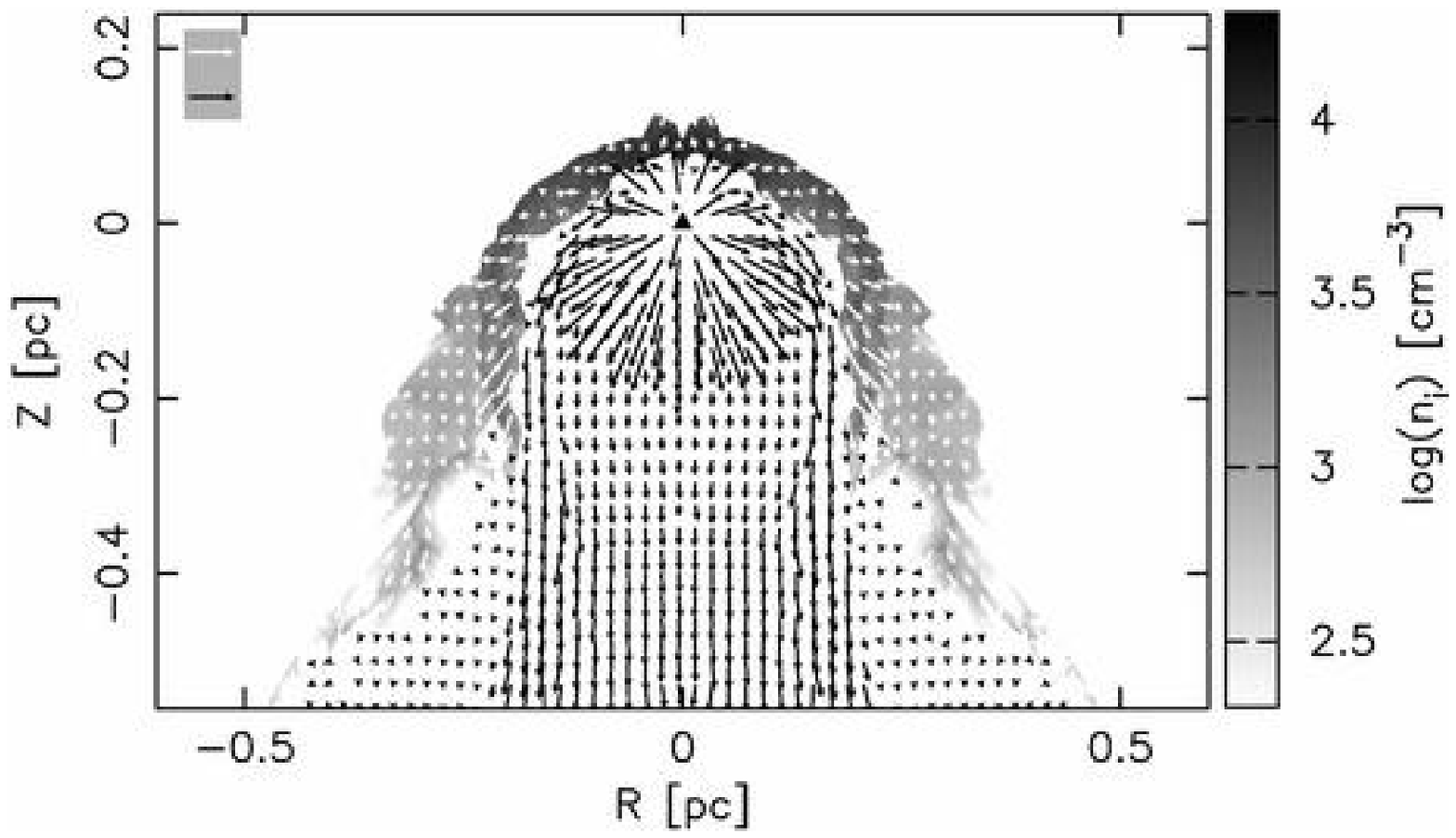}
\caption{Same as Figure~\protect\ref{fig:ev1b} but for models with
density gradient plus stellar wind plus stellar motion (from 
top to bottom) (a) Model I: $V_* =
10$~km~s$^{-1}$, $H = 0.2$~parsec after 10,000~yrs (b) Model I after 20,000~yrs of evolution.}
\label{fig:ev2y}
\end{figure}
\subsubsection{Velocity field}

Figure~\ref{fig:ev1x} shows the ionized density (greyscale) and
velocity field in the $r-z$ plane for Models~D, E and F of
Table~\ref{tab:models} after 20,000~yrs of evolution. Model~D looks
very similar to a champagne flow, with the stellar wind only affecting
a narrow region deep within the ionized zone. Models~E and F are both
cases of a champagne flow in a shallow density distribution. In these
cases, the stellar wind can reach pressure balance with the external
medium within a scale height and the inner wind shock is almost
spherical. The shocked stellar wind region is also approximately
spherical, except for in the tail region where it has broken out into
the lower density surroundings. At the head of the cometary \ion{H}{2} region there is a
thick shell of ionized gas flowing off the ionization front, which is
then diverted around the outside of the stellar wind bubble. Because
the ambient density does not fall off very quickly, the density in the
ionized gas shell is high and fairly uniform, and the ionization front
becomes trapped in the shell, unlike in Model~C, where the rapid
fall-off in density means that the ionization front can break through
the swept-up shell in the tail. In Models~E and F the shallow density
gradient allows the shadowing instability to form broad spokes in the
lowest density gas, where the pressure at the base of the spokes opens
gaps in the swept-up shell through which the shocked stellar wind can
flow. This can be seen clearly in the corresponding panels of
Figure~\ref{fig:ev1x}.

The velocities in Model~D are very similar to the simple champagne
flow model. In Models~E and F the velocities in the ionized shell are
fairly low and there are some turbulent regions due to instabilities
which are slowly being advected down the shallow density gradient. 

Figures~\ref{fig:ev2x} and \ref{fig:ev2y} show the ionized density and velocity fields
for Models G, H, and I. Model I has the shallowest density
gradient and this is reflected in its more spherical, thicker,
ionized shell. There is a lower emission measure region outside of the main
shell. Models G and H have the same steep density
gradient. Both of these models show a champagne flow as well as a
stellar wind swept-up shell. The stellar velocity in Model G is half that
of Model H and this results in the photoionized shell ahead of the
star in Model G being thicker than that in Model~H. In
Figure~\ref{fig:ev2x}, the star in Model~H has reached a denser part
of the ambient density distribution than the star in Model~G but
comparing the two models when the star is at the same position still
shows that the photoionized shell ahead of the star in Model~H is
thinner. Also, the hot, shocked stellar wind in Model~G occupies a greater
fraction of the volume than in Model~H. In fact, in the tail of Model~H, the
stellar wind does not go through a global decelerating shock as in
Model~G, but instead appears to have an oblique shock surface that
neither heats nor decelerates the wind very much. In Model~I, the
stellar wind suffers a global decelerating shock but then in the
downstream direction is focussed by the ionized shell and
reaccelerated by a nozzle effect.

The velocity fields of Models G, H, and I show that large
accelerations can occur around the sides of the flow in the ionized
gas, with velocities reaching 30~km~s$^{-1}$ or more. This is an
effect of including a density gradient, since in a simple bowshock
model there is no acceleration around the sides and velocities fall
off quickly towards the tail. Models G and H show the largest
accelerations towards the tails of the champagne flows, and this is
entirely consistent with them having the steepest density gradient. In
Model~I, the highest velocities occur in the swept up shell as it
expands into the low density region. Ahead of the star, the velocities
at the ionization front in these three models are low (i.e., lower
than the stellar velocities) and in the direction of stellar motion,
unlike the stationary star cases.

In these models, the shadowing instability only occurs
at the earliest times and is most noticeable in Models~G (due to the
low stellar velocity) and I (due to the shallow density gradient). In
both cases the instability soon disappears ($t < 3000$~yrs) because of
advection of the shadowing clumps around the shell. At later times the
dominant instabilities are Rayleigh-Taylor and Kelvin-Helmholtz type
due to the shear between the champagne flow and the shocked stellar
wind flow. 
\subsubsection{Emission measure maps}
\begin{figure*}
\centering
\includegraphics[width=0.75\textwidth]{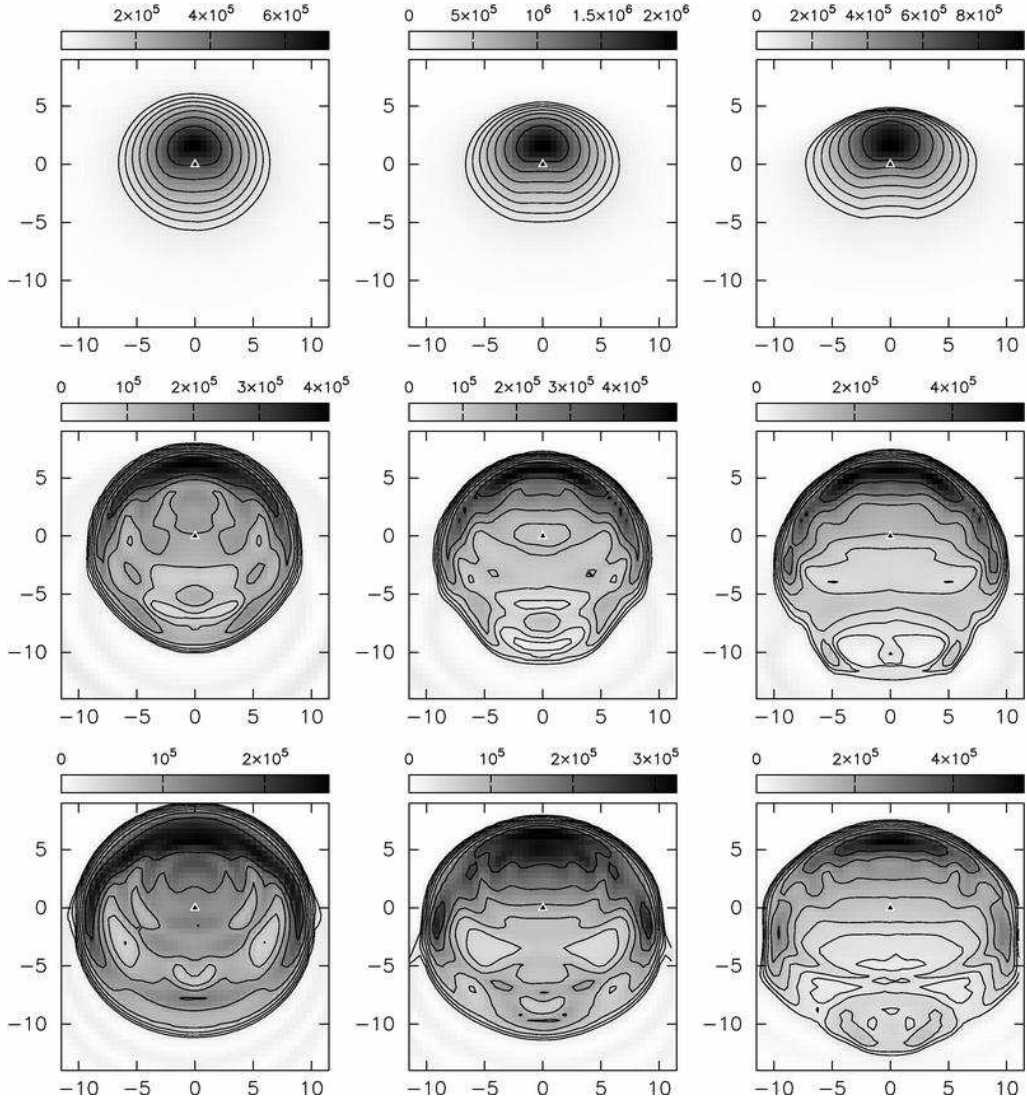}
\caption{Same as Figure~\protect\ref{fig:emissmeasure} but for
Models~D, E, and F after 20,000~years of evolution.}
\label{fig:emissmeasure2}
\end{figure*}
In Figure~\ref{fig:emissmeasure2} we show the emission measure maps at
projection angles of $i = 30\arcdeg$, $45\arcdeg$ and $60\arcdeg$ for
models D, E, and F. (Emission measure maps for other inclination angles are
available from the authors.) Model~D closely resembles the
emission measure one expects from a pure champagne flow, since the
stellar wind is not strong enough ($\dot{M} = 10^{-7}
M_\odot$~yr$^{-1}$) to confine the photoevaporation flow
from the ionization front. Subtle evidence of the wind interaction can
be seen in the contours. There is a lack of emission on the axis and a
faint brightening corresponding to the shell formed by the interaction
between the wind and the champagne flow.

Models~E and F are examples of champagne flows with stellar winds in
shallow density gradients. In these cases, the stellar wind
 is strong enough ($\dot{M} = 10^{-6} M_\odot$~yr$^{-1}$) to confine
the photoevaporated flow to a thick shell and the resultant emission
measure maps show an arclike rather than fanlike morphology. The
shallow density gradient in these models means that the emission
measure does not fall off very quickly towards the tail of the
\ion{H}{2} region and so they appear more shell-like \citep[see, e.g.,][]
{2005ApJ...624L.101D}.
\begin{figure*}
\centering
\includegraphics[width=0.75\textwidth]{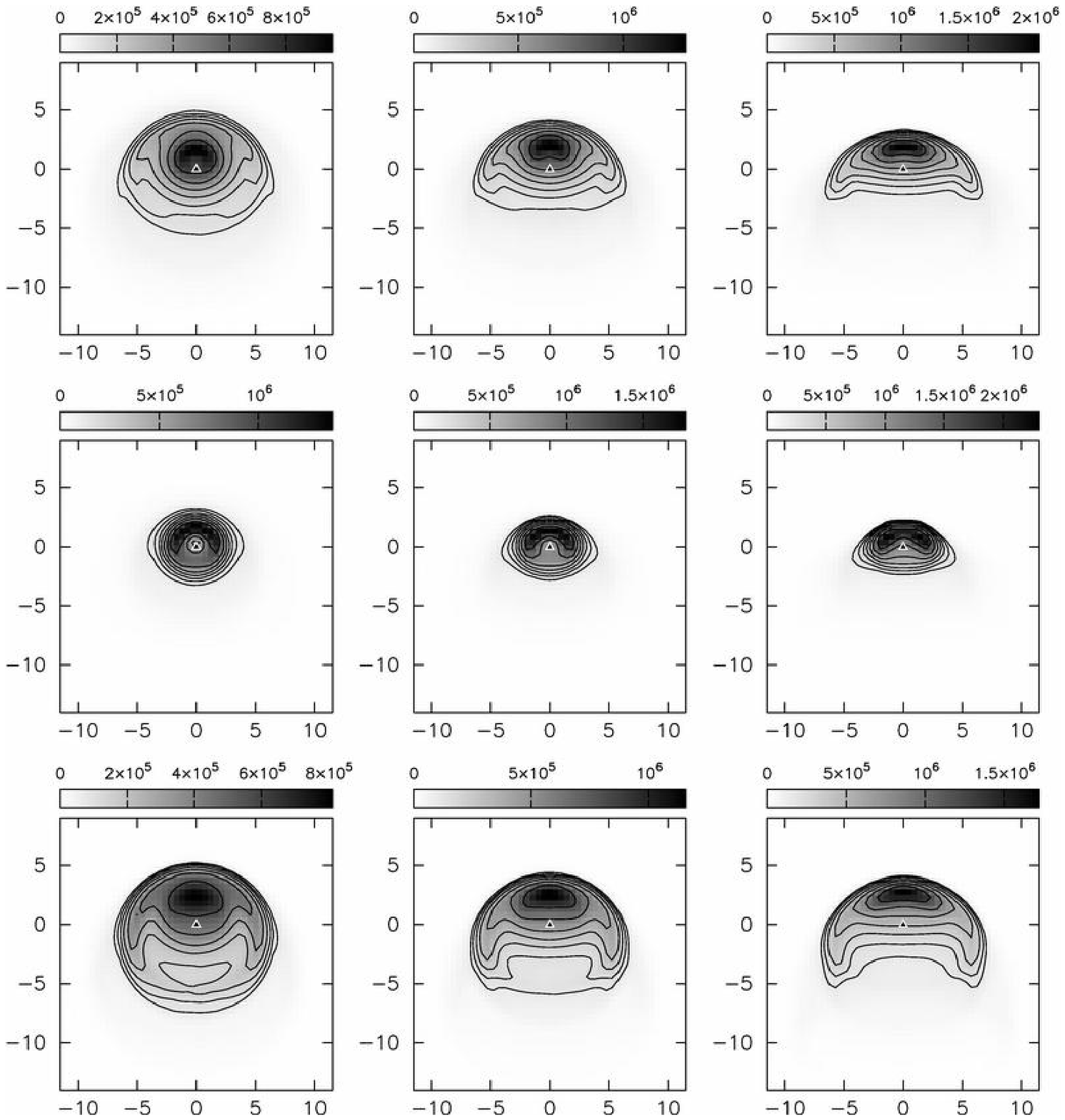}
\caption{Same as Figure~\protect\ref{fig:emissmeasure} but for
Models~G, H, and I after 20,000~years of evolution.}
\label{fig:emissmeasure3}
\end{figure*}
Models G, H, and I all show limb-brightened emission measure maps
(Fig.~\ref{fig:emissmeasure3}), with the limb-brightening being most
enhanced for the largest angle to the line of sight ($i =
60\arcdeg$). Model I, having the shallowest density gradient, has a
``filled shell'' type morphology, whereas Models G and H are
arclike. In Model H the arc of most intense emission is situated much
closer to the star than in the other models, principally because the
star has reached denser gas than in Models~G and I. The ionized shell
in Model~H (see Figure~\ref{fig:ev2x}) is much thinner than in
Models~G and I and this is reflected in the steep drop-off in
intensity both ahead and behind the emission peak.

\subsubsection{Simulated spectra}
\begin{figure*}
\centering
\includegraphics[width=0.45\textwidth]{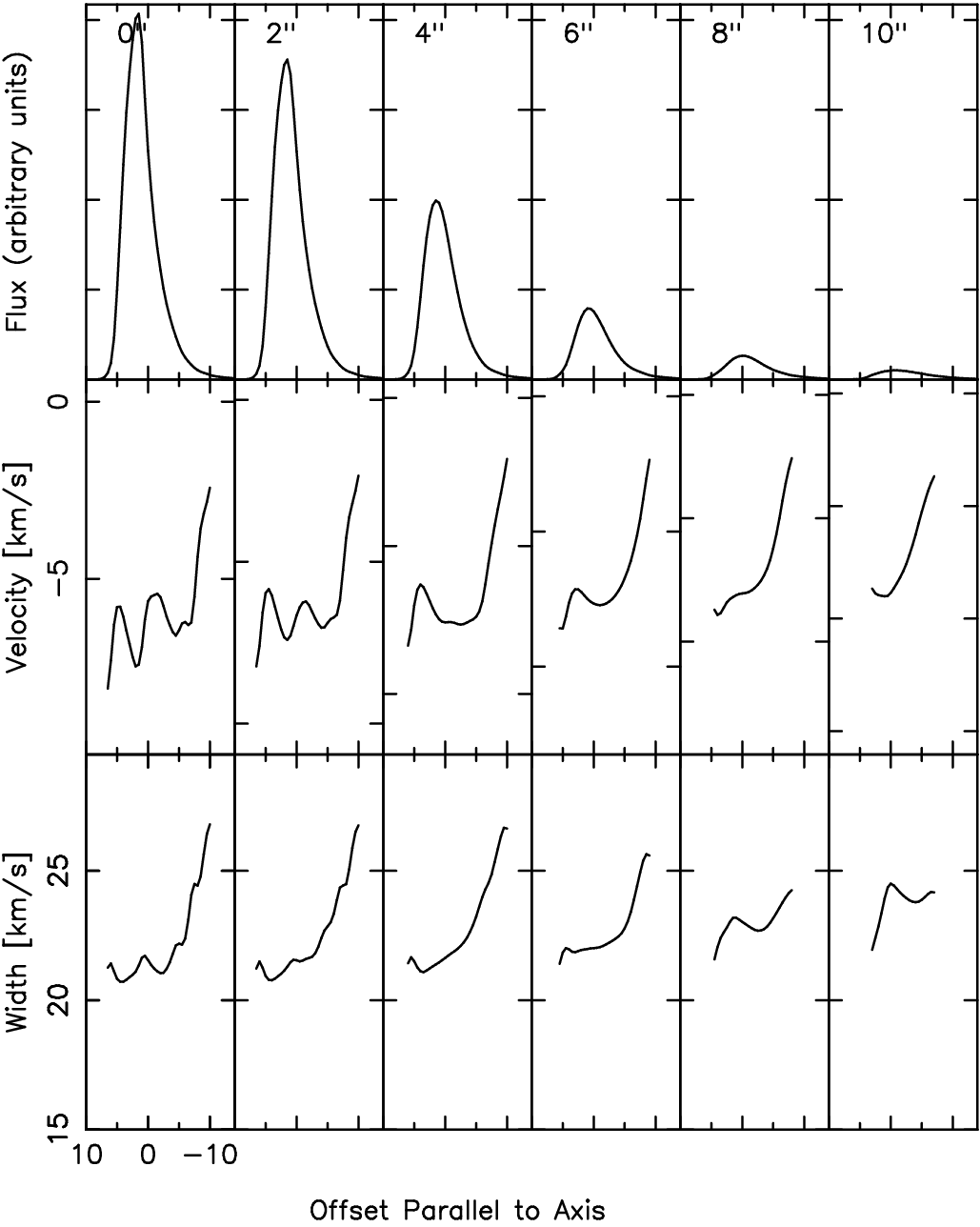}\hfill
\includegraphics[width=0.45\textwidth]{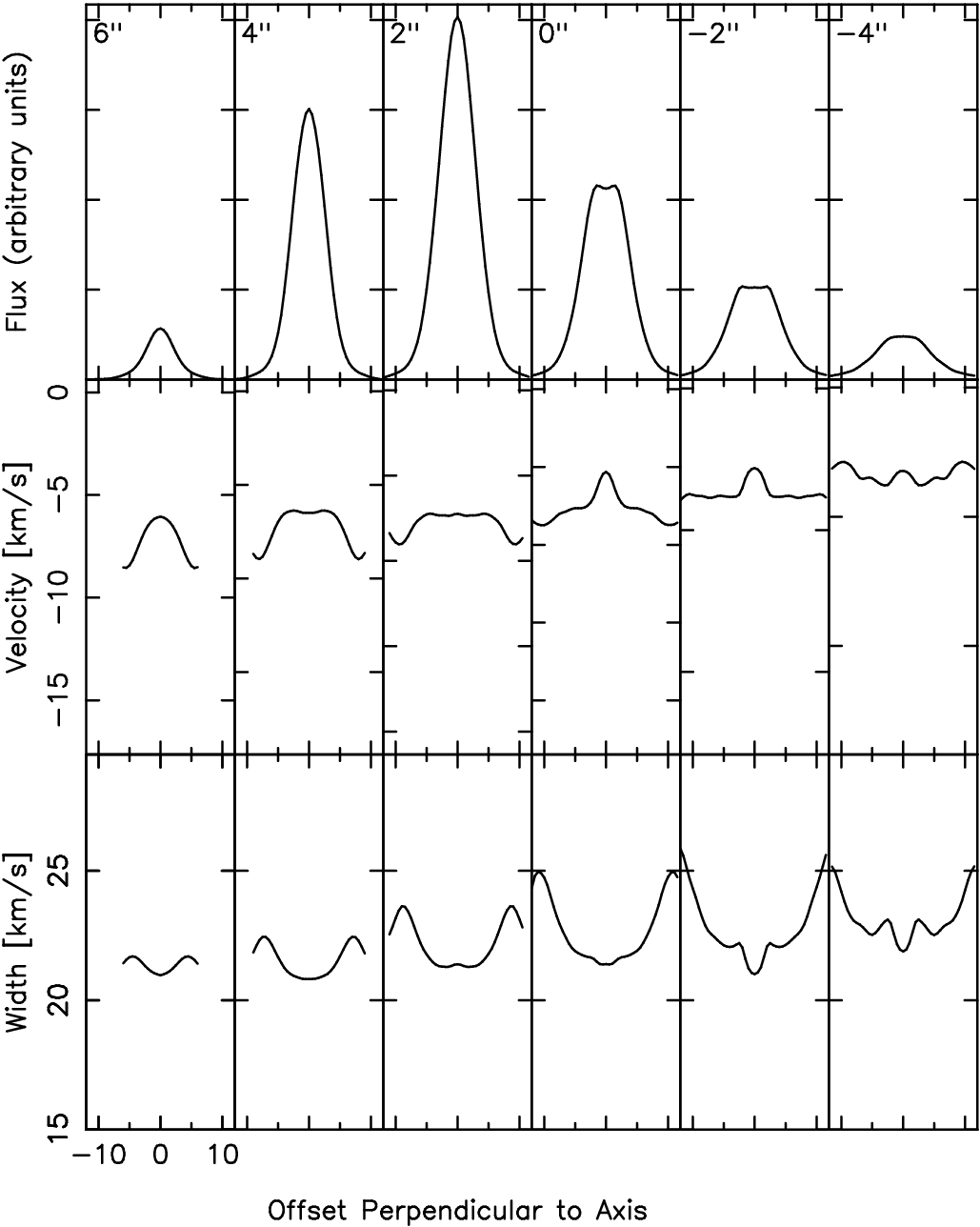}
\caption{Same as Fig.~\protect\ref{fig:specA} but for Model~D.}
\label{fig:specD}
\end{figure*}
\begin{figure*}
\centering
\includegraphics[width=0.45\textwidth]{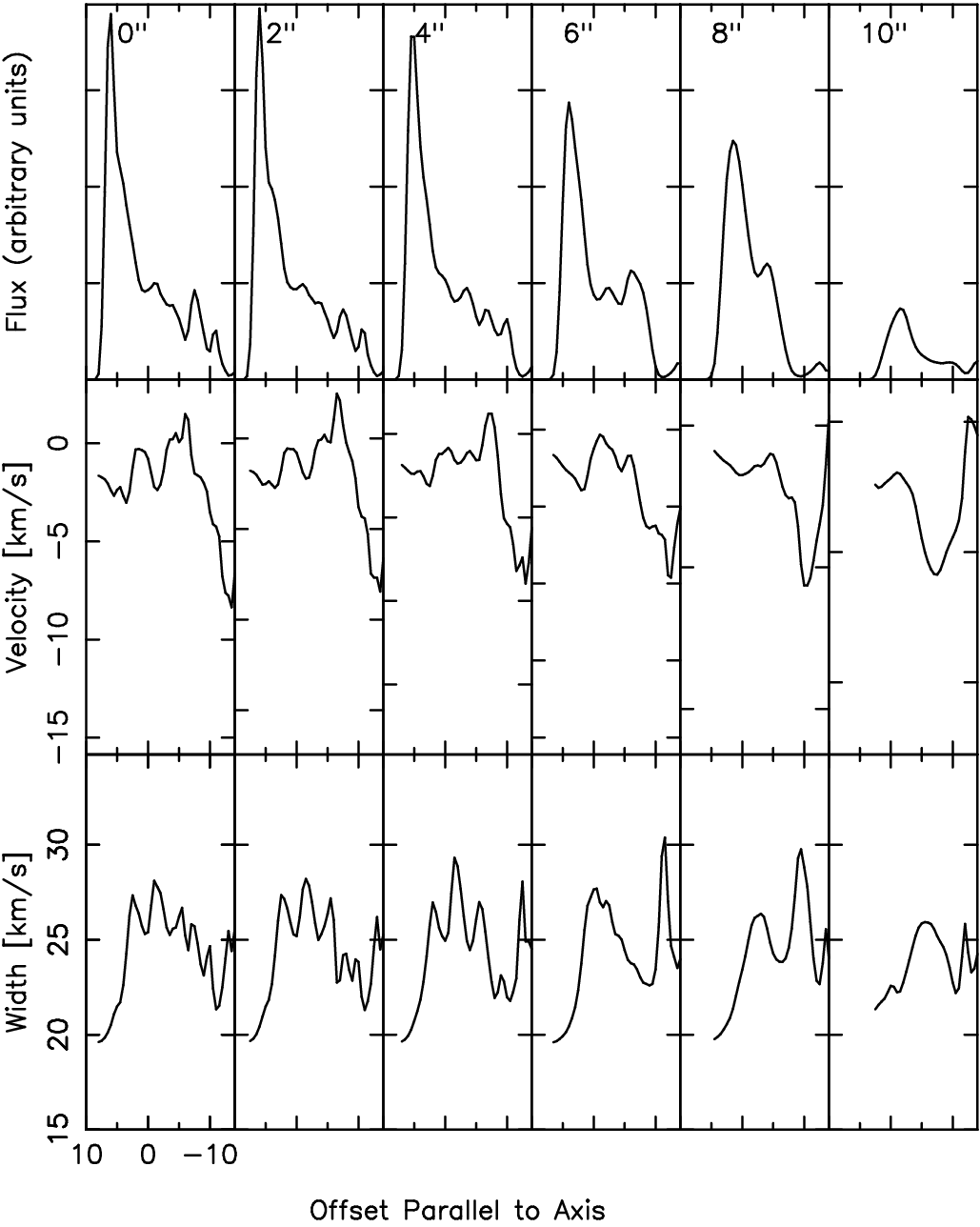}\hfill
\includegraphics[width=0.45\textwidth]{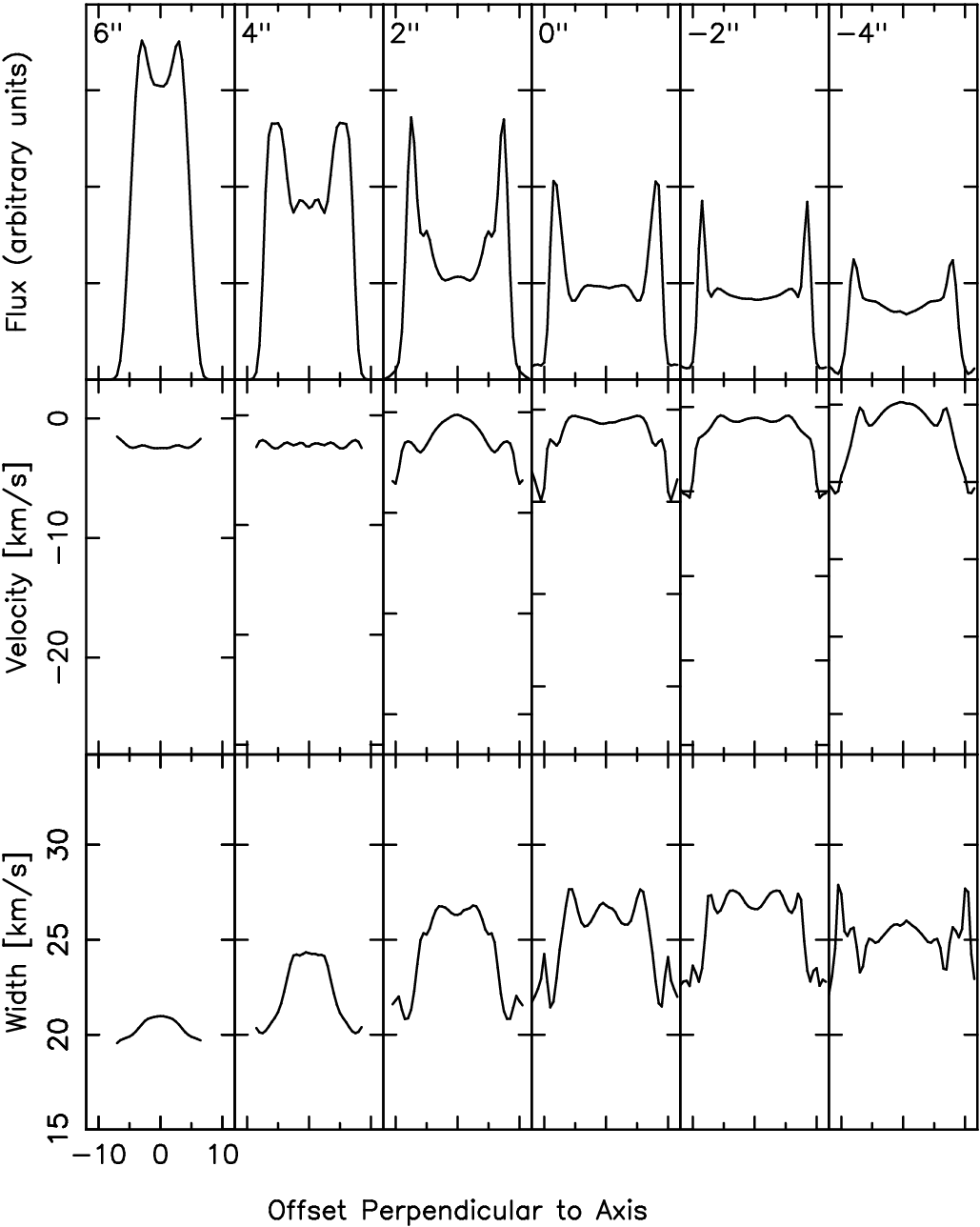}
\caption{Same as Fig.~\protect\ref{fig:specA} but for Model~E.}
\label{fig:specE}
\end{figure*}
\begin{figure*}
\centering
\includegraphics[width=0.45\textwidth]{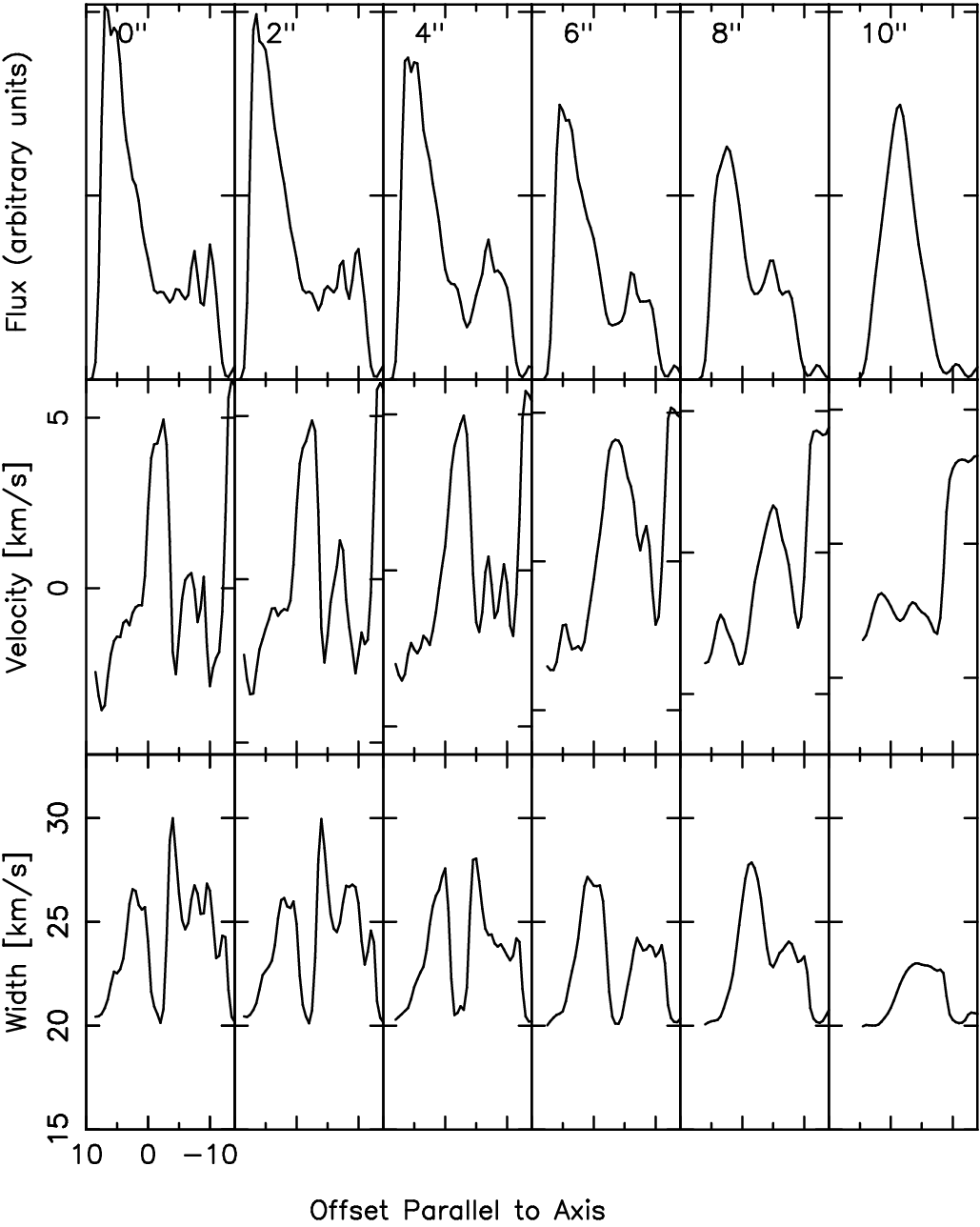}\hfill
\includegraphics[width=0.45\textwidth]{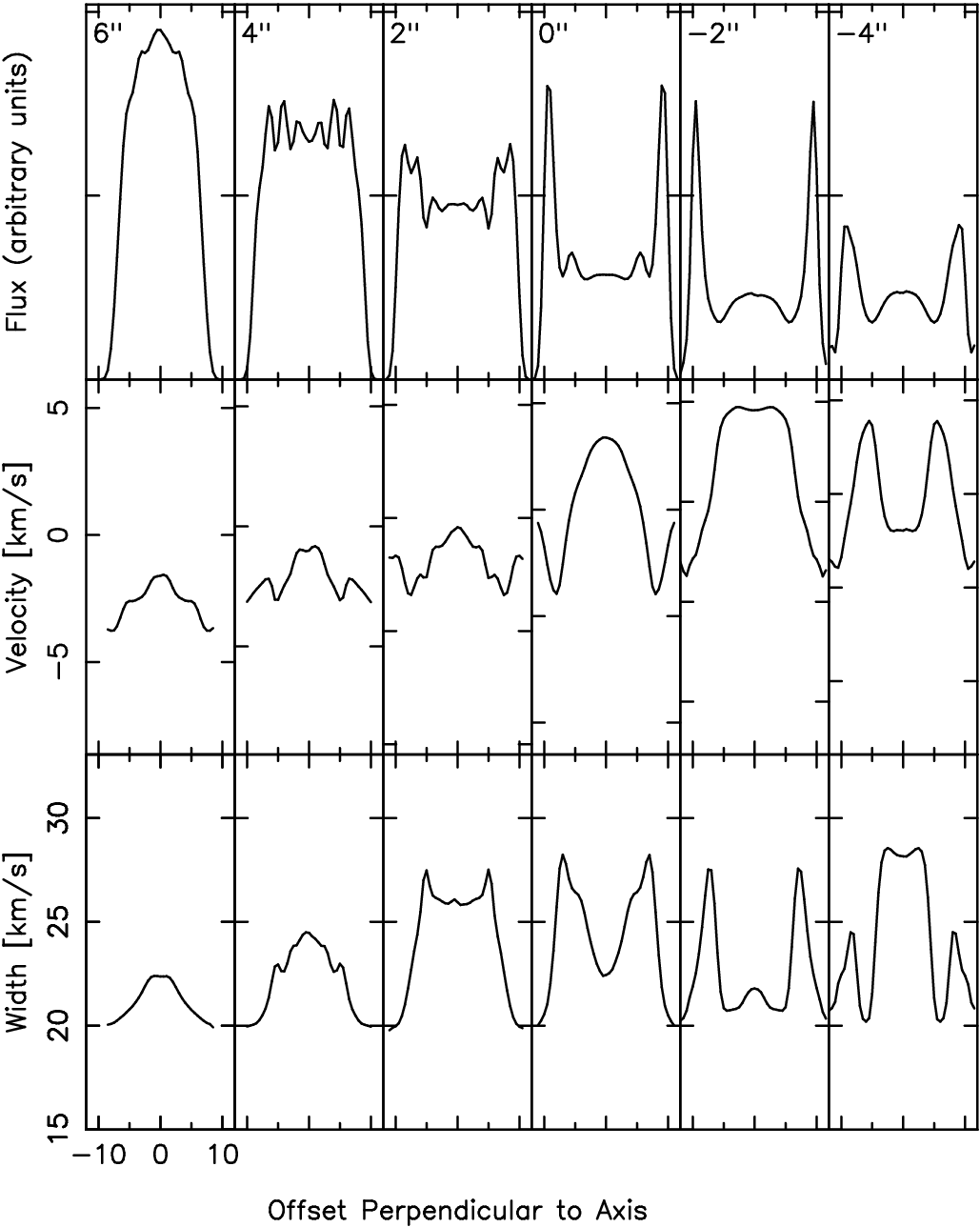}
\caption{Same as Fig.~\protect\ref{fig:specA} but for Model~F.}
\label{fig:specF}
\end{figure*}
The spectra of Model~D shown in Figure~\ref{fig:specD} do not differ too much from the standard
champagne flow spectra (Model~A). In the spectrum of Model~E (Fig.~\ref{fig:specE}) the
flux does not fall off so fast with distance from the head of the
cometary \ion{H}{2} region since the density gradient is shallower. Knots
in the swept-up wind shell cause features in the velocity plot, but
otherwise line-of-sight velocities in the ionized gas are very low ($<
5$~km~s$^{-1}$). The density gradient has little effect on the
dynamics because it is so shallow and the swept-up shell of ambient
material is neither expanding very rapidly nor has important transverse
motions. The spectrum of Model~F (Fig.~\ref{fig:specF}) is very similar, although for offsets
between $-10\arcsec$ and $ -5\arcsec$ the flux, velocity and velocity widths are dominated by
the broken shell which can be seen in Figure~\ref{fig:ev1x}. This
broken shell is a transient feature which will eventually be advected
downstream. 
\begin{figure*}
\centering
\includegraphics[width=0.45\textwidth]{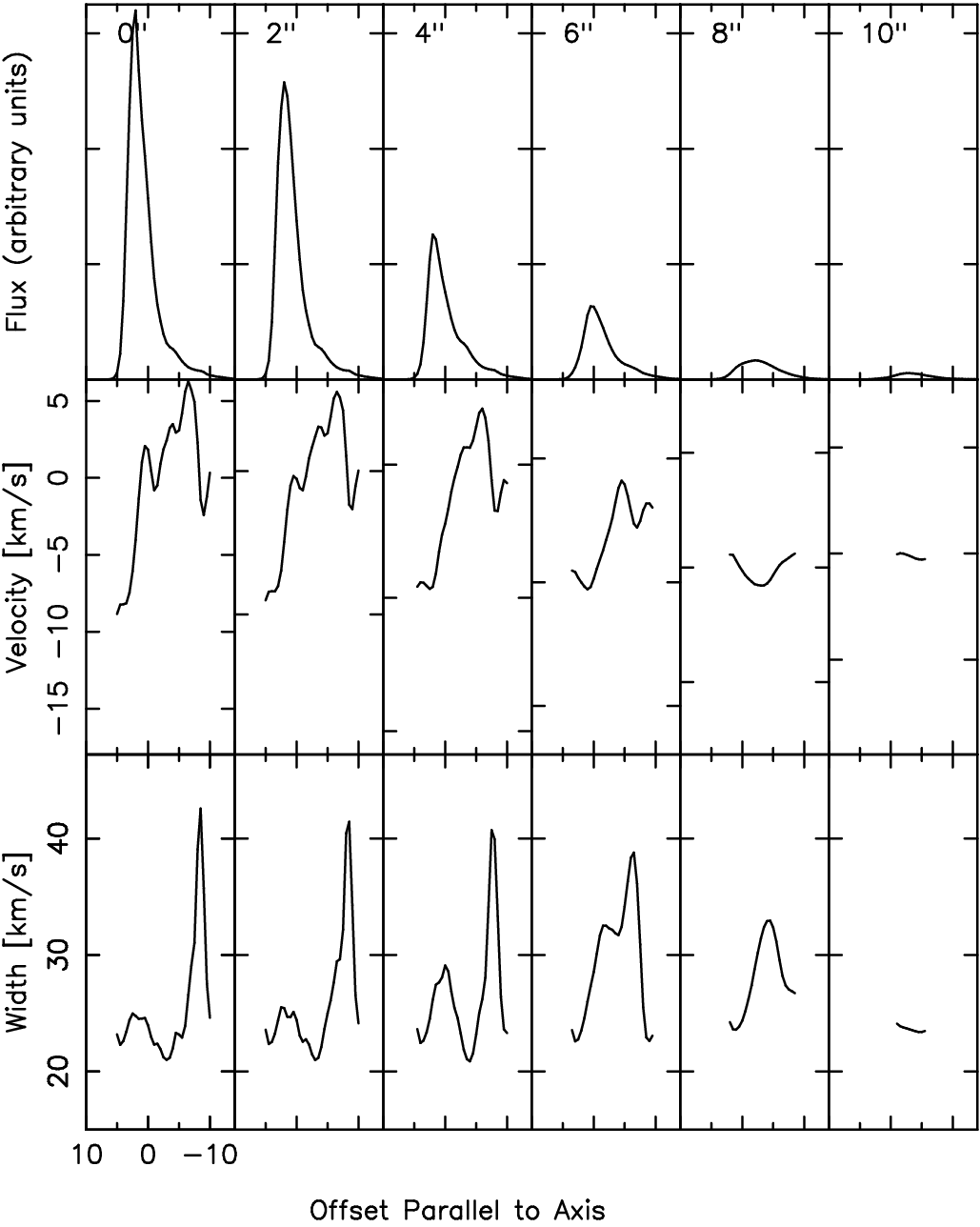}\hfill
\includegraphics[width=0.45\textwidth]{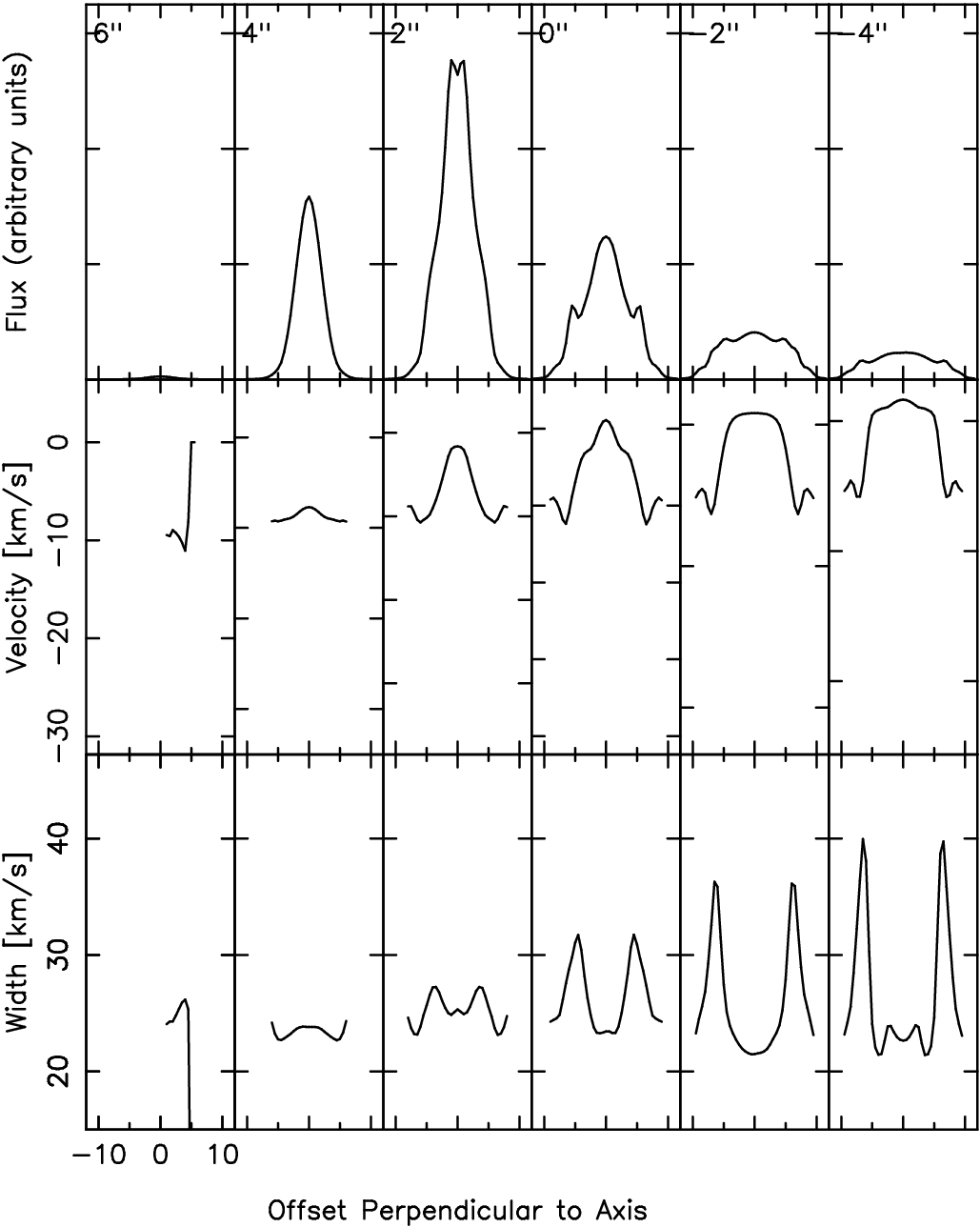}
\caption{Same as Fig.~\protect\ref{fig:specA} but for Model~G.}
\label{fig:specG}
\end{figure*}
\begin{figure*}
\centering
\includegraphics[width=0.45\textwidth]{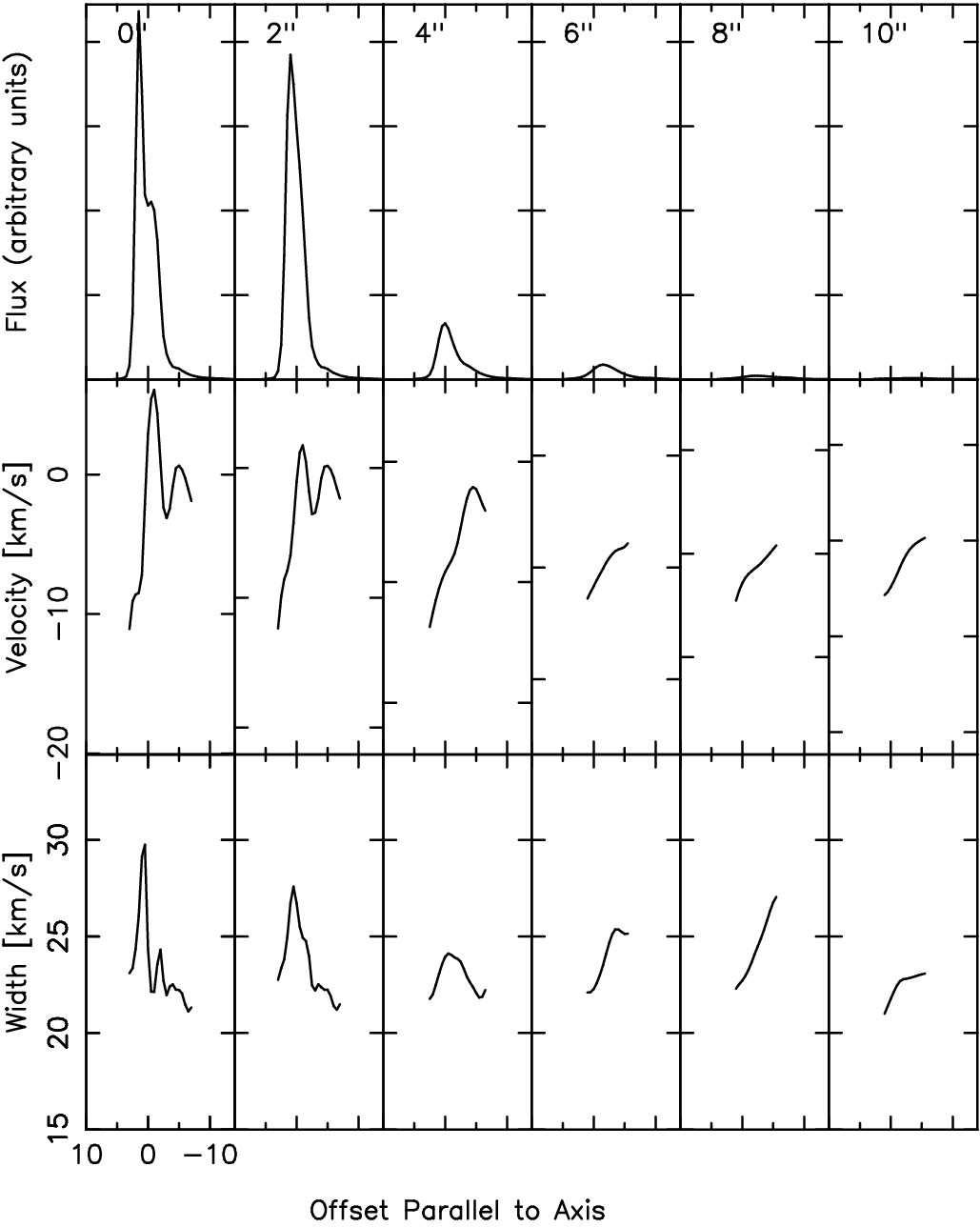}\hfill
\includegraphics[width=0.45\textwidth]{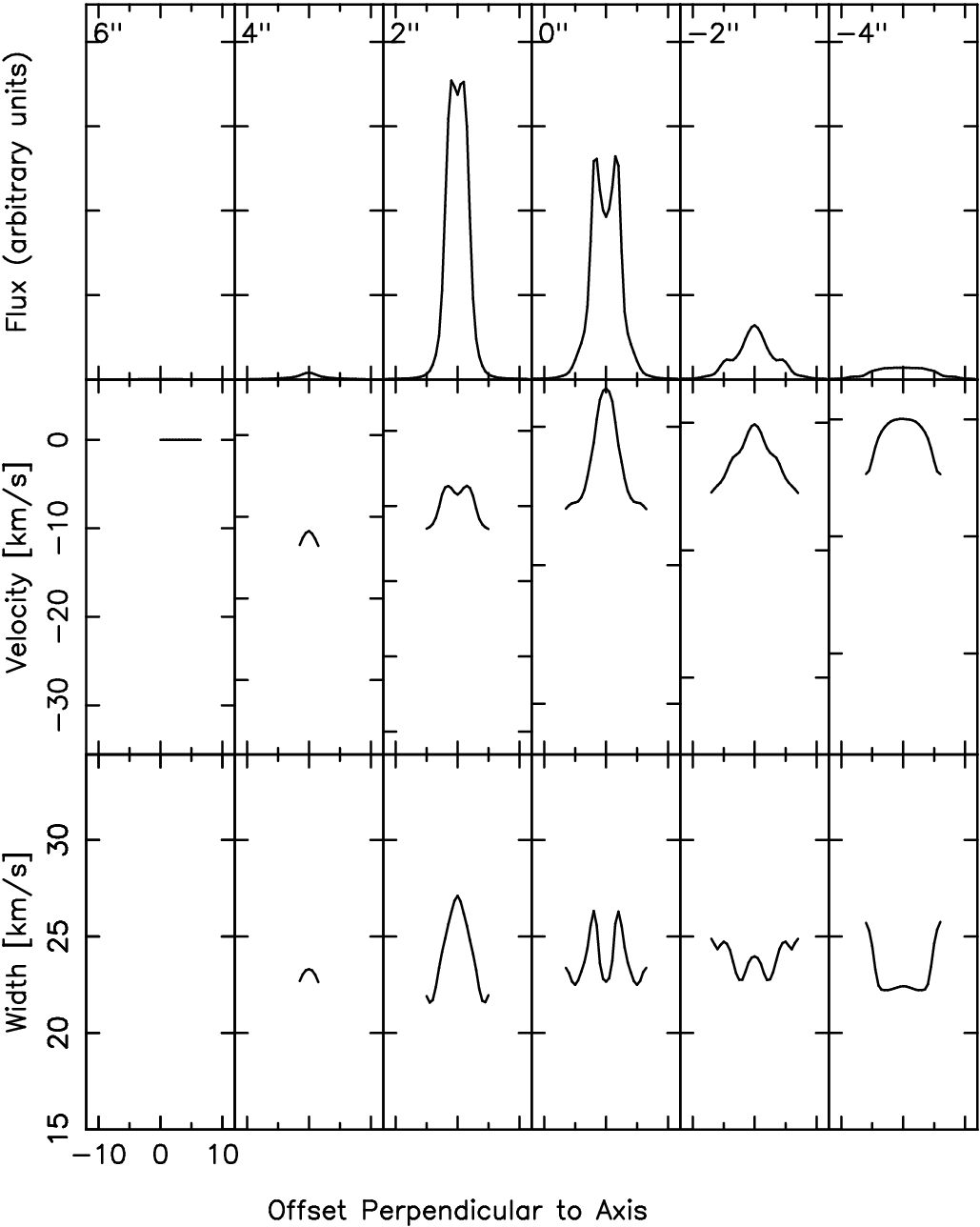}
\caption{Same as Fig.~\protect\ref{fig:specA} but for Model~H.}
\label{fig:specH}
\end{figure*}
\begin{figure*}
\centering
\includegraphics[width=0.45\textwidth]{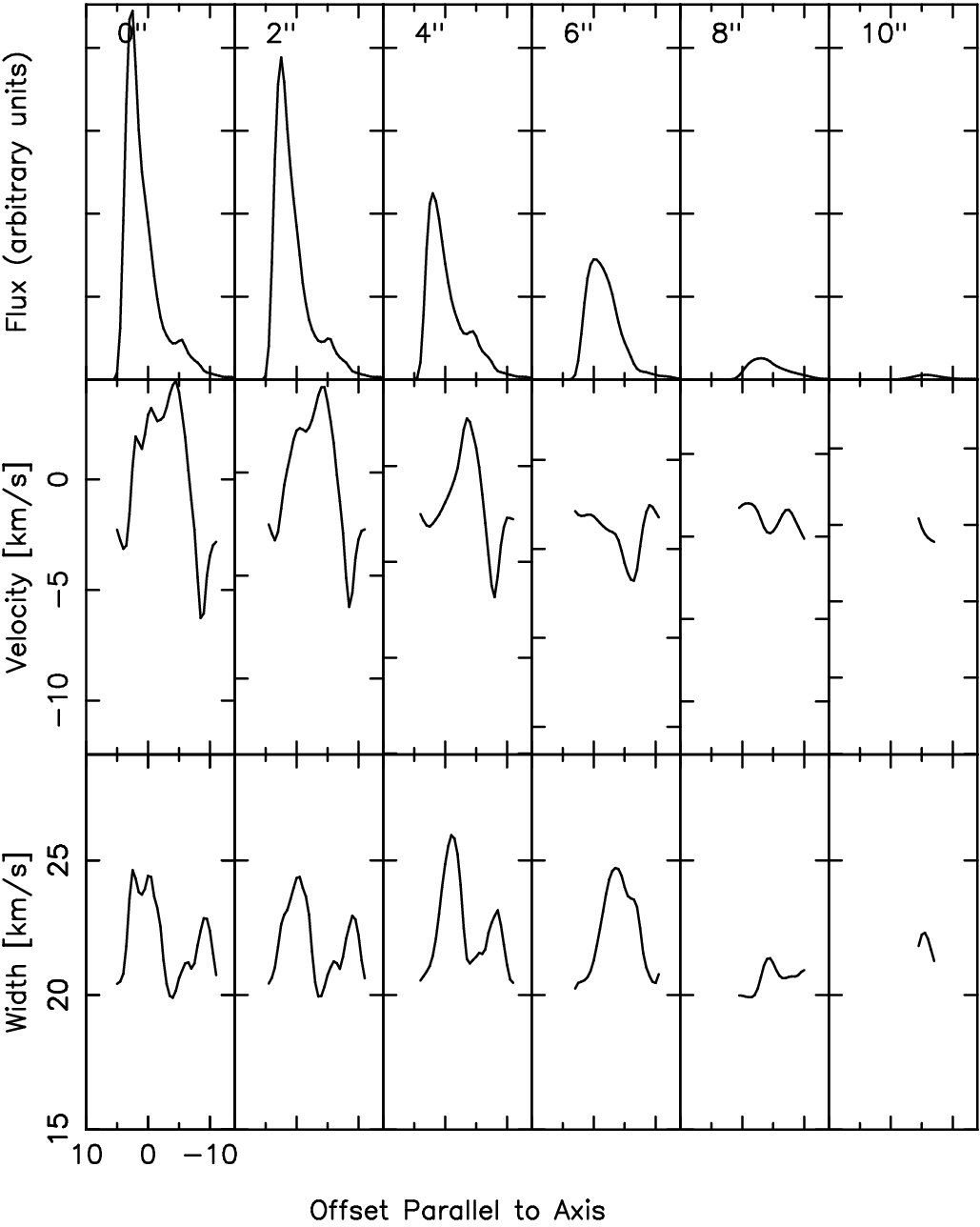}\hfill
\includegraphics[width=0.45\textwidth]{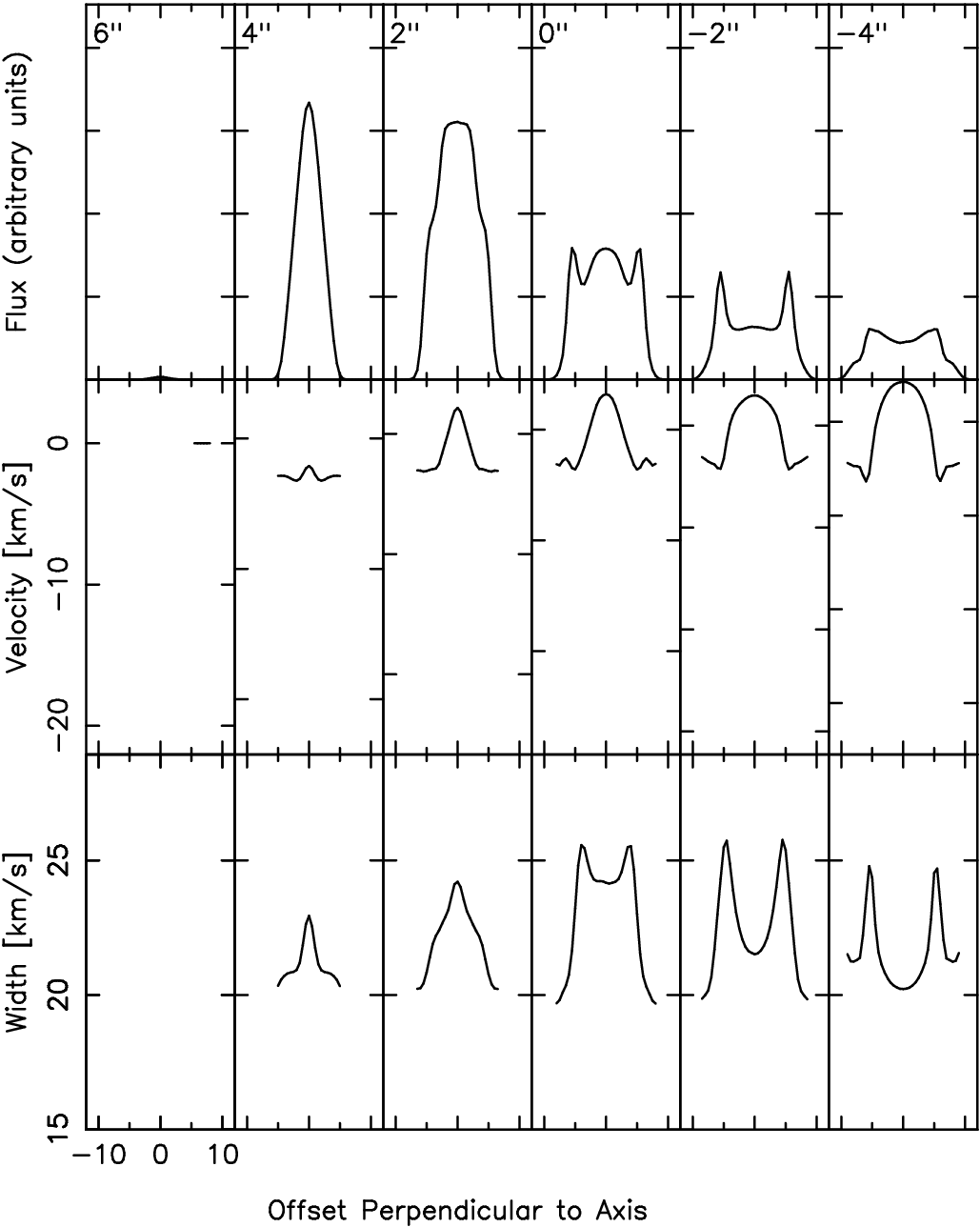}
\caption{Same as Fig.~\protect\ref{fig:specA} but for Model~I.}
\label{fig:specI}
\end{figure*}
All the Models G, H, and I  show blueshifted velocities at the head
of the cometary \ion{H}{2} region changing to  redshifted velocities
once the offset has passed the position of the star, as can be seen in
Figures~\ref{fig:specG}--\ref{fig:specI}. Even though the ionized gas
velocities ahead of the stars in these models are positive (i.e., in
the direction of stellar motion), the projection angle gives more
weight to the gas in the blue-shifted flow in the ionized shell and
champagne flow at the
nearside of the cometary \ion{H}{2} region. The ``noise'' 
seen in these spectra is due to instabilities (eddies) moving through
the ionized layer. Model I, having the shallowest density
gradients, shows the smallest range of velocities ($\sim
10$~km~s$^{-1}$), while models G and H show a range of $\sim
15$~km~s$^{-1}$ even though the velocity of the star is only
5~km~s$^{-1}$ and 10~km~s$^{-1}$ respectively. This is mainly due to
the champagne flow formed in the steep density gradient and is very
similar to what is seen in the case of Model~C
(Fig.~\ref{fig:specC}). 

The velocity widths show varied behaviour. Model~G achieves widths of
$\sim 40$~km~s$^{-1}$ in the wings of the tail. On the other hand, the
highest widths $\sim 30$~km~s$^{-1}$ of Model~H are nearer the
head. Model~I displays a small range of velocity widths, which are
also higher towards the head of the object. Some of the variation in
the line widths is due to eddies being advected through the structure.

\section{Discussion}
\label{sec:discussion}
\subsection{Champagne flow Model}
From Figure~\ref{fig:distance} we can derive approximate empirical
expansion laws for the ionization front and shock into
the neutral material for the photoevaporated champagne flow (Model~A) and champagne flow plus
stellar wind (Model~C) cases. We find that for times $2\times 10^3 < t <
2\times10^4$~yrs a good approximation to the ionization front velocity
for both Models~A and C is
\begin{equation}
	v = 3.46\ t_4^{-0.7} \mbox{ km~s$^{-1}$} ,
\end{equation}
while the shock front expansion for the same temporal range for Model~%
A can be approximated by
\begin{equation}
	v = 4.75\ t_4^{-0.64} \mbox{ km~s$^{-1}$} ,
\end{equation}
and for Model C by 
\begin{equation}
	v = 4.99\ t_4^{-0.64} \mbox{ km~s$^{-1}$} ,
\end{equation}
where $t_4$ is the time, $t/10^4$~yrs.  If we work out the shock velocities
for $t = 2\times 10^4$~yrs, we find $v_s = 3.05$~km~s$^{-1}$ for
Model~A and $v_s = 3.20$~km~s$^{-1}$ for Model~C. The shock velocity
for Model~C is slightly faster than that of Model~A because the
stellar wind shock has overtaken the ionization front shock. Although
these results are specific to the stellar wind parameters and density
distribution (and time), they are broadly consistent with observations
by
\citet{2005ApJ...626..253R} of carbon recombination lines towards the
ultracompact \ion{H}{2} region G35.20--1.74. These observations show
velocities in the neutral gas in the PDR of 2.5~km~s$^{-1}$
into the cloud, while the ionized gas has velocities away
from the cloud, as expected for a photoevaporated flow \citep[see][ for
a detailed discussion of the velocities of the different components
associated with ionization
fronts]{2005ApJ...627..813H}.

The ionization front velocity for both models at 20,000~yrs is
2.13~km~s$^{-1}$. In these models, the front is almost D-critical
 and the ionized gas leaves the
front at about half the sound speed ($v \simeq - 5$~km~s$^{-1}$),
passing through a sonic point ($v \simeq -10$~km~s$^{-1}$) by the time
the ionization fraction becomes
equal to unity \citep{2005ApJ...627..813H}. For more open (i.e., less
concave) ionization fronts the gas leaves the front even faster and
accelerates more rapidly to the sound speed. Hence the net
velocity of the ionized gas at the ionization front is $\sim - 3$~km~s$^{-1}$ in the rest frame
of the cloud and becomes more blueshifted (in this scenario) as it
flows away from the front. A popular misconception in the literature is that the
velocities at the head of a champagne flow should be the same as those
of the molecular cloud.

We note here that the analytic expressions for the shape of the
ionization front in a strong density gradient derived by, for example,
\citet{1980ApJ...236..808I} are not correct since they do not take
into account the motion of the gas, which has the effect of flattening
the density gradient. Thus, the density gradient in the
quasi-stationary champagne flow is not the same as the much steeper
density gradient in the ambient material. The density gradient within
these analytic ionized regions has no physical basis. In our numerical
models, however, the hydrodynamics and radiative transfer are solved
simultaneously, giving a physical result.

The numerical champagne flow models of Tenorio-Tagle and collaborators
\citep{{1979A&A....71...59T},{1979ApJ...233...85B},{1981A&A....98...85B},{1983A&A...127..313Y}},
and more recently by \citet{1997A&A...326.1195C} are for the different
physical setup of a spherical \ion{H}{2} region breaking out from a
dense, uniform cloud into a diffuse intercloud medium, where the
distance of the star from the sharp cloud/intercloud interface is a
determining parameter for the resultant flow. We feel that a more
realistic density distribution would not be discontinuous, as in the
above papers, but would show a smoother transition between high cloud
and low intercloud densities.  Hydrodynamical simulations of the
formation and expansion of \ion{H}{2} regions in disk-like clouds with
stratified density distributions were
done by \citet{1989RMxAA..18...65F} and \citet{1990ApJ...349..126F},
but in these models the star is placed at the midplane, which is a
special place. For this reason, these models do not give rise to
photoevaporated champagne flows. 

In the present work we adopted exponential
density distributions with different scale heights. These
distributions have the advantage that the density depends only on the
height, $z$. We also investigated a spherical density distribution
(Model~F), where the density depends on both $z$ and the cylindrical
radius $r$. However, the results in this case are not qualitatively dissimilar
from those of the exponential distribution with a large scale height
(Model~E). Models with large scale heights show the smallest
departures from idealized \ion{H}{2} regions, while small scale
heights produce the greatest differences.

\subsection{Bowshock Model}
\begin{figure}
\centering
\includegraphics{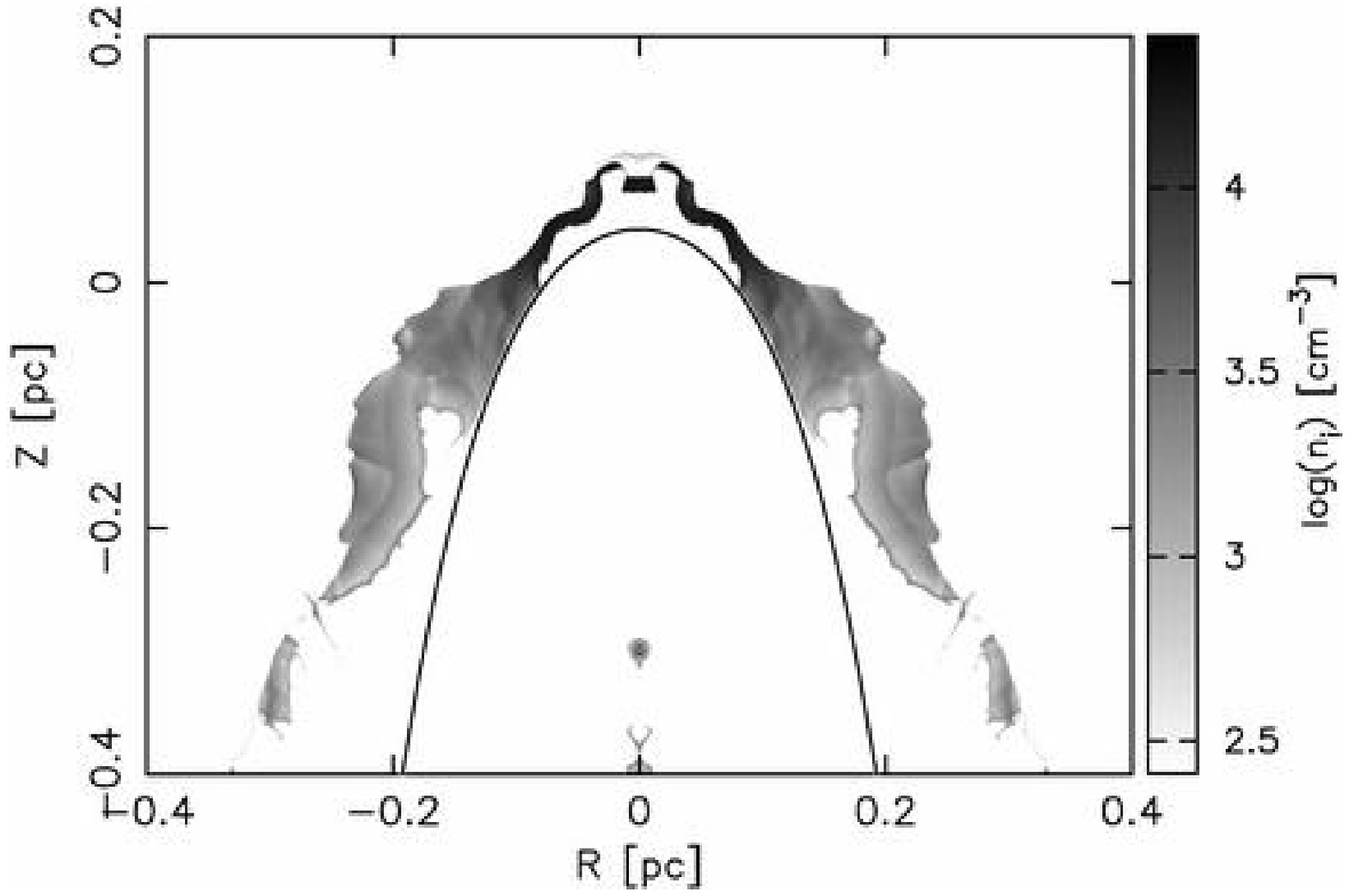}
\caption{Logarithm of ionized number density for the simple bowshock
model (Model B after 20,000~years of evolution) with the analytic solution for the shape of the thin shell from \citet{1996ApJ...459L..31W} overplotted.}
\label{fig:bow}
\end{figure}
\begin{figure}
\centering
\includegraphics[width=0.8\linewidth]{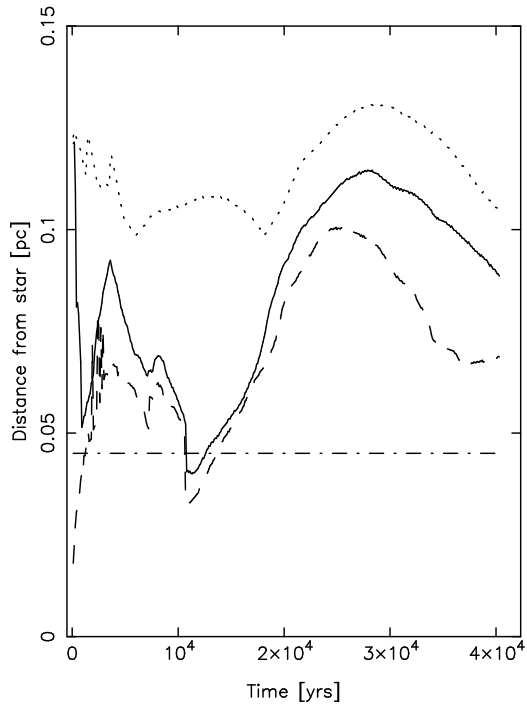}
\caption{Distance of ionization front (solid line), stellar wind
contact discontinuity (dashed line), and outer shock in neutral medium (dotted line) from
the star as a function of time for the bowshock model (Model B). The
dot-dashed horizontal line represents the position of ram-pressure
balance between the stellar wind and the ambient medium at the apex of the bowshock formed by the moving star.}
\label{fig:bowexp}
\end{figure}
In Figure~\ref{fig:bow} we compare our numerical results to the
analytical solution found by \citet{1996ApJ...459L..31W} for stellar
wind bowshocks in the thin-shell limit. The
analytical solution for the shape of the bowshock is given by 
\begin{equation}
	r(\theta) = r_0 \csc \theta \sqrt{3\left(1 - \theta\cot \theta\right)}
\end{equation}
where $r_0 = (\dot{M_w}v_w/4\pi\rho_a v_*^2)^{1/2}$ is the stand-off
distance, $\dot{M_w}$, $v_w$ are the stellar wind mass loss rate and
velocity respectively, $\rho_a$ is the ambient density and $v_*$ is
the stellar velocity. The angle $\theta$ is measured with respect to
the star from the symmetry axis. This model assumes that both the
shocked stellar wind and the swept-up shell of ambient material are
thin, i.e., that cooling is efficient in both shocked zones. The shape
of the bowshock should be compared to that of the contact
discontinuity between the shocked stellar wind material and the
swept-up ambient medium.

The numerical result differs from the analytical result for a variety
of reasons. The main reason is that the key assumption of the
analytical result is that cooling in the shocked stellar wind is
instantaneous, making this region thin. In the present models the
stellar wind shocks and a region of high temperature forms between the
free-flowing wind and the swept-up ambient medium. The thickness of
this subsonic region varies as the evolution proceeds since pressure
balance across the contact discontinuity means that it is sensitive to
what is happening behind the outer shock. 

Additionally, in the numerical model the ionizing photons and
stellar wind switch on at the same time. Thus, a Str\"omgren sphere
forms around the star almost instantaneously, while the stellar wind
bubble takes much longer to form. The stellar wind sweeps up a shell
of material, which then cools on a timescale dependent on the
postshock temperature and density.  As the
hydrodynamic flow develops, the initial photoionized region starts to expand,
preceded by a shock wave, that sets the neutral gas ahead of the
photoionized region in motion. It is into this moving medium that the
stellar wind bubble expands. The impact of the hypersonic stellar wind
on the surrounding medium sets up the familiar two shock pattern
\citep{1997pism.book.....D}, where the outer shock sweeps up the
ambient medium. This swept-up shell forms initially within the
photoionized region. However, due to the increased opacity, the
ionization front retreats until it is contained with the shell. At
this point, the shock ahead of the original ionization front is still expanding
into the ambient medium, although the previously ionized gas outside
the shocked shell now recombines. The swept-up shell expands into this moving
neutral gas and thickens. Instabilities are formed in the cooling shell, which
lead to turbulent motions in the subsonic region between the outer
shock and the inner stellar wind shock. The shell is initially
thick, and only starts to become thinner once the outer stellar wind
shock has overtaken the shock produced by the initial ionization
front. The ionization front moves outwards within the stellar wind
shell and the pressure of the photoionized gas is an additional
confinement on the shocked stellar wind. Pressure gradients across the bowshock structure and transverse
to it mean that the ram pressure balance across thin shells is inadequate to describe
the evolution of the structure at these early times ($t <
40,000$~yrs).

In Figure~\ref{fig:bowexp} the distance of the ionization front and
stellar wind contact discontinuity and the outer shock in the frame of
the star are plotted for the bowshock
model. We also indicate in this figure the position of theoretical ram
pressure balance. In this figure we see how the initial ionization
front retreats into the swept-up shell, while the shock that was sent
out into the neutral gas travels for quite a time at roughly the stellar
velocity (20~km~s$^{-1}$) before being finally overtaken by the outer stellar
wind shock. Once the ionization front is trapped in the shell, it remains
there for the rest of the evolution. The noise in the position of the
contact discontinuity occurs when instabilities form in the subsonic region
between the shocks. The first peak in the positions of the ionization
front and contact discontinuity, at about $t \sim 4000$~yrs,
corresponds to the first peak in Figure~\ref{fig:eddies}, interpreted
as the first large eddy to be shed by the bowshock. The thickness of
the shocked stellar wind region changes in response to pressure
changes across the subsonic region, either due to the formation of
instabilities or dynamics in the photoionized part of the shell.

Interestingly, the position of the contact discontinuity in
Figure~\ref{fig:bowexp} is settling down (after 40,000~yrs) to a value
close to $\approx 1.5\ r_0$, which was predicted by
\citet{1988ApJ...329L..93V} from a simple model approximating the
dynamics in the shocked wind by sonic flow in a hemispheric region
between the stellar wind shock and the contact discontinuity. However,
\citet{1990ApJ...353..570V} argued that cooling via conduction in the
shocked stellar wind would be so rapid that this layer would be very
thin and its dynamics governed by momentum conservation. We do not
include thermal conduction in the present models but suspect that it
will not play as important a role as invoked by
\citet{1990ApJ...353..570V}. The importance of conduction fronts for
the cooling of stellar wind bubbles has been cast into doubt by the
discovery of diffuse X-ray emission from high-mass star-forming
regions, which suggests that the shocked stellar winds inside
\ion{H}{2} regions do not cool efficiently \citep{2005IAUS..227..297T}. 
It is probable that conduction is greatly reduced by the presence of
magnetic fields. Our results show that instabilities at early times
in the shocked wind layer do enhance cooling and affect its
thickness. However, this is only a transient feature, and once the
initial instabilities have been advected away from the bowshock head,
the shocked stellar wind region is able to expand.

\subsection{Champagne flow plus stellar motion}\nobreak
The density gradient plays a key role as to whether the dominant flow
characteristic is champagne-flow-like or bowshock-like. In the steeper
density gradient (Models~G and H), a champagne flow sets up first,
within which the stellar wind bubble develops, as for
Model~C. However, stellar motion ensures that the outer stellar wind shock
rapidly overtakes the shock ahead of the ionization front. The
redshifted velocity of the ionized gas ahead of the star is less than it would be for a bowshock in
a neutral medium because of the photoevaporated flow set up by the
density gradient, which is blueshifted. The density
gradient in the neutral material ahead of the moving star increases
the ram pressure acting on the stellar wind and the photoionized
shell. Both the shocked stellar wind zone and the ionized shell become
thinner ahead of the star as it travels up the density gradient, with
the shocked stellar wind zone almost disappearing. This is more
pronounced in Model~H, where the faster-moving star reaches higher
density regions than the slower star of Model~G.  Because
the shocked stellar wind region ahead of the star becomes extremely thin, it is
essentially the ram pressure of
the free-flowing wind which balances that of the stellar motion through the
ambient medium. At the sides of the structure, a champagne flow is
still in evidence and the pressure of the photoionized gas confines
the stellar wind (see Figure~\ref{fig:ev2x}).

In the shallow density gradient (Model~I), on the other hand, the swept-up shell
formed by the stellar wind confines the ionization front before a
champagne flow has chance to form. The outer stellar wind shock soon
overtakes the shock ahead of the ionization front. The stellar wind
bubble structure expands laterally downstream where the densities are
lower and the swept-up shell becomes quite thick here. However, this
never becomes a champagne flow (see, Figure~\ref{fig:ev2y}). Ahead of the
star the shocked stellar wind zone becomes thinner as the star
advances up the density gradient, until it almost disappears and the
ram pressure of the free-flowing wind balances that of the stellar
motion through the ambient medium. The photoionized region ahead of
the star is quite thick.

Although the stellar velocity in Model~G is only 5~km~s$^{-1}$, the
projected velocity range shown in Figure~\ref{fig:specG} is $\sim
15$~km~s$^{-1}$, with the velocities becoming redshifted towards the tail where the
champagne flow is important. The actual velocities at the head of the
object are slightly redshifted compared to the background cloud, but the projected velocities at the head of
the object are blueshifted, unlike in the simple bowshock case (see
Figure~\ref{fig:specB}). This shows that the pressure gradients within
the shell due to the density distribution play an important
role in advecting photoionized material in the shell at the head of
the object round to the sides. Also, in projection we are seeing the
near part of the champagne flow component, which has a strong
influence on the projected velocities. The linewidths of this model also rise
sharply towards the tail, reaching 40~km~s$^{-1}$, suggestive of champagne flow. Model~H also
presents a projected velocity range of $\sim 15$~km~s$^{-1}$ for a
stellar speed of 10~km~s$^{-1}$. The peak redshifted velocities in this model are
just downstream of the stellar position, and the peak linewidths are just
ahead of it, which is more in keeping with a bowshock model. On the
other hand, the velocities ahead of the star are blueshifted, as in
Figure~\ref{fig:specG} for Model~G, and this must be due to the
influence of the density gradient. The velocities then become
blueshifted again far down the tail.  

The stellar speed in Model~I is 10~km~s$^{-1}$ and the ionized gas velocities
in the model ahead of the star are low and redshifted. However, even the shallow
density gradient in this model ensures that the projected velocities
ahead of the stellar position are blueshifted (see
Figure~\ref{fig:specI}) and become redshifted in the tail before
turning blueshifted again. The full
velocity range for this model is only about 8~km~s$^{-1}$, principally
due to the shallow density gradient. The linewidths are highest around
the projected stellar position.

These combined models show that stellar motion in a density gradient 
has important differences to the standard bowshock model. In
particular, the velocity range is larger than that for the case of
stellar motion only (e.g., Model~B) and the radial velocities change sign along the
length of the object. These models are not hugely different from the champagne flow plus
stellar wind cases, except that since there is effectively no shocked stellar wind
zone ahead of the star (at least in the cases we have studied), then
the peak emission occurs closer to star (see
Figures~\ref{fig:emissmeasure} and \ref{fig:emissmeasure2}). 

The intention of Models G, H and I is to investigate what happens to
the dynamics when density gradients, stellar winds and stellar motion
are combined. In particular, we have not been overly concerned that
the stars in our models start from low density regions. However, we
could postulate a scenario in which stars, under the influence of the
gravitational field of a molecular cloud, move in and out of local
density peaks, and that our models represent a particular stage in
this picture. Another possibility is that an external shock triggers
star formation in the outer parts of a cloud and that, once formed,
the massive star continues to coast in towards the cloud center, or is
attracted to the center by the gravity of the main cloud core. In any
case, the low density material is quickly advected off grid and
replaced by either photoevaporated flow material (Models G and H) or
material advected round the photoionized shell from the region of the
head by the pressure gradient (Model I). In either case, the details
of the initial environment are quickly lost by the model as it
advances up the density gradient.

\citet{2005arXiv.0508467.F} have recently studied the 
evolution of \ion{H}{2} regions in dense,
structured molecular clouds where the star starts from a dense core
and travels down the density gradient. They do not include the effect of stellar
winds or discuss the dynamics of the ionized gas. The velocities of the
stars  vary between 2 and 12~km~s$^{-1}$ (i.e., subsonic and
supersonic with respect to the ionized gas). It is found that
for slow moving stars a champagne flow dominates, while for the faster
moving stars a ``bow'' shape is formed when the star overtakes the
ionized gas in expansion, since the gas can only expand at the
sound speed and the star is moving supersonically. In the high
density regions the star has left behind the gas recombines. This is an interesting
scenario but unfortunately the authors do not include any analysis of
the kinematics. It is probable that the inclusion of a stellar wind will
modify the faster case, since the stellar velocity is not supersonic
with respect to the stellar wind and the direction of motion is
towards decreasing densities, so the ram pressure of the ambient
medium is decreasing (or constant). In our models, the ram pressure
increases as the star moves up the density gradient and so it is quite
possible that the star overtakes the swept-up wind shell with its
trapped ionization front if the evolution is followed for long enough.

\subsection{Effect of Eddies}
\begin{figure}
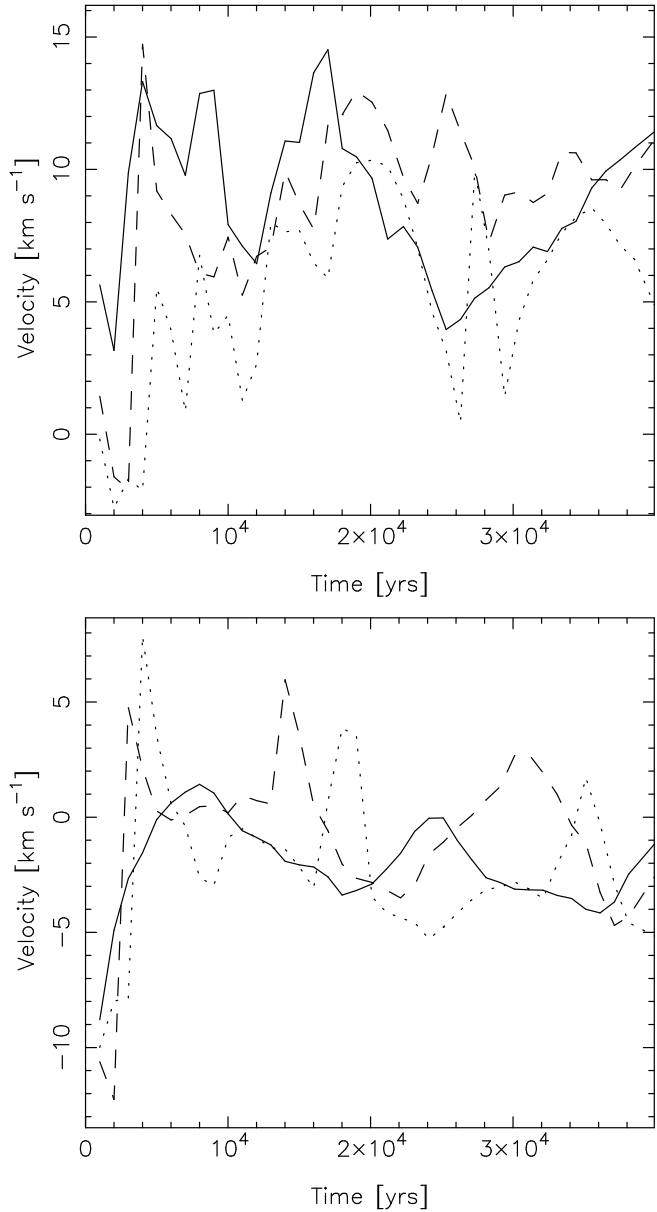

\centering
\includegraphics{f24a}\\[\medskipamount]
\includegraphics{f24b}
\caption{Time evolution of projected velocities at half peak flux (solid line),
quarter peak flux (dashed line) and one eighth peak flux (dotted line)
for inclination angle $i = 45\arcdeg$: Top panel---Model~B (bowshock);
Bottom panel---Model~C (champagne flow plus stellar wind).}
\label{fig:eddies}
\end{figure}
In Figure~\ref{fig:eddies} we examine the evolution of the
line-of-sight velocity at half, quarter and one eighth of the peak
flux along the symmetry axis of the numerical model for an inclination
angle of $i = 45\arcdeg$. The large peaks in these curves correspond
to eddies being shed from the head of the cometary \ion{H}{2} region
and travelling downstream, hence there is a phase lag between peaks on
each line. In the champagne flow plus stellar wind case an ``average''
profile can be discerned for each curve. However, in the bowshock case
the curves are far too noisy. It is not possible to draw
any conclusions from these plots to aid as a diagnostic tool for
discriminating between bowshock and champagne flow plus stellar wind
models. We note that in the bowshock case there appears to be a slight trend for
the velocities further down the slit to rise with time, while in the
champagne flow plus stellar wind case there is a trend for all
velocities to become slowly more blueshifted with time.

The spectra obtained by, e.g.,
\citet{{1996ApJ...464..272L},{1999MNRAS.305..701L}} are not as noisy
as our theoretical spectra. This could partially be due to the
resolution of the observations, but it also suggests that the eddies
we see forming in our simulations are dissipated more quickly in
reality. In order to investigate this possibility numerically, we
would have to incorporate a turbulence model, such as that outlined by
\citet{1994MNRAS.269..607F}, in our numerical scheme. This is beyond the scope of
the present paper.

\subsection{Comparison with observations}
\subsubsection{G29.96$-$0.02}
G29.92--0.02 is, perhaps, the most well studied of all cometary
ultracompact \ion{H}{2} regions. It was first observed by
\citet{1989ApJS...69..831W} at radio wavelengths. The morphology of
this region is that of a bright arc of emission at the head trailing
off into a low-surface brightness tail. The background cloud systemic
velocity is best determined from molecular line observations, which trace the bulk of the
gas. Both single dish \citep{{1992A&A...253..541C},{2003ApJ...596..362K}} and
interferometric \citep{{1999ApJ...517..799P},{2003A&A...407..225O}} observations of
optically thin CO and CS lines give velocities of 97--98~km~s$^{-1}$.

The infrared observations of
\citet{{1996ApJ...464..272L},{1999MNRAS.305..701L}},
\citet{2003A&A...405..175M}  and \citet{2005ApJ...631..381Z} provide the most
detailed kinematic results of G29.96$-$0.02 with which to compare our numerical
simulations. \citet{1996ApJ...464..272L} and \citet{2003A&A...405..175M} 
obtained long slit spectra of the Br$\gamma$ hydrogen recombination
line, while
\citet{2005ApJ...631..381Z} obtained
mid-infrared observations of the [\ion{Ne}{2}] 12.8~$\mu$m
fine-structure line. These fine-structure line studies are
particularly promising because the FWHM line width due to thermal
broadening of Ne$^+$ at $10^4$~K is only $\sim 4.8$~km~s$^{-1}$,
whereas that of H$^+$ is $\sim 21$~km~s$^{-1}$, hence the kinematics
of the ionized gas can be separated from the thermal and turbulent
broadening contributions if the velocity resolution of the
observations is good enough (as long as the Ne$^+$ ion is present in
the whole volume).

These detailed observational studies find a radial velocity range of $\sim
-15$ to $\sim +10$~km~s$^{-1}$ for the ionized gas with respect to the
molecular cloud (assuming a background cloud velocity of
98~km~s$^{-1}$), with the highest blue-shifted velocities ahead of the
emission peak of the object, red-shifted velocities behind the
emission peak and blue-shifted velocities again in the tail. The line
widths of the Br$\gamma$ data are highest at the head, as are the
[\ion{Ne}{2}] line widths. In the past, this object has been modeled
as a bowshock \ion{H}{2} region, with an inferred stellar motion of
20~km~s$^{-1}$ in order to give the velocity range
\citep[e.g.,][]{{1994ApJ...437..697A},{1996ApJ...464..272L},{2003A&A...405..175M},{2005ApJ...631..381Z}}.
However, for a slit placed along the symmetry axis of the cometary
\ion{H}{2} region, simple bowshock models give either wholly
redshifted or wholly blueshifted velocities with respect to the
background cloud, depending on the angle of inclination to the line of
sight. For this reason, several authors
\citep{{1996ApJ...464..272L},{1999MNRAS.305..701L},{2003A&A...405..175M},{2005ApJ...631..381Z}}
have argued that a simple bowshock model cannot reproduce the observed
velocity behavior.  Although a stellar velocity of 20~km~s$^{-1}$ can
give the observed velocity range, it cannot reproduce the velocity
offset with respect to the molecular cloud, which requires a lower
stellar velocity. On the other hand, simple champagne flow models
would not give a limb-brightened morphology, nor reproduce the
velocity and linewidth behavior.
\citet{1996ApJ...464..272L} suggested that
stellar winds and density gradients might be important for the
kinematics of this object, while \citet{2005ApJ...631..381Z} argued
that pressure gradients within the parabolic shell are needed to
increase the velocity range.

Our numerical models comprising a star with a stellar wind in a
density gradient (e.g., Models~C, G, H and I) can all reproduce the
blueshifted to redshifted to blueshifted velocity behavior exhibited
by G29.96$-$0.02, together with a strong, arc-like emission at the
head, assuming an inclination angle of $45\arcdeg$ to the line of
sight. However, the maximum velocity range our models achieve is only
$\sim 15$~km~s$^{-1}$ before the flux drops below 1\% of the peak
value, as against the $\sim 20$ to 25~km~s$^{-1}$ seen in the
observations. The velocity range increases as the density gradient
becomes steeper, so it is conceivable that a steeper gradient than the
ones we have investigated would give the observed velocity range. The
spatial variation of the flux of Br$\gamma$ along the symmetry axis in
the high resolution observations of
\citet{2003A&A...405..175M} bears a striking resemblance to that of
our Model~I, as does the variation of the velocity centroid of the
same line along the slit (ignoring the blue-shifted shoulder seen in
the observations, which appears to be unconnected with the \ion{H}{2}
region and is probably due to motions associated with the nearby hot core
source). Even the ``knee'' in the observed
Br$\gamma$ flux, also present in the [\ion{Ne}{2}] observations of
\citet{2005ApJ...631..381Z}, 
is seen in the model (where it is due to the line of sight cutting
through thick, dense regions of ionized shell on both sides of the
star), although this is serendipitous. Models~C, G and H, however, have a larger
velocity range than Model~I. The linewidths in Models~H and I are highest at the head of
the object for this slit, although not ahead of the peak emission, as
seen in the observations by
\citet{{1996ApJ...464..272L},{1999MNRAS.305..701L}} and
\citet{2003A&A...405..175M}. The large linewidths seen in Model~I are
most likely due to instabilities in the photoionized shell and are
thus transient. The linewidths in Model~H, on the other hand, appear
to be the result of the general flow pattern. 

The H$_2$ observed in this object comes from the region outside of the
ionization front. However, it is not likely to come from the quiescent
background cloud, rather it  arises in the region between the
shock ahead of the ionization front in the neutral medium and the
ionization front, probably due to fluorescence, and thus will have a red-shifted velocity with
respect to the cloud and the ionized gas. This is suggested by the
observations reported by \citet{2003A&A...405..175M}, where the H$_2$
is redshifted by $\sim 2$~km~s$^{-1}$ with respect to the background
cloud velocity as determined from molecular line observations. Our models
show that a shock front advances slowly into the neutral medium with
velocities of order a few km~s$^{-1}$, which is consistent with this
velocity redshift.

The parameters of the numerical models have not been tailored to fit
G29.96$-$0.02, but the qualitative agreement is encouraging.  We have
shown that models including a strong stellar wind and a density
gradient are able to reproduce the morphology, velocity behavior, and
to some extent the line-width behavior of this object. Stellar motion
appears to modify the line-width behavior of an essentially champagne-type
flow and so we suggest that G29.96$-$0.02 could be the \ion{H}{2} region
formed by  a star with a strong stellar wind in a density
gradient, possibly with the star moving up the gradient. 

\subsubsection{Hydrogen radio recombination line studies}
A detailed examination of H92$\alpha$ radio recombination line
observations of three cometary \ion{H}{2} regions was presented by 
\citet{1994ApJ...429..268G}. They found velocity ranges of up to
12~km~s$^{-1}$  between the head and tail regions of these
objects. Two of these objects were interpreted as champagne flows,
mainly because the velocities at the head were approximately equal  to
the background cloud velocities as measured from observations of
NH$_3$. The third object was interpreted as a bowshock flow since the
velocities were all redshifted with respect to the background
cloud. However, even in a simple photoevaporated champagne flow model such as that
presented in Model~A, we would expect
to find a velocity shift between the ionized gas at peak emission and
the background cloud (unless the object is in the plane of the sky),
since the gas leaves the ionization front with a velocity about half
the sound speed and then accelerates  (see, e.g., Figure~\ref{fig:specA}). Also,
in the champagne flow objects, the linewidth variations
between head and tail are not entirely consistent with the champagne
flow interpretation: in one object the linewidths decrease towards the
tail and in the other they are roughly
constant. 

\citet{1994ApJ...429..268G} discuss the possibility that the
stellar wind modifies the flow, but their analytic treatment requires
the star to be at the center of a radially decreasing density
distribution. In our models we are able to place the star and its
stellar wind at arbitrary points in density distributions with steep
and shallow gradients. We find that the combination of a stellar wind
and a density gradient changes the linewidth behavior, since the
ionized flow is forced to flow around the outside of the stellar wind
bubble (e.g., Models~C, E and F).

\citet{2003ApJ...596..344C} have recently reported hydrogen
recombination line radio observations of
dual cometary \ion{H}{2} regions in DR~21. The radial velocities of
the ionized gas suggest a bowshock interpretation since they are
higher at the head. However, for one of
the \ion{H}{2} regions, it is reported elsewhere
\citep{1989A&A...222..247R} that velocities down the tail become
increasingly redshifted with respect to the background cloud, while
velocities at the head are blueshifted. Furthermore, the linewidths
for both objects suggest that there is an increase towards the tail,
which is not expected for a bowshock model. Models incorporating
stellar winds and density gradients, such as our Models~C, F and G,
can demonstrate this spatial variation in velocity and linewidth. The
``hybrid'' picture developed by \citet{2003ApJ...596..344C} to explain
the motions they see in DR~21 has some similarities with our models.

From observations of the neutral, atomic
and molecular gas associated with the cometary \ion{H}{2} region
G111.67$+$0.37, \citet{2001ApJ...560..806L}  find that both the
neutral and ionized gas exhibit velocity gradients from the head to
the tail of the cometary region, but that the neutral gas has a lower
velocity range than the ionized gas and is detected principally
towards the tail where the PDR is thickest. Our numerical models  
also demonstrate this behavior. Although we do not show results for the
neutral gas here, it will be the subject of a future paper. 

\section{Conclusions}
\label{sec:conclusion}
We have performed numerical simulations of cometary \ion{H}{2} regions forming
in media with density gradients and where the ionizing star may
additionally have either or both a strong stellar wind and a supersonic motion with
respect to the neutral gas. Analysis of our results and a comparison
with observations from the literature show that:
\begin{enumerate}
\item Photoevaporated champagne flow models with stellar winds can have
limb-brightened morphologies, which simple champagne flows
cannot. This is because the stellar wind bubble acts as an obstacle to
the photoevaporated flow, which must divert around it.
\item Depending on the steepness of the density gradient, the
morphologies of the champagne flow plus stellar wind models can appear
more limb-brightened (steep gradient) or more shell-like (shallow
density gradient).
\item In the absence of thermal conduction, numerical bowshock
simulations (in uniform media) are not well described by momentum-conserving, ram
pressure balance analytical models. This is because the shocked
stellar wind does not cool efficiently and its thermal pressure must also be
taken into account.
\item In both champagne type models and bowshock models a shock runs ahead
of the ionization front in the neutral material. Typical velocities in
the neutral gas behind these shocks are a few km~s$^{-1}$. These
velocities could influence H$_2$ and carbon recombination line
measurements of the systemic cloud velocity, since this emission is
most likely to come from the photon dominated region, which is a thin
region just outside the ionization front and thus wholly contained in
the zone behind the shock in the neutral gas.
\item Models in which a star with a strong stellar wind advances
supersonically up a density gradient are champagne-flow dominated if
the gradient is steep enough and stellar wind shell dominated if the
density gradient is shallow. These models produce limb-brightened
\ion{H}{2} regions. In these models, the shocked stellar wind region
is extremely thin due to the increasing ram pressure as the star
advances up the density gradient, and the fact that the shocked wind
material is quickly advected away from the region of the head due to
the density gradient.
\item Models of stars with stellar winds moving in density gradients
result in the peak emission being closer to the star than for similar
models without stellar motion because the shocked stellar wind zone is
extremely thin and so the photoionized shell is closer to the star.
\item The radial velocity range of $\sim 20$~km~s$^{-1}$ seen in observations,
coupled with a limb-brightened morphology, can be reproduced either by a
fast-moving $\sim 20$~km~s$^{-1}$ star or by a star with a stellar
wind in a steep density gradient. The key observational discriminant
is the radial velocity pattern along the object with
respect to the ambient material. For an inclination angle of $i =
45\arcdeg$ to the line of sight, for example, the velocities of the
bowshock model will all be redshifted with respect to the ambient
cloud, whereas the champagne flow plus stellar wind model gives
blueshifted velocities at the head, redshifted velocities around the
stellar position, and blueshifted velocities again in the
tail. Similar velocity behavior is seen in the models that also
include stellar motion, showing that for the modest stellar velocities
considered (5 and 10~km~s$^{-1}$), the density gradient is more
important than the star's speed for determining the velocities in the
photoionized shell.
\item The linewidths in simple champagne flow models are highest in the tail
of the cometary \ion{H}{2} region, while those of bowshock models are
highest ahead of the star. In our models combining density gradients,
stellar winds and stellar motion we find that the champagne flow
dominated flows (slow-moving star in steep density gradient) have
larger velocity widths towards the tail, while for faster moving stars
in shallow density gradients the largest line widths occur further up
the cometary structure at the stellar position. Such intermediate
linewidth information is also reported in observations.
\end{enumerate}
Our results show that the most likely explanation for the observed
morphologies and kinematics of cometary compact \ion{H}{2} regions is
the outcome of ionizing stars with strong stellar winds in density
gradients, possibly with the star moving up the gradient. It is
important to obtain a reliable determination of the systemic cloud
velocity in order to be able to discriminate between possible models.

\acknowledgments
We thank Will Henney for valuable comments and a critical reading of
the manuscript. SJA would like to acknowledge support from
DGAPA-PAPIIT grants IN115202 and IN112006-3, and for financial support
from DGAPA-UNAM during a year's sabbatical visit to the University of
Leeds. Some of the numerical simulations in this paper were performed
on a linux cluster at CRyA funded by CONACyT grant 36571-E to Enrique
V\'azquez-Semadeni. We thank the anonymous referee for constructive
comments which improved the presentation of this paper. This work has
made extensive use of NASA's Astrophysics Abstract Data Service and
the astro-ph archive.

\end{document}